\documentclass[a4paper,11pt]{article}
\pdfoutput=1
\usepackage{jcappub}
\usepackage{journal_abbrev}
\usepackage{xparse,verbatim,color}
\usepackage{xspace,ulem,etoolbox,cprotect}
\xspaceaddexceptions{]\}}
	\usepackage{bm}
	\usepackage{hyperref}

\usepackage{graphicx}
\usepackage{adjustbox}
\usepackage{siunitx}
\usepackage{enumitem}
\usepackage{lmodern}

\makeatletter
\newlength{\apb@width}
\newcommand{\autoparbox}[2][c]{\settowidth{\apb@width}{#2}\parbox[#1]{\apb@width}{#2}}
\newcommand{\includegraphicsbox}[2][]{\autoparbox{\includegraphics[#1]{#2}}}
\makeatother

\newcommand{\sect}[1]{Sec.~\ref{#1}\xspace}

\newcommand{\app}[1]{Appendix~\ref{#1}\xspace}

\DeclareDocumentCommand{\eqs}{m m m o o}{%
	\IfNoValueTF {#4} {%
		Eqs.~(\ref{#1}){\xspace #2} (\ref{#3})%
	}{%
		Eqs.~(\ref{#1}){\xspace #2} (\ref{#3}){\xspace #4} (\ref{#5})%
	}%
}

\newcommand{\equ}[2][]{\begin{equation}\label{#1}#2\end{equation}}

\newcommand{\se}{\ensuremath{\sigma_\mathrm{8}}\xspace}
\newcommand{\AS}{\mathcal{A}_{s}}
\newcommand{\hse}{\ensuremath{\hat{\sigma}_\mathrm{8}}\xspace}

\newcommand{\specialcell}[2][c]{%
	\begin{tabular}[#1]{@{}c@{}}#2\end{tabular}}

\newcommand{\pprof}{\ensuremath{P^\mathrm{prof}}\xspace}

\newenvironment{referee}{\bf}{}
\newcommand{\bref}{\begin{referee}}
\newcommand{\eref}{\end{referee}}

\def\ba#1\ea{\begin{align}#1\end{align}}
\def\bea{\begin{eqnarray}}
\def\eea{\end{eqnarray}}
\def\be{\begin{equation}}
\def\ee{\end{equation}}
\def\d{\delta}
\def\s{\sigma}
\def\({\left(}
\def\){\right)}
\def\[{\left[}
\def\]{\right]}
\def\<{\left\langle}
\def\>{\right\rangle}
\def\lapl{\nabla^2}

\newcommand{\perm}[1]{ \expandafter\ifstrempty\expandafter{#1} {\mbox{perm.}} {\mbox{$#1$ perm.}} }

\def\thH{\Theta_\text{H}}

\def\nbarh{\bar{n}_h}

\newcommand{\vs}{\nonumber\\}
\def\d{{\delta}}
\def\eps{{\varepsilon}}

\def\NLO{\text{NLO}}
\def\otd{\text{td}}

\renewcommand{\v}[1]{\bm{#1}}

\def\vx{\v{x}}

\def\vk{\v{k}}

\def\vp{\v{p}}

\def\knl{k_\text{NL}}
\def\blapl{b_{\lapl\d}}

\def\bt{b^{h}}

\def\dirac#1{\delta_{\mathrm{D}}^{(#1)}}

\def\fsum#1{\sum_{#1\neq 0}^{k_{\rm max}}}

\def\Mpch{\,h^{-1}\text{Mpc}}
\def\iMpch{\,h\,\text{Mpc}^{-1}}
\def\Msunh{\,h^{-1} M_\odot}
\def\Plin{P_\text{L}}

\def\L{\Lambda}
\def\Lin{\Lambda_\text{in}}
\def\kmax{k_\text{max}}

\newcommand{\refeq}[1]{Eq.~(\ref{eq:#1})}
\newcommand{\refeqs}[2]{Eqs.~(\ref{eq:#1})--(\ref{eq:#2})}
\newcommand{\reffig}[1]{Fig.~\ref{fig:#1}}

\newcommand{\reftab}[1]{Tab.~\ref{tab:#1}}
\newcommand{\refsec}[1]{Sec.~\ref{sec:#1}}

\newcommand{\refapp}[1]{Appendix~\ref{app:#1}}

\def\emph#1{\textit{#1}}

\title{\emph{Unbiased} Cosmology Inference from Biased Tracers using the EFT~Likelihood}

\author[a]{Fabian Schmidt,}
\author[a]{Giovanni Cabass,}
\author[b]{Jens Jasche,}
\author[c]{and Guilhem~Lavaux}

\emailAdd{fabians@mpa-garching.mpg.de}
\emailAdd{gcabass@mpa-garching.mpg.de}
\emailAdd{jens.jasche@fysik.su.se}
\emailAdd{guilhem.lavaux@iap.fr}

\affiliation[a]{Max--Planck--Institut f\"ur Astrophysik, Karl--Schwarzschild--Stra\ss e 1, 85748 Garching, Germany}
\affiliation[b]{The Oskar Klein Centre, Department of Physics, Stockholm University, Albanova University Center, SE 106 91 Stockholm, Sweden}
\affiliation[c]{Sorbonne Universit\'e, CNRS, UMR 7095, Institut d'Astrophysique de Paris, 98 bis bd Arago, 75014 Paris, France}


\arxivnumber{2004.06707}

\abstract{We present updates on the cosmology inference using the effective field theory (EFT) likelihood presented previously in Schmidt et al., 2018, Elsner et al., 2019 \cite{paperI,paperII}. Specifically, we add a cutoff to the initial conditions that serve as starting point for the matter forward model. We show that this cutoff, which was not employed in any previous related work, is important to regularize loop integrals that otherwise involve small-scale, non-perturbative modes. We then present results on the inferred value of the linear power spectrum normalization \se from rest-frame halo catalogs using both second- and third-order bias expansions, imposing uniform priors on all bias parameters. Due to the perfect bias-\se degeneracy at linear order, constraints on \se rely entirely on nonlinear information. The results show the expected convergence behavior when lowering the cutoff in wavenumber, $\L$. When including modes up to $k \leq \Lambda = 0.1 \iMpch$ in the second-order case, \se is recovered to within $\lesssim 6\,\%$ for a range of halo masses and redshifts. The systematic bias shrinks to $4\,\%$ or less for the third-order bias expansion on the same range of scales. Together with additional evidence we provide, this shows that the residual mismatch in \se can be attributed to higher-order bias contributions. We conclude that the EFT likelihood is able to infer unbiased cosmological constraints, within expected theoretical systematic errors, from physical biased tracers on quasilinear scales.}

\begin{document}

\maketitle

\flushbottom

\section{Introduction}
\label{sec:intro}

State-of-the-art approaches for the analysis of large-scale structure
(LSS) data typically make use of summary statistics like the two-point
correlation function to compare theoretical models to observational
data. 
Alternative approaches have been developed however that take a
more ambitious avenue to cosmological signal
inference. Instead of focusing on summary statistics, they aim
directly at reconstructing the three-dimensional underlying matter
density field from observations of astrophysical tracers like galaxies
\cite{1989ApJ...336L...5B, 1994ApJ...423L..93L, 1995MNRAS.272..885F,
	1999AJ....118.1146S, 2004MNRAS.352..939E, 2010MNRAS.406...60J,
	2010MNRAS.407...29J, 2010MNRAS.409..355J, 2010MNRAS.403..589K,
	2012MNRAS.420...61K, 2013MNRAS.432..894J, Wang:2014hia, 2015MNRAS.446.4250A,
	Wang:2016qbz, 2017MNRAS.467.3993A} (see \cite{2017JCAP...12..009S,schmittfull/etal:2018,Modi:2019hnu} for closely related approaches). 

The general approach works as follows. Starting from a set of initial conditions at high redshift, drawn from a multivariate Gaussian distribution consistent
with the measurements of cosmic microwave background radiation
experiments, nonlinear effects of
gravitational collapse are taken into account via approximate
semi-analytical or numerical methods to compute the corresponding
evolved density field at low redshift that is to be compared to
observations. Then, one samples the initial conditions, as well as 
cosmological and nuisance parameters, to obtain the desired posterior for
the initial phases and cosmological parameters given the observed density field.

However, what we observe is not the evolved matter density field itself, but
rather biased tracers of this field such as
galaxies, quasars, galaxy clusters, the Lyman-$\alpha$ forest, and 
others (see \cite{biasreview} for a review). These are complex
nonlinear objects whose formation happens over long time scales and involves
extremely small-scale physical processes compared to cosmological length scales.
The effective field theory (EFT) of LSS allows for a controlled 
incorporation of the effects of fully nonlinear structure formation 
on small scales in the framework of cosmological perturbation theory 
\cite{baumann/etal:2012,carrasco/etal:2012}. This is especially important when 
attempting to infer cosmological information from observed biased tracers,
since we currently have no way of simulating the formation of 
such tracers from first principles, at least not to the required accuracy.
Hence, approaches 
which rigorously abstract from this imperfect knowledge of the small-scale 
processes involved in the formation of observed tracers are highly 
valuable.

While the calculation of galaxy clustering observables in the EFT has 
largely been restricted to correlation functions so far, Refs.~\cite{paperI,cabass/schmidt:2019} recently presented a derivation of the likelihood of the 
entire galaxy density field $\delta_g(\vx,\tau) = n_g(\vx,\tau)/\bar n_g(\tau)-1$ given the nonlinear, 
evolved matter density field, in the context of the EFT.
Here, $n_g(\vx,\tau)$ is the comoving rest-frame galaxy density, while $\bar n_g(\tau)$ denotes its time-dependent mean.
This result offers several advantages over the previous results restricted to correlation functions, among them the fact that it does not rely on a perturbative expansion of the matter density field. Rather, the 
likelihood is given in terms of the fully nonlinear density field, which can be predicted for example using 
N-body simulations, and thus isolates the truly uncertain aspects of the observed galaxy density.
This conditional likelihood of the galaxy density field given the evolved matter density 
field is precisely the key ingredient required in the Bayesian inference approaches mentioned above, and can be employed there directly \cite{paperI,paperII}. 

Our goal for this paper, as a followup to \cite{paperII}, is to demonstrate
unbiased inference of the linear matter power spectrum normalization \se from halo
catalogs in real space.\footnote{Strictly speaking, we are really performing an inference of the primordial amplitude $\mathcal{A}_s$, rather than \se. However, to conform with standard convention in the LSS literature as well as the previous papers in this series, we continue to use \se. The conversion between the two parameters is unambiguous since we keep all other cosmological parameters fixed.} 
The degeneracy between the linear bias $b_1$ and \se, which is perfect in linear theory,
can be broken when including nonlinear information, even when marginalizing over all relevant nonlinear bias terms. In particular, the fact that the displacement term contained in the second-order
matter density is also multiplied by $b_1$, coupled with the fact that the second-order
matter density scales differently with \se than the linear-order one, breaks
the degeneracy. 
Thus, fundamentally, the
possibility of estimating \se in this way is due to the equivalence principle,
which ensures that galaxies move on the same trajectories as matter on
large scales; that is, the equivalence principle requires that the second-order displacement term is multiplied
by the same bias coefficient as the linear-order density field (see also
Sec.~2 of \cite{biasreview}). At higher orders in perturbations, such
as when going to third order, 
more terms that are protected by the equivalence principle appear,
and the EFT likelihood will consistently capture those as well \cite{cabass/schmidt:2019}.

The results presented in \cite{paperII} however showed a residual bias
in the inferred \se value that was on the order of $10\,\%$--$20\,\%$. Here,
we present a key modification in the construction of the likelihood presented
in \cite{paperII} in form of a cutoff $\Lin$ applied in the initial conditions
in Fourier space, and show that this lowers the bias in the inferred \se
value by a factor of several. Moreover, it restores the expected convergence
behavior as a function of the cutoff scale. 
We also extend the bias expansion to third order, which further improves
the accuracy in the determination of \se significantly, again demonstrating
the expected convergence behavior of the EFT when applied on scales where
perturbation theory is valid.

The paper is organized as follows. In \sect{sec:like} we briefly
review the statistical framework to analyze LSS data developed in
\cite{paperI,paperII}, which forms the basis of our analysis.
We then discuss the significance of the cutoff in the initial conditions
for the EFT likelihood in \refsec{MAP}. After describing the
numerical implementation in \refsec{num}, we turn to the results
in \refsec{results}. We conclude in \refsec{concl}. The appendices
contain some additional calculations and implementation details relevant
for \refsec{MAP} and \refsec{num}.

\section{The EFT likelihood}
\label{sec:like}

The full posterior for cosmological parameters given an observed density
of biased tracers is obtained by marginalizing the likelihood
$P(\delta_h | \delta_{\rm in}, \theta, \{ b_O, \sigma_a \})$
of observing
the density field\footnote{Since our numerical results below are for halos, we refer to the data as $\delta_h \equiv n_h/\nbarh-1$ throughout; the general approach described in this section applies to any LSS tracer however.} $\delta_h$ given the initial conditions $\d_{\rm in}$, and cosmological as well
as nuisance parameters ($\theta$ and $\{ b_O, \sigma_a \}$, respectively), over the initial phases of the matter density field,
weighted by the Gaussian prior for the latter. The physics of the formation 
and evolution of biased tracers is contained in the likelihood 
$P(\delta_h | \delta_{\rm in}, \theta, \{ b_O, \sigma_a \})$, which can be broken down into three components: 
\begin{enumerate}[leftmargin=*]
\item The deterministic forward model for matter, $\d_{\rm fwd}[\d_{{\rm in},\Lin}]$. In this paper, we show results for two forward models: second-order Lagrangian perturbation theory (2LPT), and full N-body simulations. Crucially, and in contrast to the results presented in \cite{paperI,paperII}, initial density perturbations with wavenumbers $k > \Lin$, where $\Lin$ is an initial cutoff, are set to zero. We denote the resulting filtered initial density field as $\d_{{\rm in},\Lin}$. We return to this in \refsec{MAP}.
\item The bias relation, which yields the prediction for the halo density field in a mean-field sense, which we write as
\equ[eq:dhdet]{
\d_{h,\rm det}(\vx) = \sum_O b_O O[\d_\L](\vx)\,\,,
}
where the set of operators is ranked according to orders in perturbations and spatial derivatives, following the general bias expansion or equivalently EFT approach \cite{senatore:2015,MSZ,biasreview}. Note that the bias expansion accounts for the time evolution (formation history) of tracers, even though it is written at a fixed time. This is possible by virtue of including operators corresponding to convective time derivatives, and by restricting to a fixed order in perturbation theory (see Sec.~2.5 of \cite{biasreview}).
In \refeq{dhdet}, $\d = \d_{\rm fwd}[\d_{{\rm in},\Lin}]$ is obtained from the forward model for matter, and $\d_\L$ denotes the sharp-$k$ filtered version of $\d$, where all modes with $k > \L$ are set to zero. Notice that the filter is applied \emph{before} constructing the bias fields $O$. In this paper, we will use second- and third-order bias expansions, as explained later in this section.
Throughout, we drop the time argument on fields for clarity. All fields,
including halos, operators, and matter density, are evaluated at the same
epoch here.
\item The distribution of the likelihood around the mean-field halo density field, which, as derived in \cite{paperI,cabass/schmidt:2019}, can be written as a Gaussian in Fourier space with diagonal covariance that is given as a power series in $k^2$. Specifically, we have 
\equ[eq:PcondGFT]{
	\ln P\left(\d_h\Big|\d, \{ b_O, \sigma_a \}\right) =
	- \frac{1}{2} \fsum{\vk} \left[ \ln [2\pi \s^2(k)] + \frac{1}{\s^2(k)}
	\left| \d_h(\vk) - \d_{h,\rm det} [\d, \{b_O\}](\vk) \right|^2 \right] \,\,.
}
We parametrize $\s^2(k)$ as
\equ[eq:noise_var]{
	\s^2(k)
	= \left( \s_{\eps} + k^2 \s_{\eps,2} \right)^2\,\,.
}
The parametrization is chosen so that $\s^2(k)$ is positive definite.
$\s_\eps^2$ can be interpreted as the amplitude of halo stochasticity in the large-scale limit ($k \to 0$). 
$\s_{\eps,2}^2$ is the leading scale-dependent correction to the halo stochasticity. This term scales as $k^2$, rather than some lower power of $k$, since it captures the backreaction of small physical scales in real space, and thus has to correspond to a local operator in real space (see Sec.~2.7 of \cite{biasreview} for a discussion).\footnote{As argued in \cite{cabass/schmidt:2019,cabass/schmidt:2020} the higher-derivative stochastic term is actually subleading compared to the modulation of the stochasticity by large-scale density perturbations; however, even the latter is less relevant than any of the
deterministic bias terms that we include, which go up to third order. We leave an
exploration of the field-dependent stochasticity to future work.} In fact, we find that the inclusion of $\sigma_{\eps,2}$ has a negligible impact on our results.
\end{enumerate}
In the actual implementation, all fields are discretized on a uniform cubic grid. 
We employ the discrete Fourier transform in our equations, so that fields in Fourier space are dimensionless as well.
In the following, we will refer to the conditional probability in \refeq{PcondGFT} simply as ``likelihood,'' since it is the part of the overall likelihood of biased tracers that is relevant for the study presented in this paper. 

The likelihood involves, in principle, all three distinct cutoffs, $\Lin$, $\L$ [\refeq{dhdet}], and $\kmax$ [\refeq{PcondGFT}], 
that we introduced in points 1 to 3 of the previous paragraph, respectively. From the EFT perspective, the cutoff $\Lin$ on the initial conditions is the relevant scale \cite{carroll/etal,cabass/schmidt:2019}, while the latter two are choices made in the numerical implementation. In this paper, we will set $\kmax = \L = \Lin$ throughout. We return to this in \refsec{results}.

Following Ref.~\cite{paperI}, we show results for the bias expansion up to second
order in perturbations, i.e., we restrict ourselves to the following set of bias operators:
\equ[eq:operators]{
	O \in \{ \d, \ \d^2, \ K^2, \ \lapl\d \} \,\, ,
}
where $\d$ is the fractional matter density perturbation, and
\be
K^2 \equiv (K_{ij})^2 = \left( \left[\frac{\partial_i\partial_j}{\lapl} - \frac13 \d_{ij} \right] \d \right)^2
\ee
is the tidal field squared. The corresponding bias parameters are denoted as $b_O$; we also denote $b_1 \equiv b_\d$.

We have also extended our bias model to third order, in which case
the set of bias operators now comprises
\be
O \in \{ \d,\ \d^2,\ K^2,\ \lapl\d,\ \d^3,\ \d K^2,\ K^3,\ O_\text{td} \}\,\,, 
\label{eq:Ocubic}
\ee
where 
\ba
O_{\otd} \equiv \frac{8}{21} K_{ij}
\frac{\partial_i\partial_j}{\lapl} 
\left( \d^2 -\frac32 K^2 \right)\,\,.
\ea
At third order, this operator can be interpreted in a variety of different
ways, for example as the convective time derivative of the tidal field squared,
or the difference between tidal and velocity shear (see Sec.~2.4 of \cite{biasreview}). 

Notice that the bias expansion \refeq{dhdet} is an expansion in two
small parameters, essentially in perturbations and spatial derivatives
(see Sec.~4.1 of \cite{biasreview} for a detailed discussion).
In \refeq{operators} and \refeq{Ocubic},
we assume that both small parameters are comparable, which leads us
to include terms up to second or third order in perturbations, as well as the
leading higher-derivative operator $\lapl\d$ in \refeq{operators}. The
reasoning behind this is discussed in greater detail in \cite{paperI}.
We emphasize that it is extremely simple to add additional higher-derivative
bias terms in the EFT likelihood, and numerically efficient as well once the
analytical marginalization over bias parameters is employed following the procedure described in \cite{paperII}.

For all results apart from those in \refsec{rk}, we marginalize over all bias parameters apart from $b_1$ analytically. In the second-order bias case, we thus marginalize over three parameters, while in the third-order case, the marginalization is over seven parameters. This substantially speeds up the numerical search for the maximum-likelihood point, since, for either bias expansions, the parameter set is reduced to $\{ b_O, \sigma_a\} \to \{ b_1, \s_\eps, \s_{\eps,2}\}$.

Ref.~\cite{paperI} describes a renormalization procedure for the operators
that ensures that their coefficients match the bias parameters that would be
inferred from the large-scale statistics of halos, such as the power spectrum
and bispectrum. Since the results on \se which we focus on in this paper
are independent of the renormalization, we do not employ it here.

One important difference in \refeq{noise_var} as compared to that presented
in \cite{paperI,paperII} is the removal of the term $b_1 \s_{\eps\eps_m,2} k^2$
in $\s^2(k)$. This would represent, at the level of correlation functions, the leading contribution to 
the halo-matter power spectrum that is analytic in $k^2$.
As long as uniform priors are employed on the parameters
$b_1$ and $\s_{\eps,2}$, this term can be absorbed by a $b_1$-dependent redefinition of $\s_{\eps,2}$ and thus does not influence the maximum-likelihood point.
As we show in \refapp{MAPdelta}, the use of a cutoff in the initial conditions
in fact removes the justification for this term that was put forward in \cite{paperI}.

To summarize, the differences to the likelihood presented in \cite{paperI,paperII}
are $(i)$ the use of a cutoff $\Lin$ in the initial conditions; $(ii)$
the removal of a $b_1$-dependent term in the variance of the likelihood
\refeq{PcondGFT}; $(iii)$ the extension of the bias expansion to third order.

As for the results reported in \cite{paperI,paperII}, we do not sample
the initial phases but rather fix them to the values used to initialize the N-body
simulations within which the halos were identified. Since this removes cosmic variance
to the largest extent possible,
and thus shrinks the error bars significantly over the case where phases are allowed to vary,
this is likely to be the most stringent possible test of the conditional likelihood $P(\d_h | \d, \{ b_O, \sigma_a \})$. 
We use the profile likelihood \cite{wilks1938} introduced in
\cite{paperII} in order to estimate the maximum-likelihood value for
\se. For a
probability distribution $P(\se, \{b_1, \sigma_a\}|\d_h)$, the profile likelihood for
the parameter $\se$ is defined as the maximum probability within the parameter space that is being profiled over:
\equ[eq:prof_def]{
	\pprof(\se) = \underset{\{b_1, \sigma_a\}}{\max} [ P(\se, \{b_1, \sigma_a\} | \d_h) ] \,\, .
}
Here, the set of parameters $\{b_1, \sigma_a\}$ has been profiled
out. In practice, we interpolate the profile likelihood evaluated on a
predefined grid in \se centered about the fiducial value of the
simulation. The details of this procedure are the same as described in
\cite{paperII}.

\section{The cutoff on the initial conditions and the maximum-a-posteriori point}
\label{sec:MAP}

The main change to the EFT likelihood implementation presented here over
the previous results in \cite{paperI,paperII} is the imposition of a
wavenumber (or momentum) cutoff in the initial conditions. Let us now discuss the
significance of this cutoff.

In order to determine whether the EFT likelihood in \refeq{PcondGFT} leads to unbiased estimates of the parameters of interest, which in the present case are $\se$ and the set of bias parameters, we study the maximum-a-posteriori (MAP) relation of the likelihood at fixed phases. Since the likelihood depends nonlinearly on the phases (via the forward model $\d_{\rm fwd}$ as well as the nonlinear bias operators), it is extremely difficult to study the likelihood with varying phases analytically. Fortunately, the \emph{ensemble average over the phases of the fixed-phase MAP} relation can be derived analytically \cite{paperI}; this is summarized in \app{app:MAP_review}. This relation is the relevant quantity to compare with the MAP value of \se obtained below in the application to halos in N-body simulations, which should follow the analytical relation in the limit of infinite simulation volume (to the precision of perturbation theory at the order and value of the cutoff considered).

We can in fact further restrict to the MAP relation for the bias parameters. The relation for $\se$ can be obtained via a generalization of this relation, since each term in the likelihood has a definite scaling with $\se$ in the context of perturbation theory \cite{paperI}. Taking the derivative of the logarithm of the likelihood with respect to a bias parameter $b_O$, and neglecting any priors on the bias parameters, we obtain (\app{app:MAP_review}):
\ba
\fsum{\vk} \frac1{\s^2(k)} \< \d_h(\vk) O(\vk') \> = \fsum{\vk} \frac1{\s^2(k)} \sum_{O'} b_{O'} \< O'(\vk) O(\vk')\> 
\qquad\forall\ O\,\,,
\label{eq:maxlikeG}
\ea
where the sum runs over all operators in the deterministic bias expansion.
As discussed above, and shown in \cite{paperI}, these relations need to be fulfilled in order to ensure
an unbiased inference of cosmological parameters and in particular \se.

The left-hand side of \refeq{maxlikeG} involves the cross-correlation of the data, $\delta_h$, with an operator constructed from the forward-evolved density, $O$, while the right-hand side only involves operators constructed from the forward-evolved density field. Both sides can be evaluated, to a given order in perturbations, by means of the EFT of large-scale structure, which allows us to establish whether the construction of the likelihood in fact leads to unbiased inference. Each side contains loop integrals which come from three sources of nonlinearities: the nonlinear forward model for matter, the nonlinearity in the bias operators, and the nonlinear evolution of the actual halos, which enters on the left-hand side. Thus, establishing the identity \refeq{maxlikeG} is not trivial. In the following, we study the properties of these loop integrals in detail, and show under what conditions the MAP relation holds within the EFT. In this section, we present general derivations; concrete examples for specific operators are given in \refapp{MAPdelta}.

Restricting \refeq{maxlikeG} to a single $k< \kmax$ to be specific, we thus have
\ba
\< \d_h(\vk) O(\vk') \> = \sum_{O'} b_{O'} \< O'(\vk) O(\vk') \> \qquad\forall\ O\,\,.
\label{eq:maxlikeG2}
\ea
Notice that we expect the EFT approach to apply for any shape of the linear power spectrum 
(although the reach of perturbation theory will depend on this shape). For this reason, 
it is necessary that the equality \refeq{maxlikeG2} hold for individual $k$. 
Our goal is to investigate this relation in the context of perturbation theory. 
Thus, we expand the halo density field in a set of renormalized bias operators, multiplied by bias coefficients \cite{biasreview}: 
\be
\d_h(\vk) = \sum_O \bt_O [O](\vk)\,\,.
\label{eq:renormbias}
\ee
Notice the crucial difference between the operators $[O]$ appearing here and
those appearing explicitly in \refeq{maxlikeG2}. The former are assumed to be 
constructed from the evolved density field without any cutoffs:
\be
[O] = \big[O[\d_{\infty}]\big]\,\,,
\ee
where $\d_{\infty}$ denotes the full forward-evolved density field without
cutoffs. 
That is, halos in regular N-body simulations, or actual observed galaxies,
evolve together with the nonlinear matter distribution without any cutoffs
on the initial or final density fields imposed.
On the other hand, the operators appearing in the likelihood are constructed from the
filtered evolved density field
\be
O = O[\d_\L]\,\,,\quad\mbox{where}\quad
\d_\L(\vk) = W_\L(\vk) \d_{\rm fwd}\left[ \d_{{\rm in},\Lin} \right](\vk)
\label{eq:d_fwd_def}
\ee
is the evolved density field filtered on the scale $\L$, starting from
initial conditions $\d_{{\rm in},\Lin}$ \emph{filtered on the scale $\Lin$}.
The effect of removing this cutoff can be obtained by sending $\Lin\to\infty$. 
In all cases, we adopt an isotropic sharp-$k$ filter,
\be
W_{\L}(\vk) \equiv \thH(\L-|\vk|)\,\,,
\label{eq:Wdef}
\ee
where $\thH$ is the Heaviside function. 

We assume that the set of operators are linearly independent, and that they form a complete basis of local 
observables at a given order in perturbation theory and derivatives (this is the case for the list of operators in 
\refeq{operators} and \refeq{Ocubic} for example). In order for 
\refeq{maxlikeG2} to hold for \emph{any} tracer, it thus has to hold individually 
for all bias operators $O$ and $O'$. This finally leads us to compare the two correlators 
\be
\< \big[O'[\d_\infty]\big](\vk ) O[\d_\L](\vk') \>
\quad\mbox{and}\quad
\< O'[\d_\L](\vk) O[\d_\L](\vk')\>\,\,.
\label{eq:corrvs}
\ee
The correlator on the left can be written as
\ba
&\< \big[O'[\d_\infty]\big](\vk ) O[\d_\L](\vk') \> =
\int_{\vp_1,\ldots,\vp_n}\!\!\!\! S_{O'}(\vp_1,\ldots\vp_n) (2\pi)^3 \dirac3(\vk-\vp_{1\cdots }) \label{eq:corr1}\\
&\qquad \times
\int_{\vp'_1,\ldots,\vp'_m}\!\!\!\! S_{O}^\L(\vp'_1,\ldots\vp'_m) (2\pi)^3 \dirac3(\vk'-\vp'_{1\cdots }) \< \d_\infty(\vp_1)\cdots\d_\infty(\vp_n)\d_{\rm fwd}(\vp'_1) \cdots \d_{\rm fwd}(\vp'_m)\> \vs
&\qquad + \mbox{counterterms}\,\,, 
\nonumber
\ea
where we have taken $O'$ ($O$) to be constructed out of $n$ ($m$) density fields. 
We will frequently denote this as $O' = O'^{[n]}$ ($O = O^{[m]}$). 
The kernels $S_{O}$, $S_{O'}$ are specific to each operator; for example
for the operators in the list \refeq{operators} we have\footnote{
In general, the kernels for bias operators at leading order in derivatives are homogeneous (of degree $0$), rational functions of linear combinations of the momenta. Kernels for higher-derivative operators are homogeneous with degrees $2, 4,\ldots$.}
\be
S_\d(\vp) = 1; \quad
S_{\d^2}(\vp_1,\vp_2) = 1; \quad
S_{K^2}(\vp_1,\vp_2) = \frac{(\vp_1\cdot\vp_2)^2}{p_1^2 p_2^2} - \frac13; \quad
S_{\lapl\d}(\vp) = -p^2.
\ee
Further, we have defined the kernels with cutoff as
\be
S_O^\L(\vp_1,\ldots,\vp_n) \equiv W_\L(\vp_1)\cdots W_\L(\vp_n) S_O(\vp_1,\ldots,\vp_n),
\ee
and denoted $\vp_{1\cdots n} \equiv \vp_1+\ldots+\vp_n$. 
Finally, we continue to denote the evolved matter density field without any
cutoffs as $\d_\infty$, while $\d_{\rm fwd}$ denotes the matter density field
evolved with a cutoff $\Lin$ in the initial conditions, cf.~Eq.~\eqref{eq:d_fwd_def}. 
The last line in \refeq{corr1} contains the counterterms which we will discuss below.
The second correlator in \refeq{corrvs} follows an analogous expression, with
$S_{O'}\to S_{O'}^\L$ and $\d_\infty(\vp_i) \to \d_{\rm fwd}(\vp_i)$, and no counterterms.
Due to the cutoffs $\L$ and $\Lin$, this correlator only involves modes with momenta (wavenumbers)
of order $\L,\Lin$ or less. 

The unfiltered density fields appearing in $[O'[\d_\infty]]$ lead to loop integrals whose momenta run to infinity. These need to be regularized by adding counterterms \cite{2014JCAP...08..056A,senatore:2015,MSZ}, which we can describe at the level of the operators as consisting of linear combinations of equal- or lower-order operators $\widetilde{O}$ which are subtracted: 
\be 
\label{eq:ctr_def}
[O'](\vk) = O'(\vk) - \sum_{\widetilde{O}} \sigma_{O',\widetilde{O}}^2\, \widetilde{O}(\vk)\,\,,
\ee
where the constants $\sigma_{O',\widetilde{O}}^2$ involve loop integrals (as well as finite
contributions in general), and can have either sign. Inserting this relation into \refeq{corr1}, it is then clear that
the counterterms on the last line can be described in terms of similar correlators
as the first, ``bare'' contribution.
Specifically, Ref.~\cite{2014JCAP...08..056A} argued that renormalization should ensure that
correlators involving $[O']$ and $l$ linear density fields
$\d^{(1)}(\vp_1),\ldots,\d^{(1)}(\vp_l)$ asymptote to tree-level results as
the external momenta become small (see also Sec.~2.10 of \cite{biasreview} for an extended discussion):
\be
\lim_{\{p_i\}\to 0} \frac{ \< [O'](\vk) \d^{(1)}(\vp_1)\cdots\d^{(1)}(\vp_l) \> }
	{ \< O'(\vk) \d^{(1)}(\vp_1)\cdots\d^{(1)}(\vp_l) \>_{\rm LO } } = 1\,\,,
	\label{eq:renormcond}
\ee
where $l=1,2,\ldots$, and the subscript on the correlator in the denominator indicates the leading-order (LO) expression in perturbation theory. Notice that, for $l=n$, this correlator is directly related to the kernel $S_{O'}$ through
\be
\< O'^{[n]}(\vk) \d^{(1)}(\vp_1)\cdots\d^{(1)}(\vp_n) \>'_{\rm LO }
= n!\, S_{O'}(\vp_1,\cdots,\vp_n) \Plin(p_1)\cdots\Plin(p_n)\,\,,
\ee
where a prime on a correlator denotes that the momentum-conserving Dirac delta is removed,
and we have assumed that the kernel $S_{O'}$ is fully symmetrized in its arguments. In the following, we will work with the linearly evolved density field $\d^{(1)}(\vp) \propto \d_{\rm in}(\vp)$ instead of $\d_{\rm in}$ itself, as is common in perturbation-theory calculations, where the two fields are simply related by the linear growth factor.

The first correlator in \refeq{corrvs} can be represented diagrammatically as 
\begin{equation}
\label{eq:D1}
\begin{split}
&\< \big[O'[\d_\infty]\big](\vk ) O[\d_\L](\vk') \> = 
\raisebox{-0.0cm}{\includegraphicsbox[scale=0.25,trim={0.75cm 3.75cm 0.75cm 3.75cm},clip]{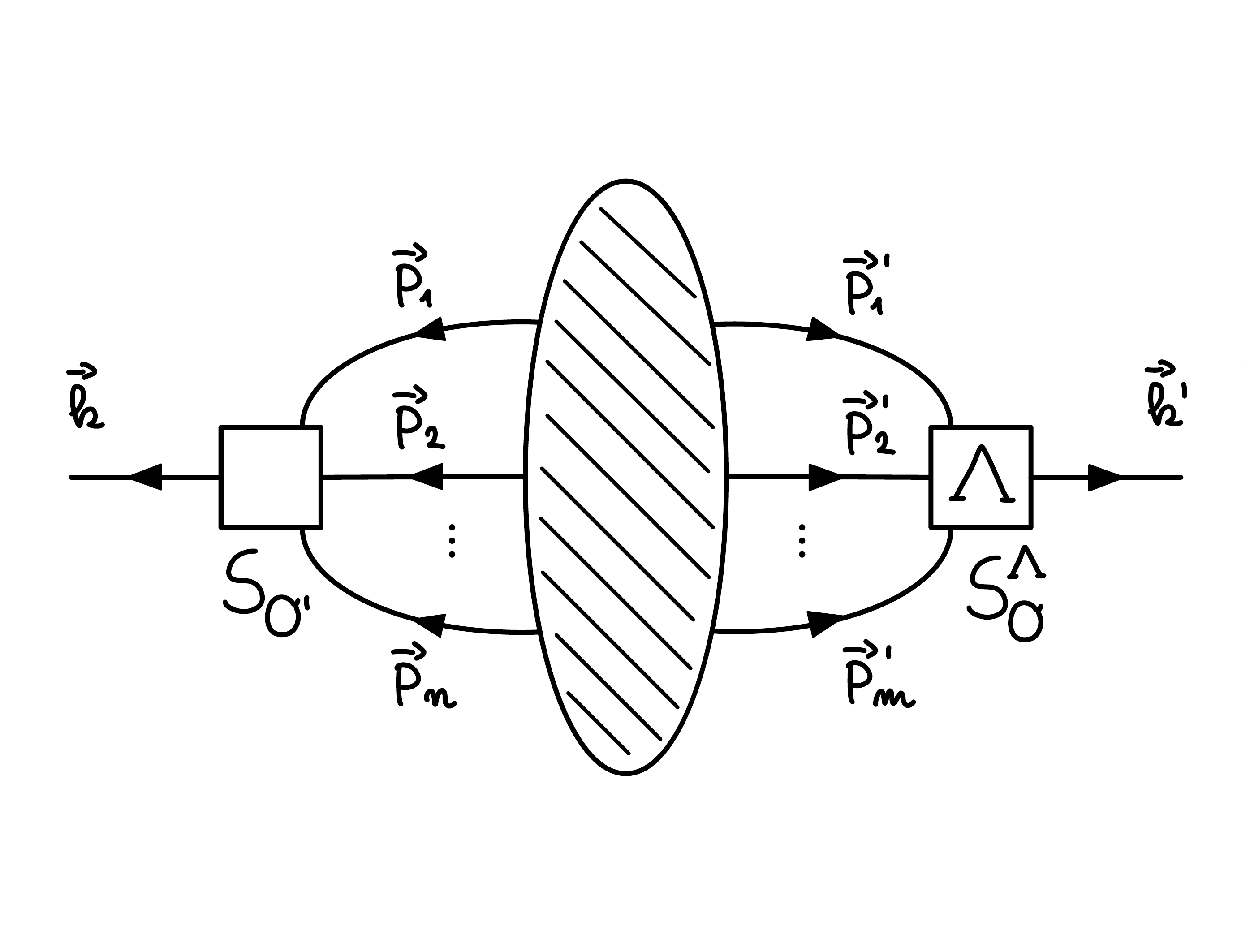}} \\ 
&+\raisebox{-0.0cm}{\includegraphicsbox[scale=0.25,trim={0.75cm 3.75cm 0.75cm 3.75cm},
clip]{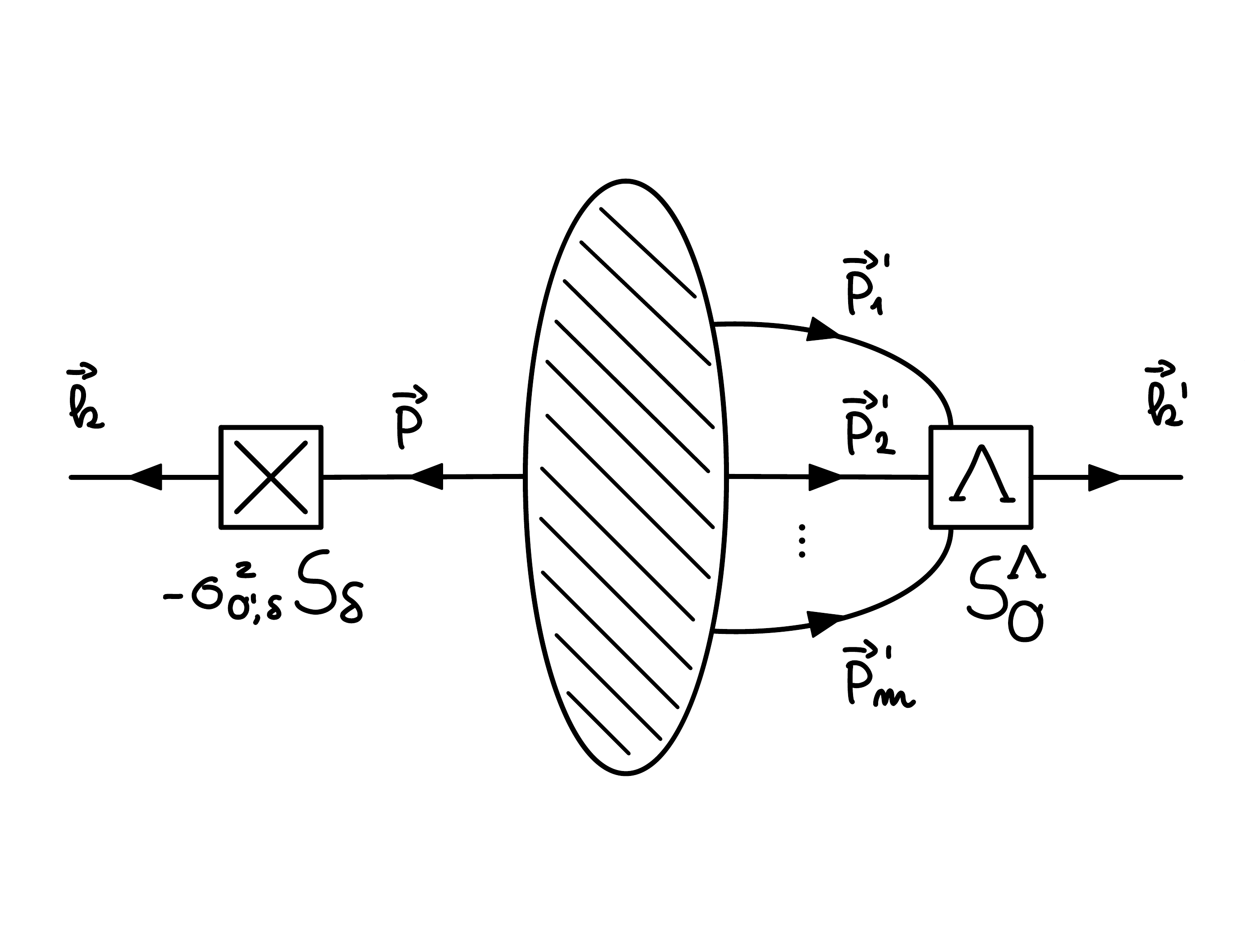}} + \text{higher-order counterterms}\,\,,
\end{split}
\end{equation} 
where the first diagram gives the contribution of the first two lines of \refeq{corr1}, 
while the second illustrates one of the counterterms, more precisely the one where $\widetilde{O}$ is equal to $\delta$ in Eq.~\eqref{eq:ctr_def}. 

In the following, we will consider specific cases in diagrammatic
form. Let us thus state the relevant Feynman rules:
\begin{enumerate}[leftmargin=*]
\item We employ the same notation as adopted in \cite{abolhasani/mirbabayi/pajer:2016} and App.~B of \cite{biasreview}. The external vertices $S_O, S_{O'}$ are denoted as squares; for an $n$-th order operator, the corresponding vertex has $n$ ingoing lines and 1 outgoing line. The perturbation-theory kernels $F_n$ (see \cite{LSSreview}) are denoted as open circles, likewise with $n$ ingoing and a single outgoing line. Each vertex further contains a momentum-conserving Dirac delta (e.g., $(2\pi)^3 \dirac3(\vk-\vp_{1\cdots n})$ in case of the $S_{O'}$ vertex).
\item Two types of linear power spectra appear: those without cut, $\Plin(p)$ (denoted with a dot), and those cut at $\Lin$, $W_{\Lin}(p)\Plin(p)$ (denoted with a crossed circle). The rule governing which to choose is that \emph{any linear power spectrum connected to a final outgoing line going to the right} (i.e.~toward $S_{O}^\L$) is cut at $\Lin$. If a power spectrum is only connected to outgoing lines ending up on the left (i.e.~at $S_{O'}$), then it is not cut. This follows from the fact that the outgoing lines connecting to the right correspond to evolved fields with an initial-condition cutoff, and that $\langle \d^{(1)}_\infty \d^{(1)}_{\Lin} \rangle' = \langle \d^{(1)}_{\Lin} \d^{(1)}_{\Lin}\rangle' = W_{\Lin} \Plin$.
\item Any unregularized loop integral appearing in a diagram for an $n$-th order operator $O'$ on the left is to be removed by a 
corresponding counterterm for the operator $O'$, whose vertex we denote
by a crossed square (in the following diagrams we will omit the labels on these vertices for simplicity of notation). 
The counterterm is obtained by cutting at most $n$ soft lines in the diagram (i.e.~lines with momenta at most of order $\L$, $\Lin$). 
\end{enumerate} 

The rule for the identification of counterterms is equivalent to the renormalization conditions in \refeq{renormcond}, since it isolates loop integrals that are fully connected with the left-hand side of the diagram, i.e.~with $S_{O'}$, in the same way as
\refeq{renormcond} ensures that loops are fully connected with $O'$ by considering only
correlators of the operator with powers of $\d^{(1)}$. However, here
we will only include loop momenta above $\Lin$ in our counterterms, since the
contribution from modes below the cutoff is matched by the corresponding correlator
involving $O'[\d_\L]$ in \refeq{corrvs}. Thus, the counterterms given below differ by a finite contribution from those
commonly defined when computing correlation functions,
which can be interpreted as adopting a different renormalization scale ($\Lin$ rather than
the large-scale limit $0$).

Let us now consider the lowest-order operator correlators,
evaluating each of them up to next-to-leading order (NLO, or $1$-loop). This will involve diagrams including
up to three linear power spectra. Let us begin with $O' = O = \d$, such that $S_{O'}=1$
and $S_{O}^\L(\vk) = W_\L(k)$, corresponding to the correlator $\< \d_\infty(\vk) \d_\L(\vk')\>$. The diagrams are 
\begin{equation}
\label{eq:delta_delta}
\begin{split}
&\< [\d_\infty](\vk ) \d_\L(\vk') \> = 
\raisebox{-0.0cm}{\includegraphicsbox[scale=0.25,trim={2.25cm 7cm 2.25cm 7cm},clip]{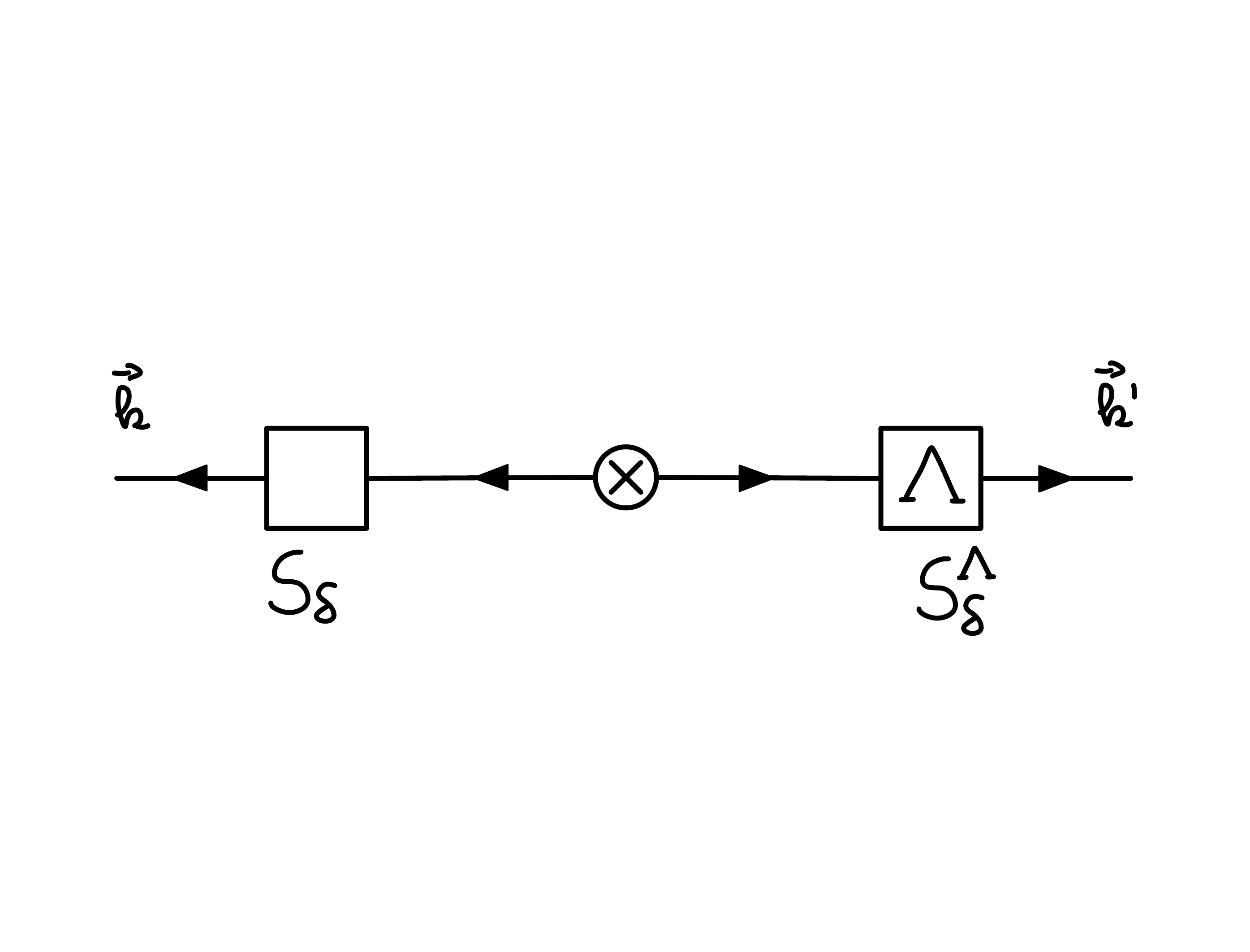}} \\ 
& + \raisebox{-0.0cm}{\includegraphicsbox[scale=0.25,trim={0.75cm 7cm 0.75cm 7cm},clip]{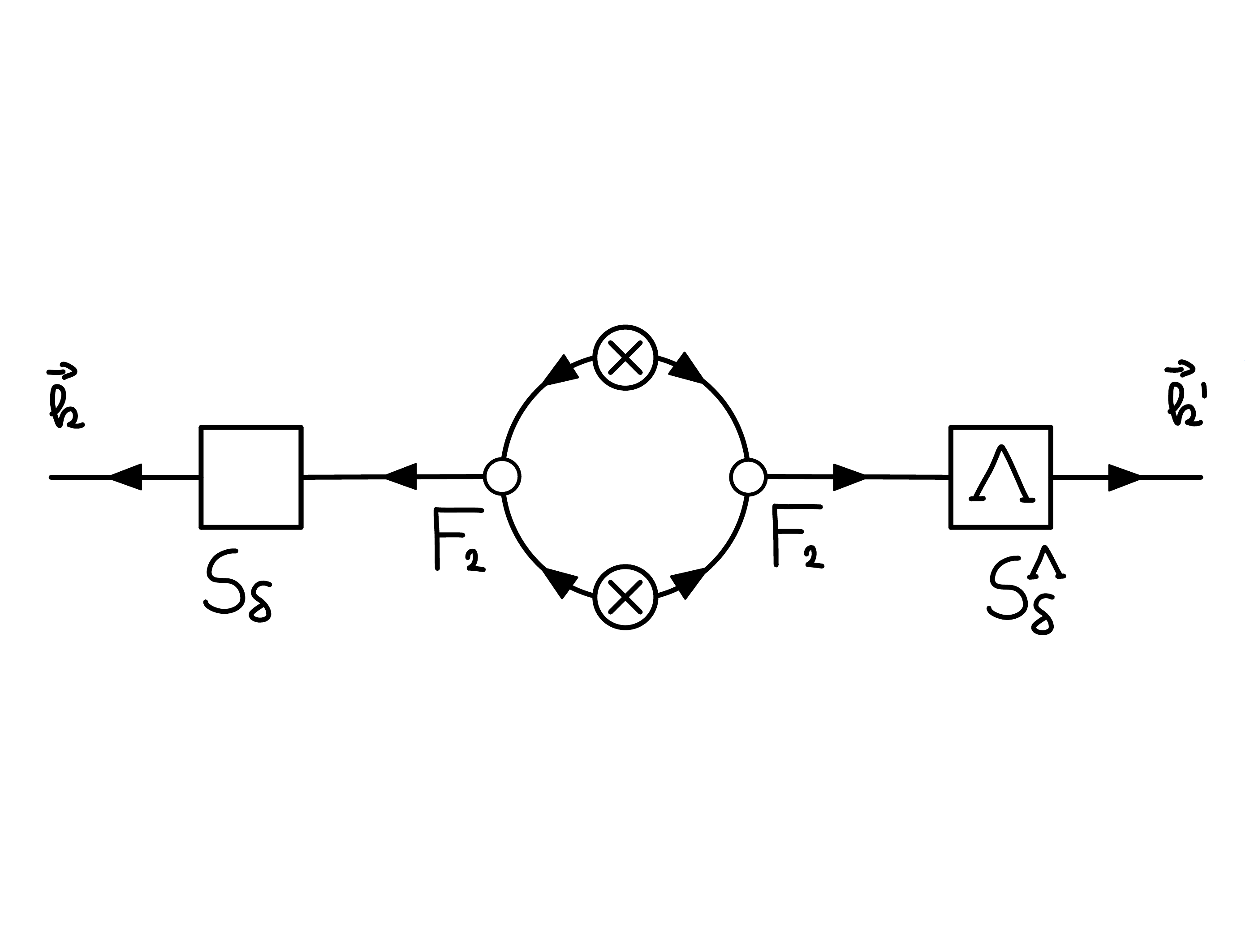}} 
+ \raisebox{+0.25cm}{\includegraphicsbox[scale=0.25,trim={0.75cm 6.5cm 0.75cm 4.5cm},clip]{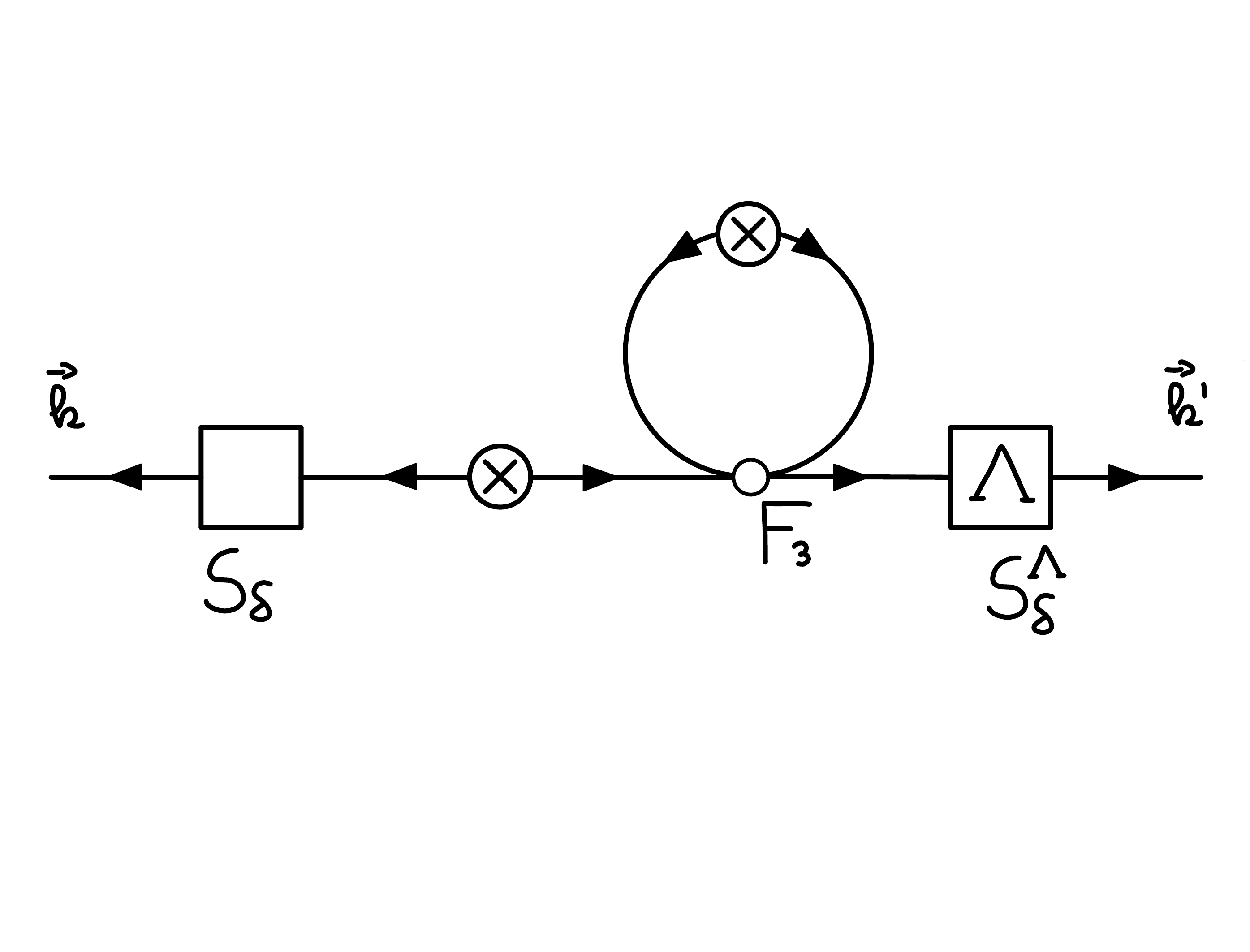}} \\
&+ \raisebox{-0.0cm}{\includegraphicsbox[scale=0.25,trim={0.75cm 4.5cm 0.75cm 4.5cm},clip]{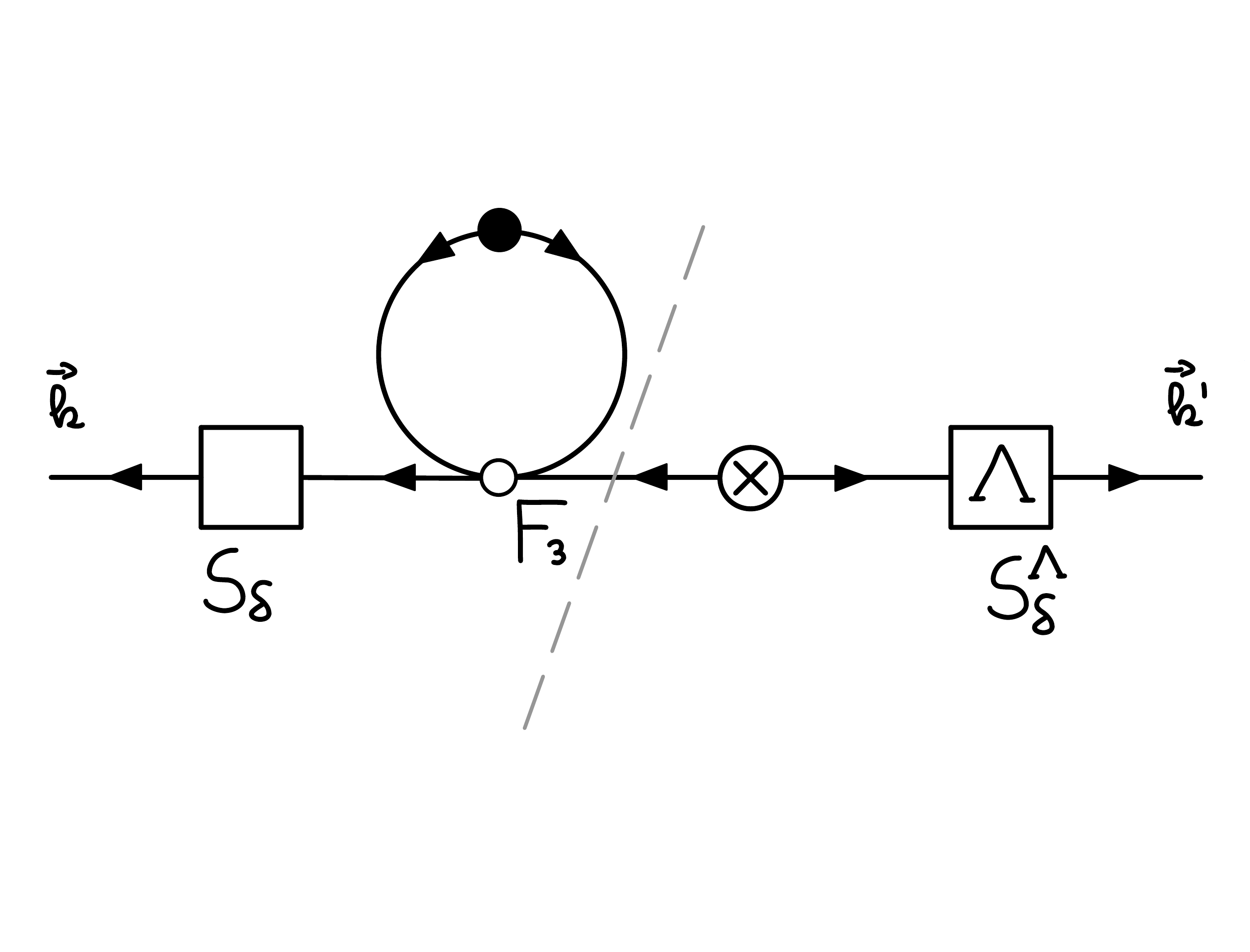}} 
+ \raisebox{-0.0cm}{\includegraphicsbox[scale=0.25,trim={2.25cm 7cm 2.25cm 7cm},clip]{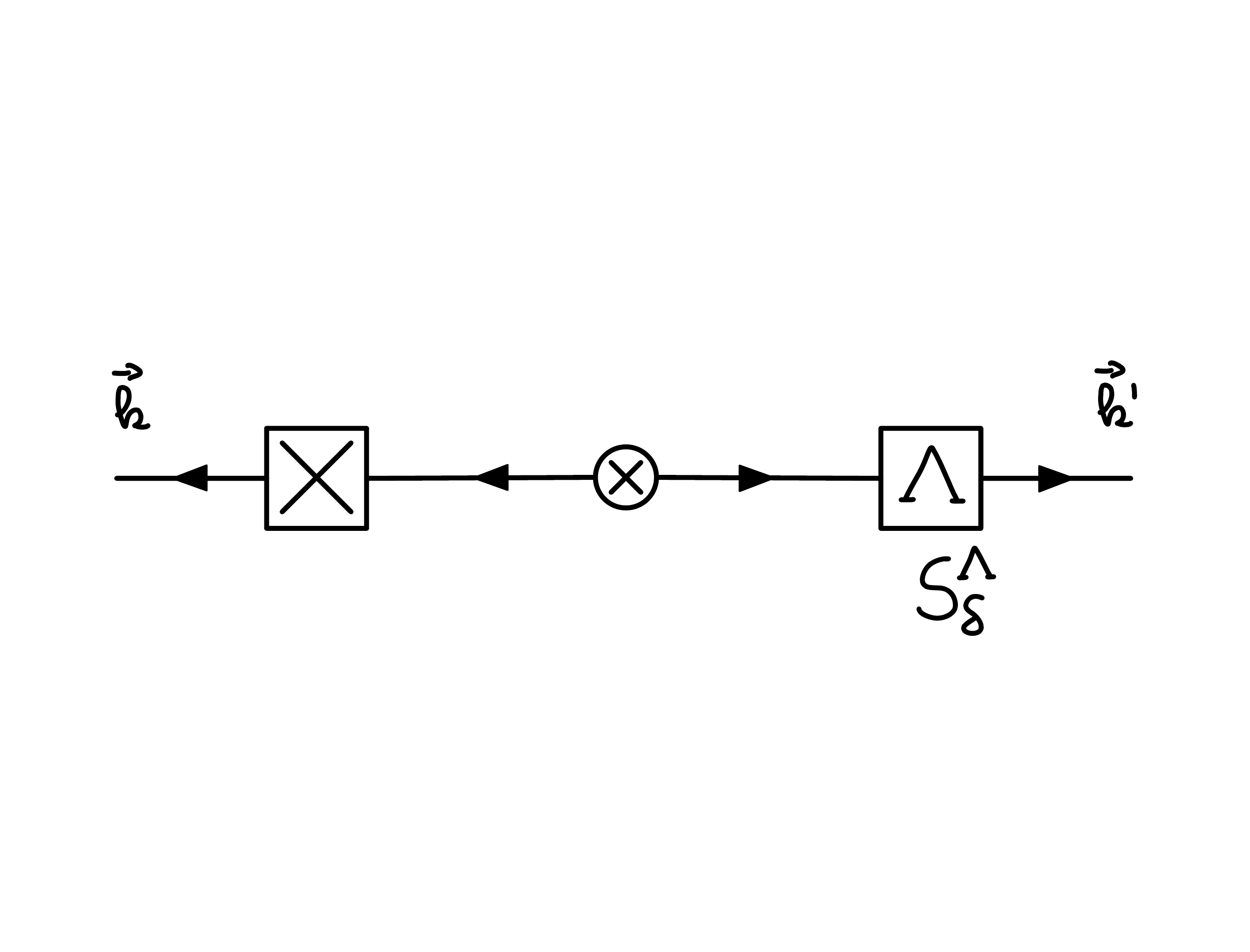}} \,\,, 
\end{split}
\end{equation}
where $F_2$ and $F_3$ are the second- and third-order perturbation-theory kernels, respectively 
(in the following diagrams we will omit the label on the perturbation-theory kernels for simplicity of notation). 
There are three NLO contributions, only one of which has a loop integral 
that runs to infinity; the others are regularized by $\Lin$. 
It is straightforward to see that the counterpart of this correlator 
in the MAP relation, i.e. the right correlator in \refeq{corrvs} 
$\< \d_\L(\vk) \d_\L(\vk')\>$, differs only through this loop contribution, 
which instead of being unregularized is now also regularized by $\Lin$. 
The unregularized loop integral, and hence the mismatch between the correlators in \refeq{corrvs} 
for $O'=O=\d$, is to be absorbed by a counterterm to $[\d]$, corresponding 
to the part of the diagram left of the dotted line, so that 
\be
[\d](\vk) = \left[1 - 3 \int_{|\vp|>\Lin} F_3(\vp,-\vp,\vk) \Plin(p) - C_s^2(\Lin) \frac{k^2}{\knl^2} \right] \d(\vk)\,\,,
\label{eq:ct1}
\ee
where the second term absorbs the loop integral while the last term is
the finite contribution of unknown size, whose coefficient is the effective
sound speed of matter $C_s^2$ \cite{baumann/etal:2012,carrasco/etal:2012},
defined with respect to the scale $\Lin$. 
Here, $\knl$ is the nonlinear scale defined through $\knl^3 \Plin(\knl)/2\pi^2 = 1$ ($\knl(z=0)\simeq 0.25\iMpch$ in the fiducial cosmology); with this definition, $C_s^2$ is of order unity. 

Next, consider $O=\d$ correlated with a second-order operator $O' = O'^{[2]}$ in the halo field, i.e.~ 
\be
\begin{split}
&\< \big[O'^{[2]}[\d_\infty]\big](\vk ) \d_\L(\vk') \> = 
\raisebox{+0.25cm}{\includegraphicsbox[scale=0.25,trim={3.5cm 6.5cm 0.75cm 4.5cm},clip]{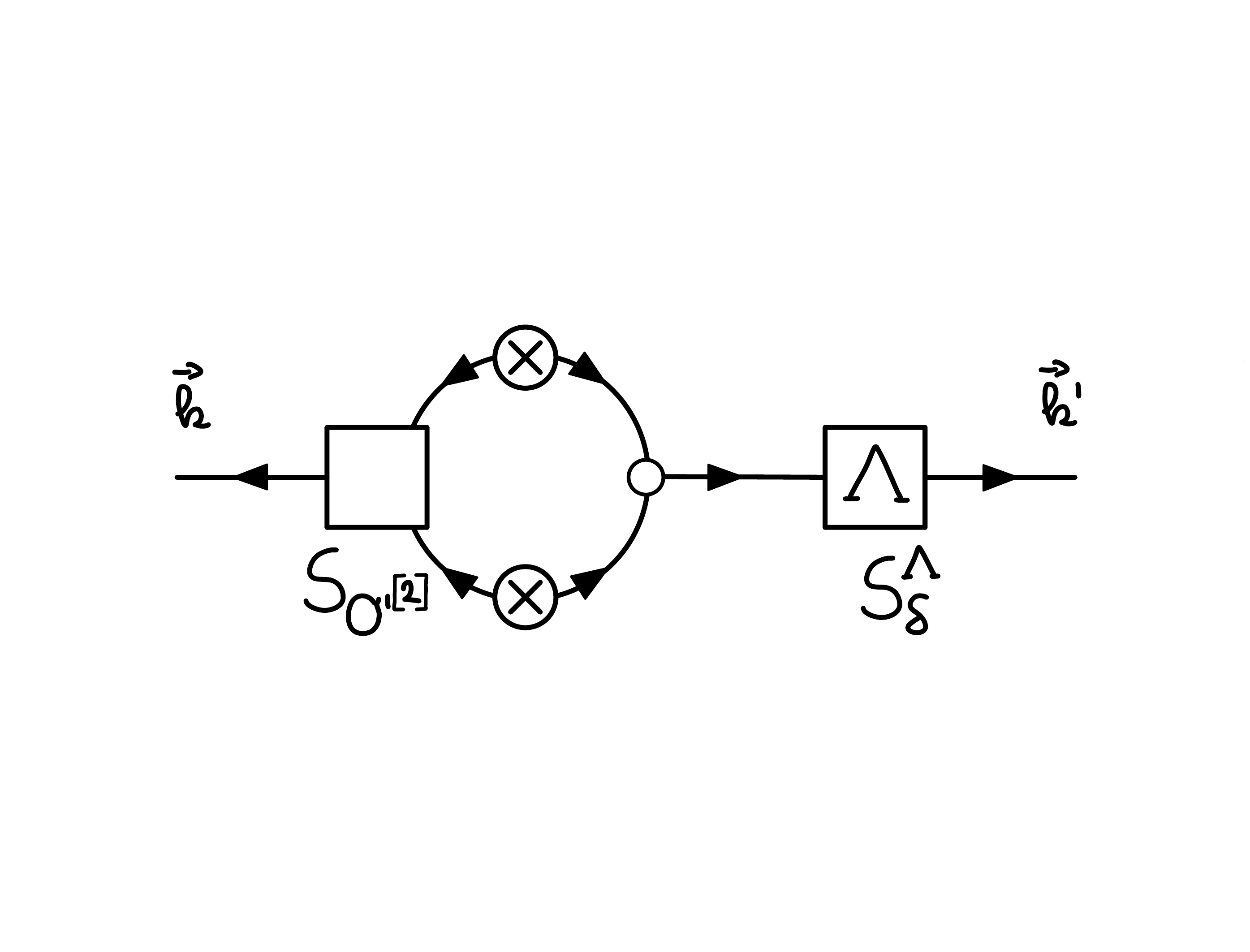}} \\
& + \raisebox{-0.0cm}{\includegraphicsbox[scale=0.25,trim={2cm 4.5cm 0.75cm 4.5cm},clip]{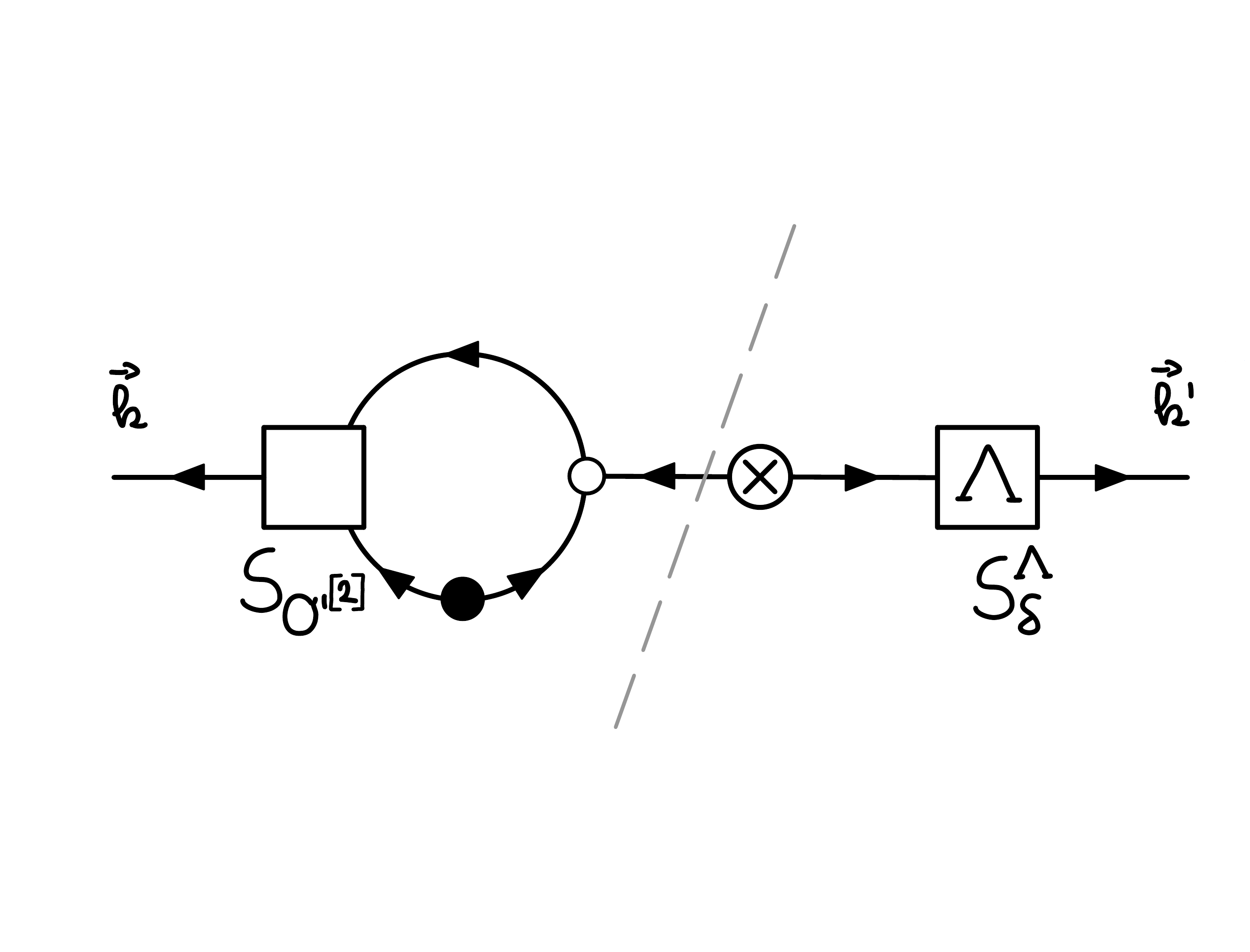}} 
+ \raisebox{-0.0cm}{\includegraphicsbox[scale=0.25,trim={2cm 7cm 2.25cm 7cm},clip]{plots/delta_delta_rhs-5.pdf}}\,\,.
\label{eq:D3}
\end{split}
\ee
Here, there are two contributions, one of which
involves an unregularized loop. This contribution is absorbed
by a counterterm to $O'$ that is $\propto \d$ (the part of the diagram left of the dotted line),
\be
[O'^{[2]}](\vk) = O'^{[2]}(\vk) -2 \int_{|\vp|>\Lin} S_{O'^{[2]}}(\vp,\vk-\vp) F_2(-\vp,\vk)\Plin(p) \, \d(\vk)\,\,. 
\label{eq:ct2}
\ee
By absorbing the term in the correlator that scales as $\Plin(p_1)$ as $p_1 \to 0$, 
\refeq{ct2} ensures that the renormalization conditions in \refeq{renormcond}
are satisfied for $O'^{[2]}$ and $l=1$ \cite{2014JCAP...08..056A}. 
Consider $O'=\d^2$. In this case, \refeq{ct2} evaluates to a formally divergent
constant multiplied by $\d$ which is simply subtracted. 
In case of $O'=K^2$, there is a contribution with nontrivial scaling in $k$,
which in the limit of $k\ll p$ however becomes analytic with a leading
contribution $\propto k^2$, so that this contribution in \refeq{ct2} is
effectively absorbed by subtracting a higher-derivative counterterm $k^2 \d(\vk)$
(see also \refapp{MAPdelta}). Once \refeq{ct2} is employed,
one again finds agreement between the two correlators in \refeq{corrvs}
for $O'=O'^{[2]}$, $O=\d$.

Notice that the first contribution in Eq.~\eqref{eq:D3} would also involve an unregularized loop integral if one were
to send $\Lin\to\infty$, leading to a mismatch in the MAP relation, as
the corresponding loop is cut at $\L$ in the second correlator in \refeq{corrvs}. 
Unlike the second contribution in Eq.~\eqref{eq:D3}, however, this loop integral
cannot be absorbed by a counterterm to $[O']$ due to its different
structure. Hence, the cutoff in the initial conditions is essential. 

Next, for $O'=O'^{[3]}$ we have 
\be
\begin{split}
\< \big[O'^{[3]}[\d_\infty]\big](\vk ) \d_\L(\vk') \> &= 
\raisebox{-0.0cm}{\includegraphicsbox[scale=0.25,trim={2cm 5cm 2cm 4cm},clip]{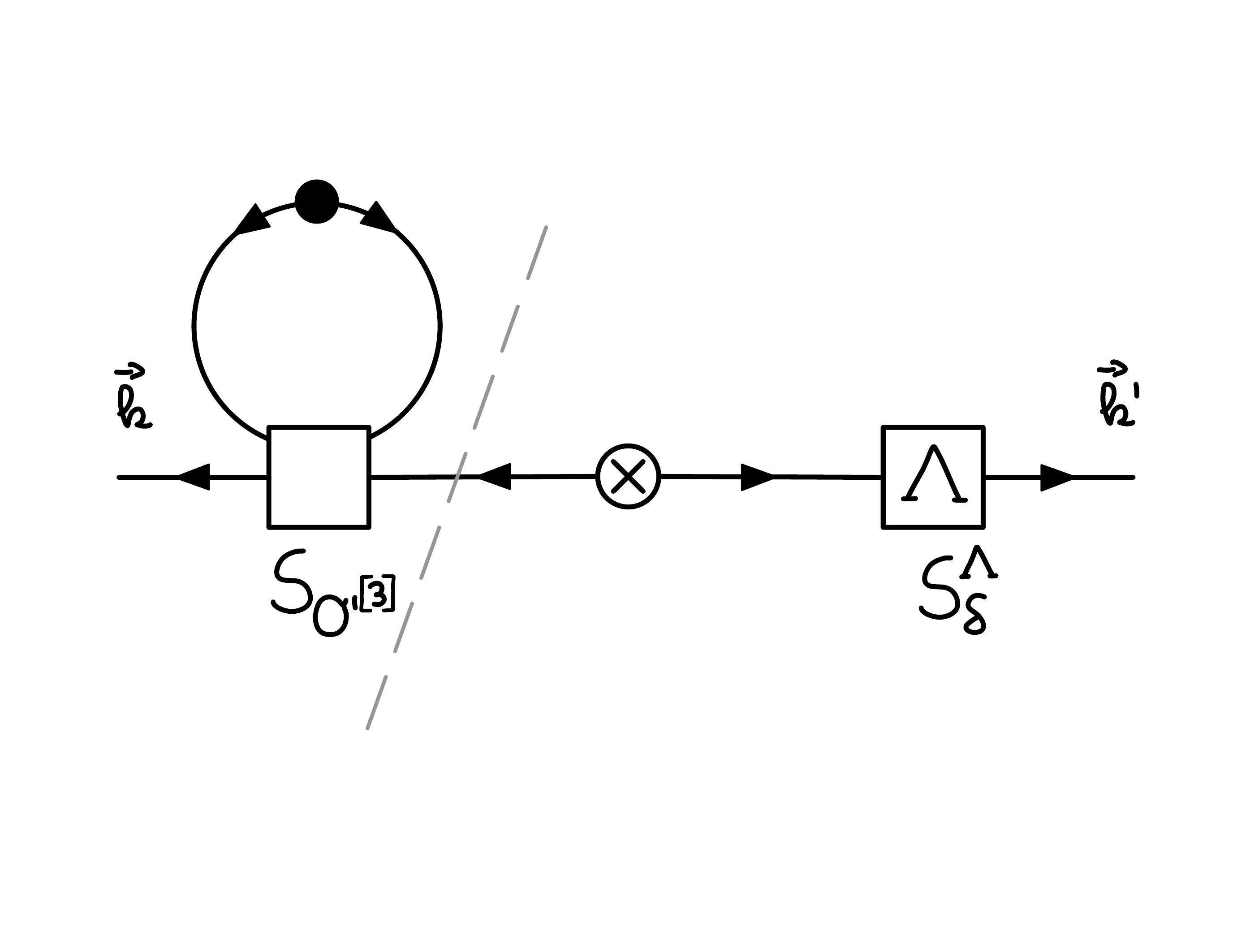}} \\
&\;\;\;\; + \raisebox{-0.0cm}{\includegraphicsbox[scale=0.25,trim={2cm 7cm 2.25cm 7cm},clip]{plots/delta_delta_rhs-5.pdf}}\,\,. 
\label{eq:D4}
\end{split}
\ee
Here, one similarly obtains a counterterm 
\be
[O'^{[3]}](\vk) = O'^{[3]} (\vk) -3 \int_{|\vp|>\Lin} S_{O'^{[3]}}(\vp,-\vp,\vk) \Plin(p)\,\d(\vk)\,\,,
\label{eq:ct3}
\ee
which again either is given by $\d$ multiplied by a formally divergent constant
(for $O'^{[3]} \in \{ \d^3, \d K^2, K^3 \}$) or analytic terms (for $O'^{[3]} = O_\otd$).
This counterterm absorbs the contribution $\propto \Plin(p_1)$ in
$\< O'^{[3]}(\vk) \d^{(1)}(\vp_1)\>$ (\refeq{renormcond} for $O'^{[3]}$ and $l=1$),
which would violate the tree-level scaling of $0$.

Finally, we turn to the cross-correlation between two quadratic operators. 
There is a single contribution at leading order, i.e.~ 
\be
\< \big[O'^{[2]}[\d_\infty]\big](\vk ) O^{[2]}[\d_\L](\vk') \>\Big|_{\rm LO} = 
\raisebox{-0.0cm}{\includegraphicsbox[scale=0.25,trim={2.5cm 6cm 2.5cm 6cm},clip]{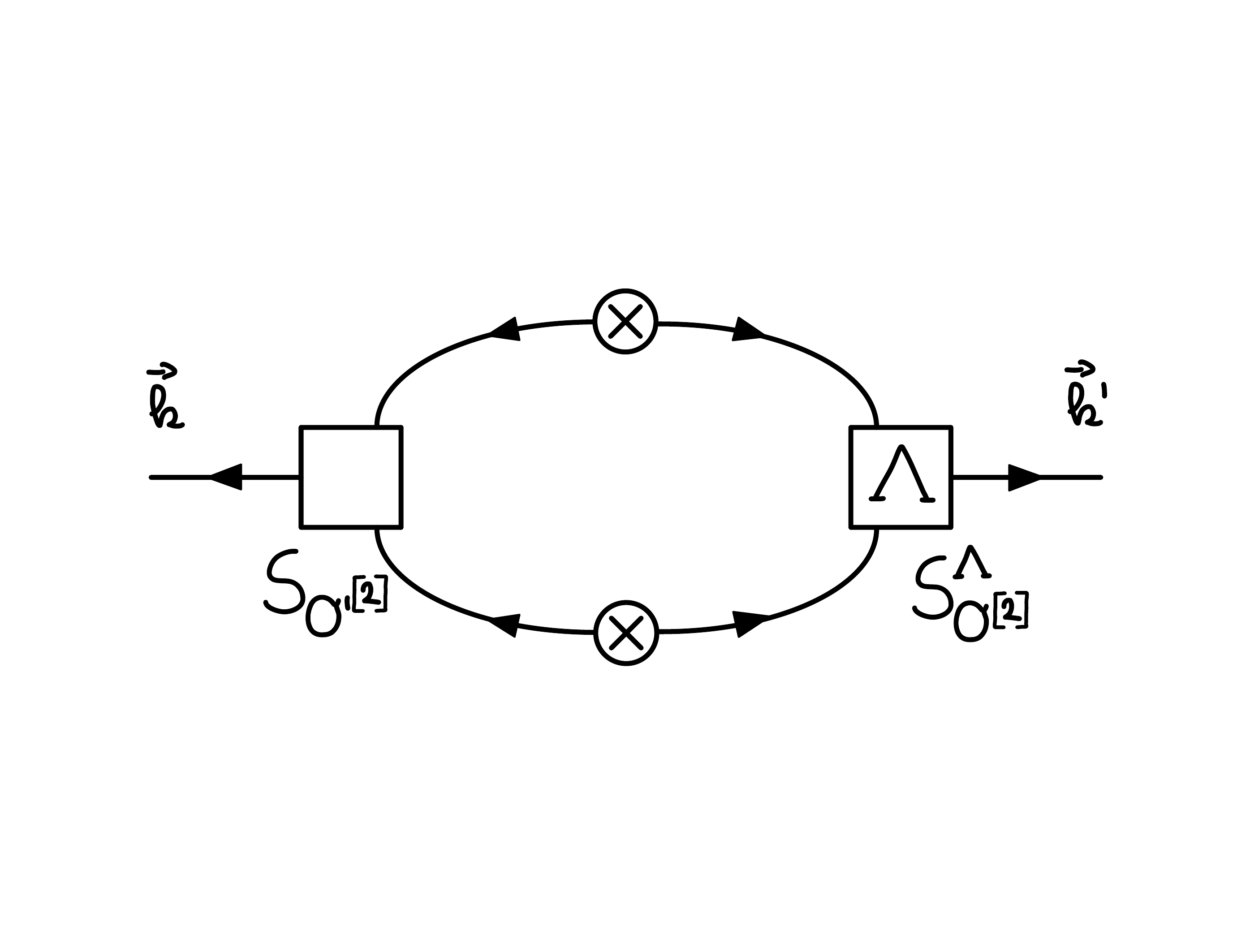}}\,\,.
\label{eq:D5-LO}
\ee
This contribution does not
involve any unregularized loop integrals. At NLO, we can distinguish eight contributions. 
Four of these do not have to be regularized: they are 
\be
\begin{split}
&\< \big[O'^{[2]}[\d_\infty]\big](\vk ) O^{[2]}[\d_\L](\vk') \>\Big|_{\rm NLO} \supset \\
&\raisebox{-0.0cm}{\includegraphicsbox[scale=0.25,trim={2.5cm 4.2cm 2.5cm 3.75cm},clip]{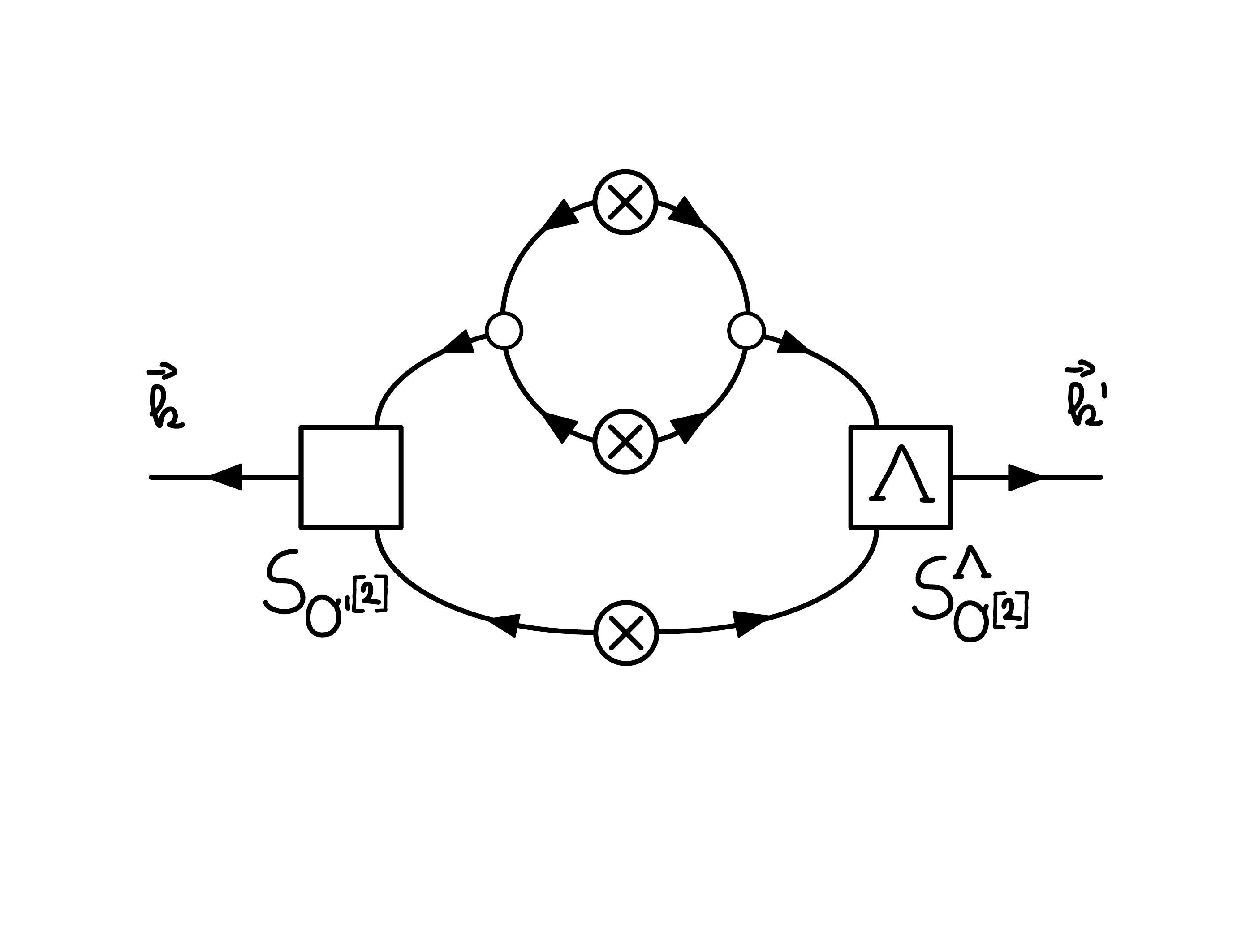}} 
+ \raisebox{0.575cm}{\includegraphicsbox[scale=0.25,trim={2.5cm 4.2cm 2.5cm 3.75cm},clip]{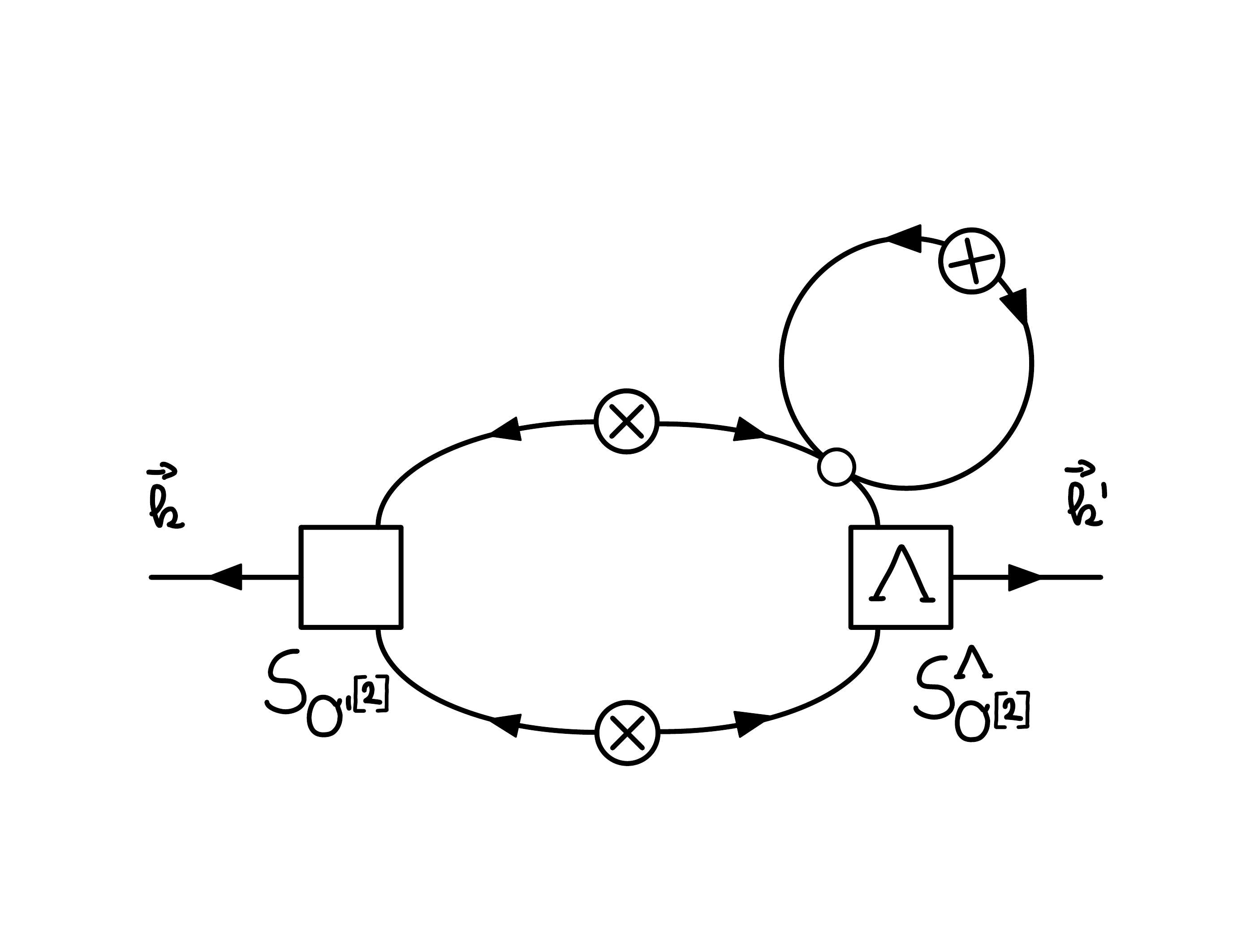}} \\
&+ \raisebox{-0.0cm}{\includegraphicsbox[scale=0.25,trim={2.5cm 5.75cm 2.5cm 5.75cm},clip]{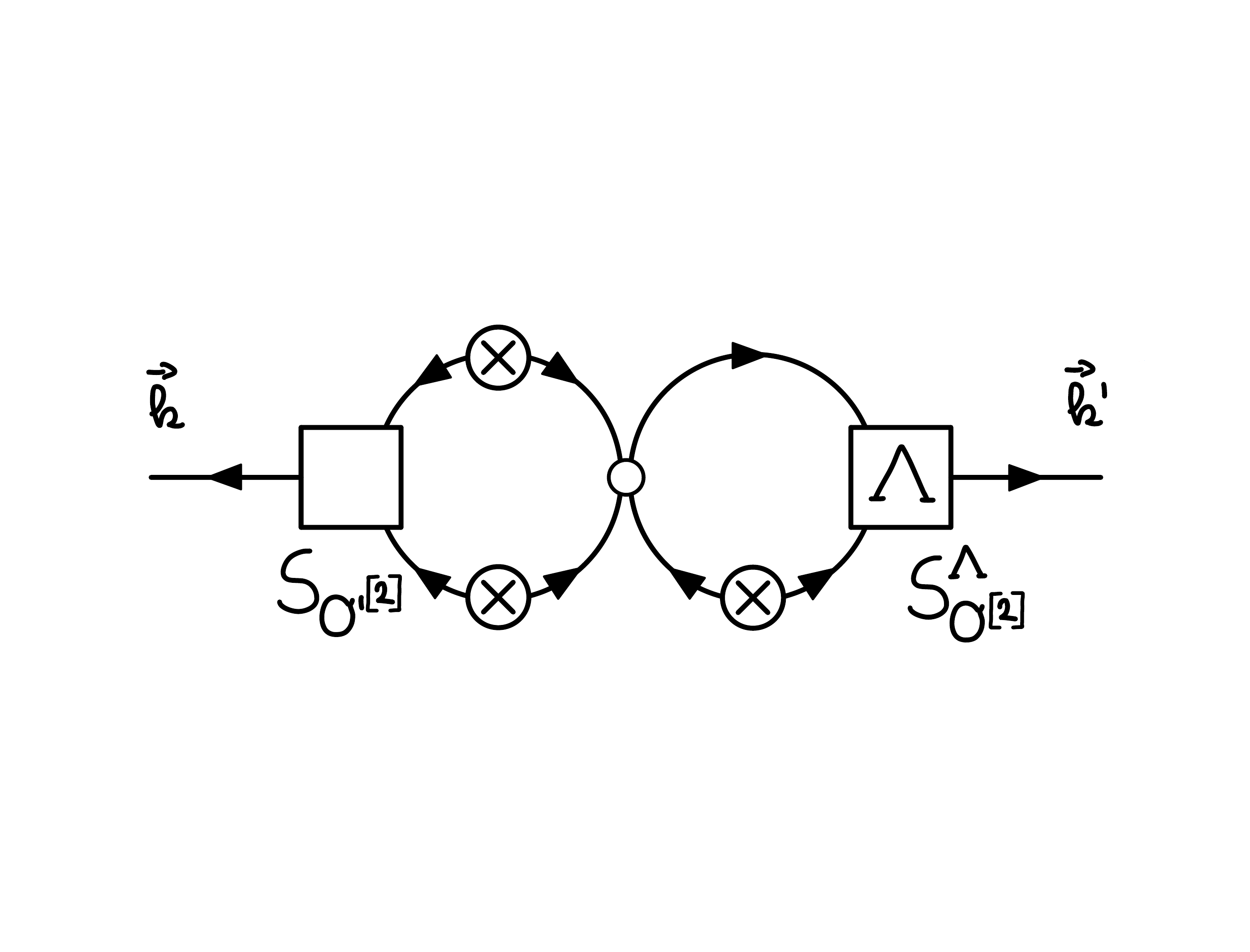}} 
+\, \raisebox{-0.0cm}{\includegraphicsbox[scale=0.25,trim={0cm 5.75cm 0cm 5.75cm},clip]{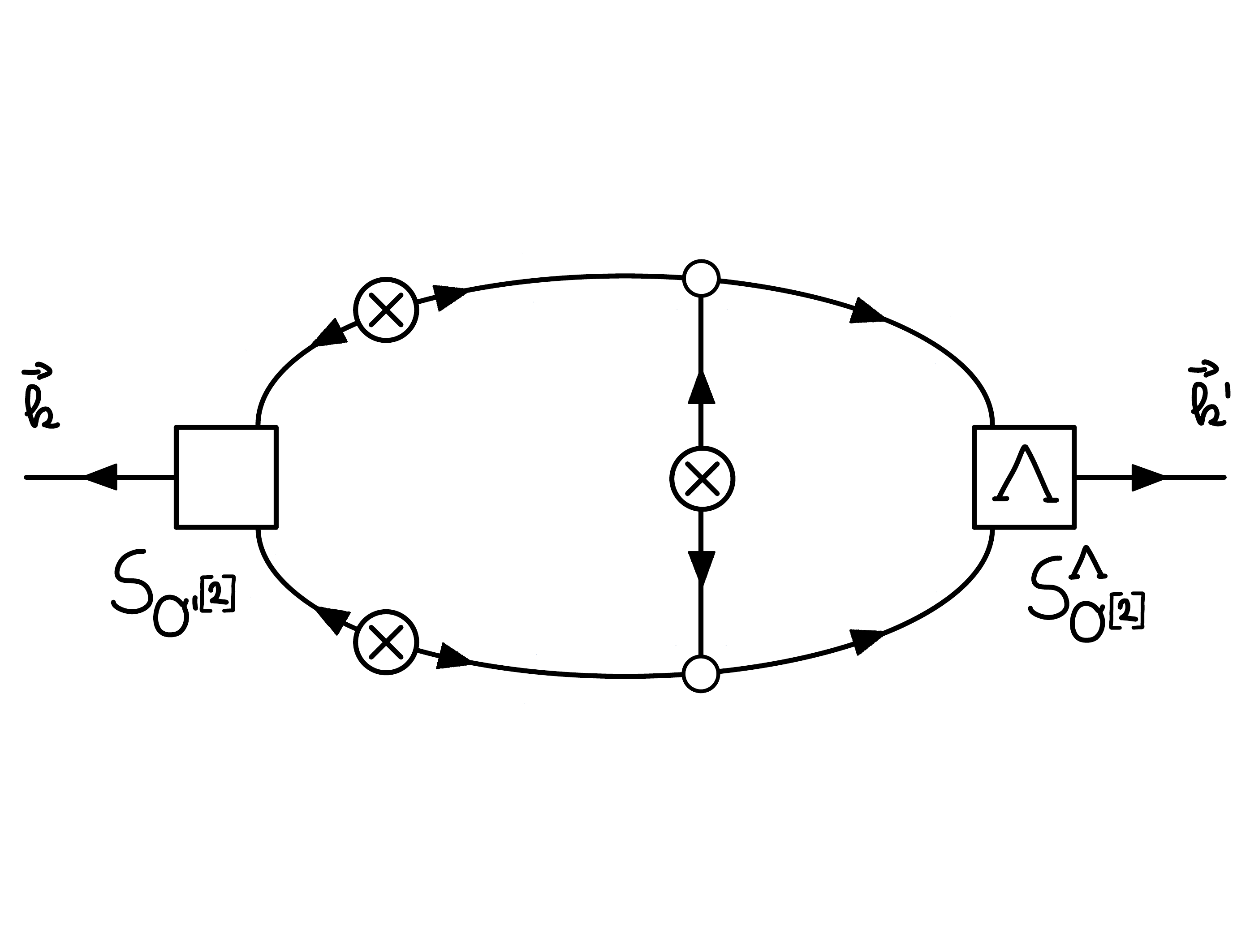}}\,\,. 
\label{eq:D5-LO-1}
\end{split}\ee
Notice that if one were to send $\Lin\to\infty$, all diagrams here would contain unregularized loop integrals. Of these, the two diagrams on the last line would moreover lead to a mismatch in \refeq{corrvs} which cannot be absorbed 
by counterterms, analogously to the first contribution in Eq.~\eqref{eq:D3}
discussed above.

The other four NLO diagrams need to be regularized for any value of $\Lin$. The parts to be regularized are indicated with dotted lines, where 
the counterterms (which we do not write for simplicity) absorb the part of the diagram 
to the left of the dotted line. They are 
\be
\begin{split}
&\< \big[O'^{[2]}[\d_\infty]\big](\vk ) O^{[2]}[\d_\L](\vk') \>\Big|_{\rm NLO} \supset \\
&\raisebox{+0.615cm}{\includegraphicsbox[scale=0.25,trim={2.5cm 2.75cm 2.5cm 3.75cm},clip]{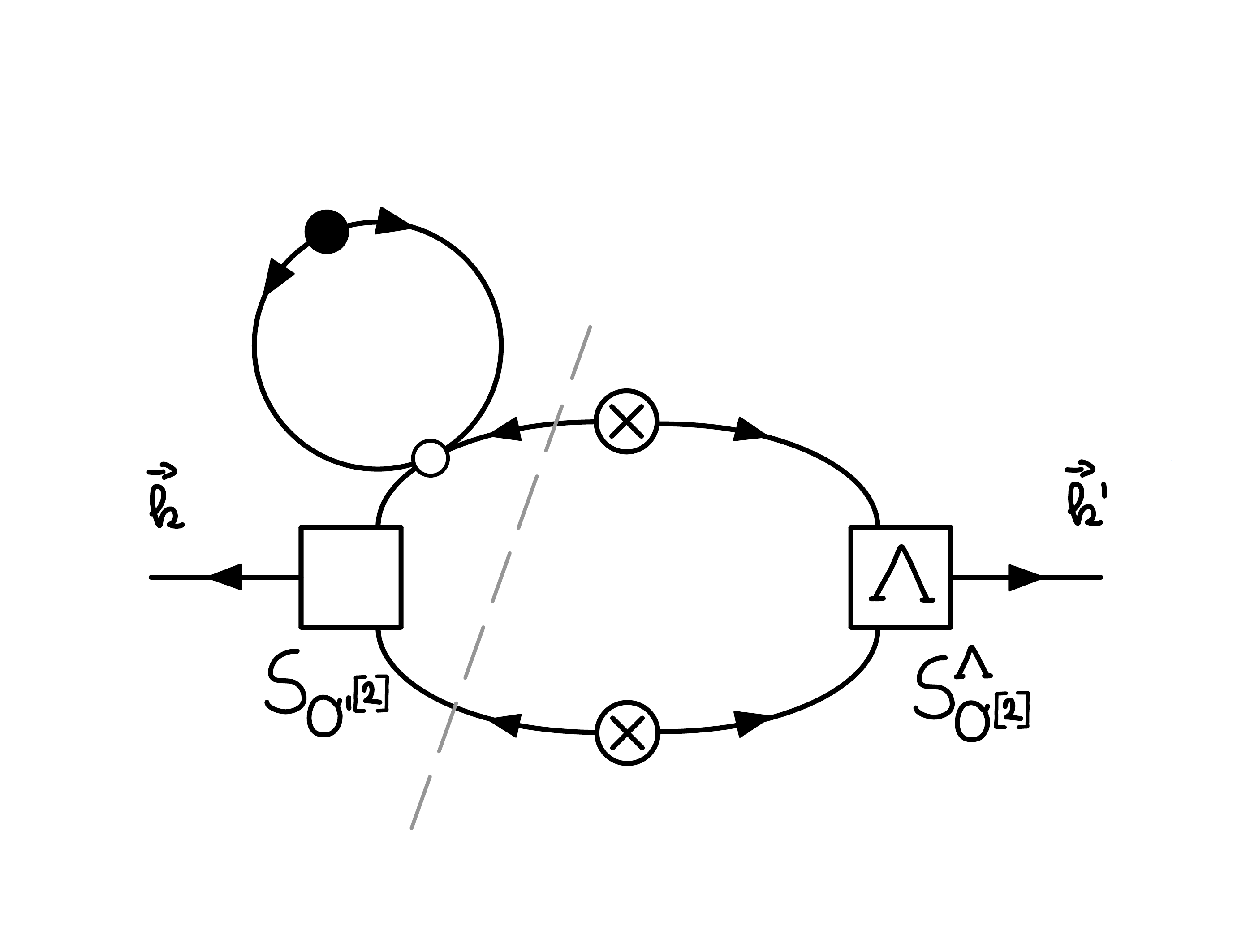}} 
+ \raisebox{+0.075cm}{\includegraphicsbox[scale=0.25,trim={2.5cm 4.2cm 2.5cm 3.75cm},clip]{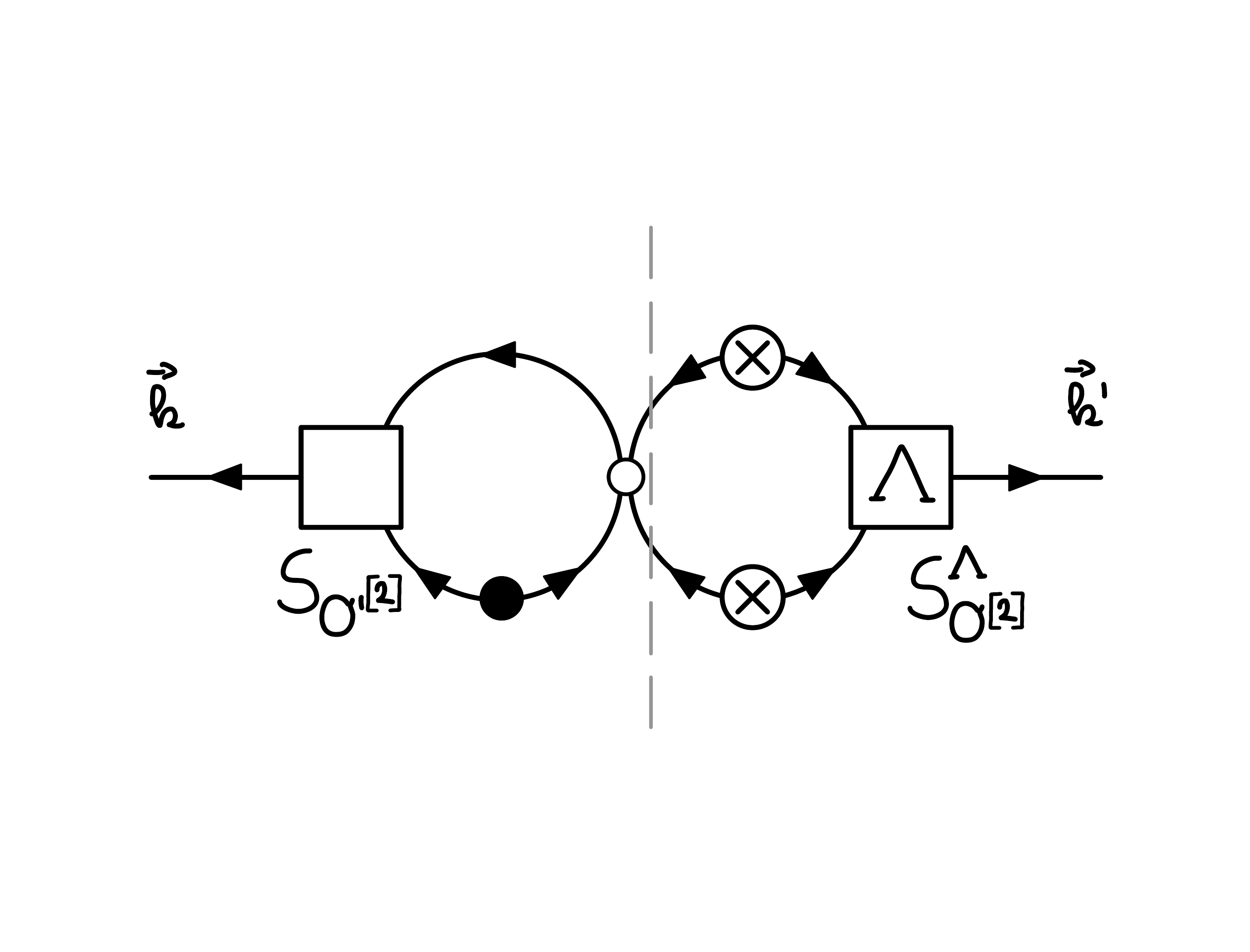}} \\
&+ \raisebox{-0.0cm}{\includegraphicsbox[scale=0.25,trim={0.2cm 5cm 0.2cm 5.1cm},clip]{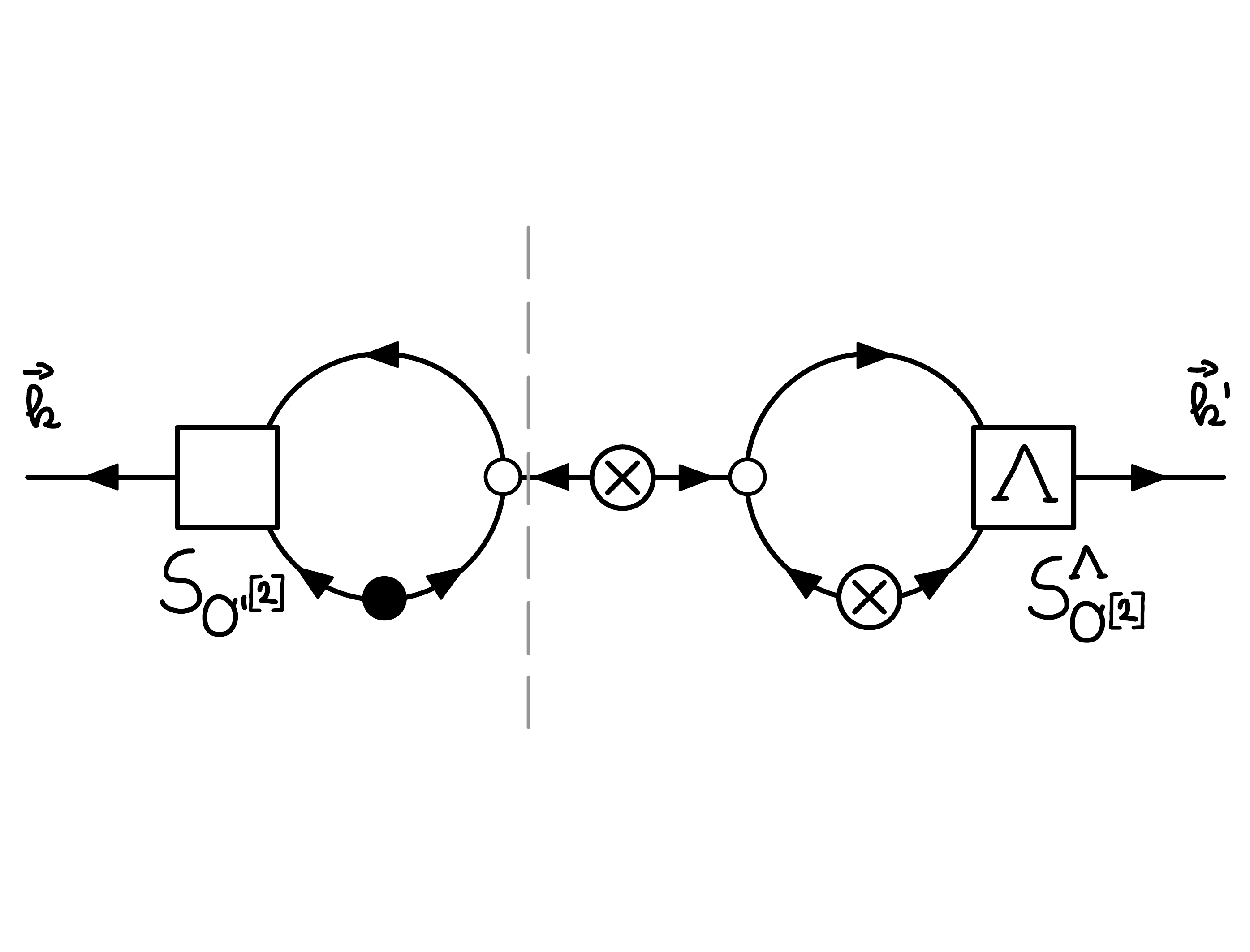}} 
+ \raisebox{-0.0cm}{\includegraphicsbox[scale=0.25,trim={0.2cm 5cm 0.2cm 5.1cm},clip]{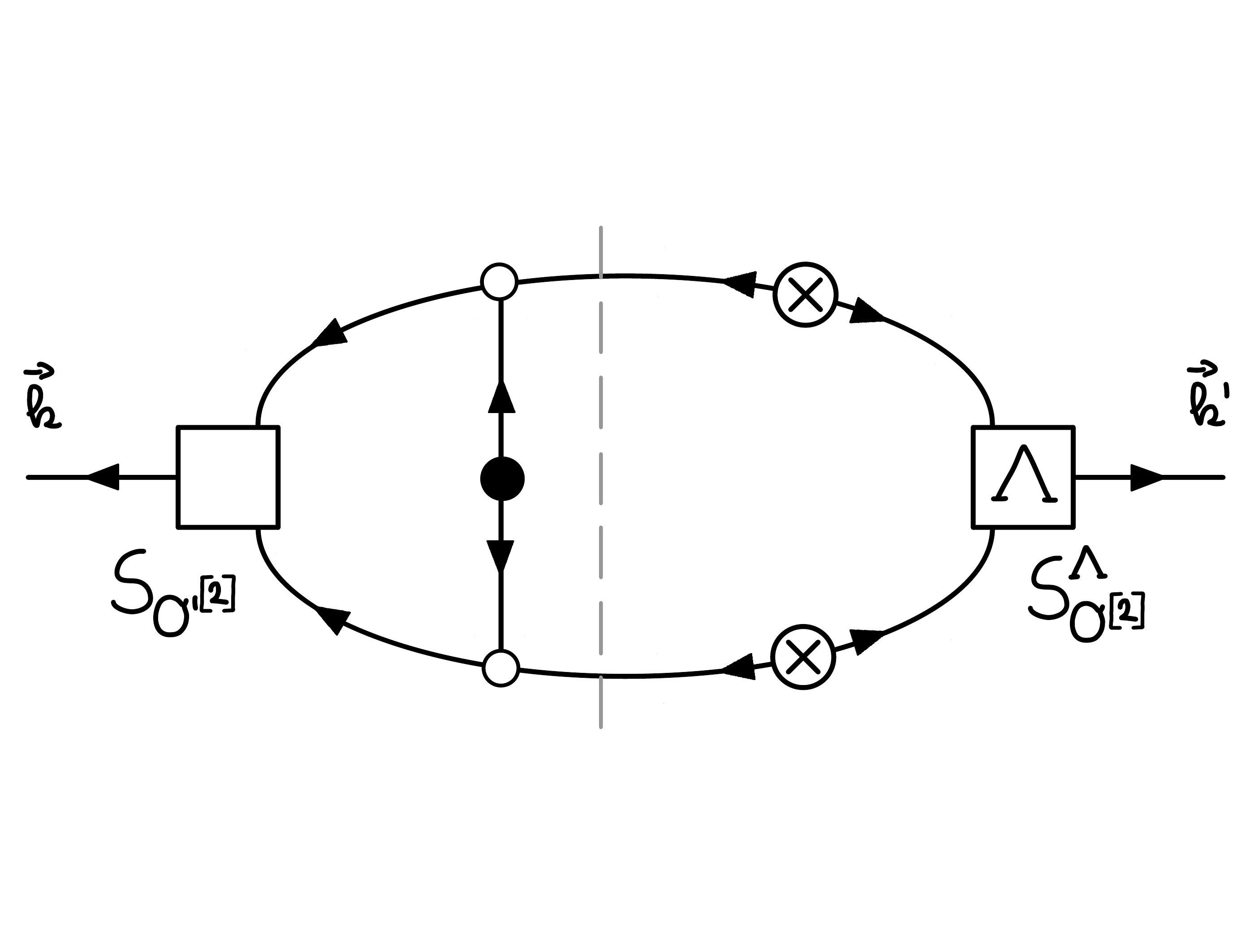}}\,\,. 
\label{eq:D5-LO-2}
\end{split}
\ee
The third diagram in this equation is already absorbed by the counterterm in \refeq{ct2}.
The remaining three are removed by counterterms to $[O'^{[2]}]$ that are proportional to
second-order operators $\widetilde{O}^{[2]}$ (the first two of which have similar
structure as \refeq{ct1} and \refeq{ct2}, respectively). These three counterterms
ensure the renormalization conditions
\refeq{renormcond} for $O'=O'^{[2]}$ and $l=2$ \cite{2014JCAP...08..056A}. 

This reasoning can be extended analogously to higher orders. The conclusion
is that, once a cutoff in the initial conditions is imposed,
all differences between the two types of correlators appearing
on the left- and right-hand sides [\refeq{corrvs}] of the maximum-a-posteriori point of \refeq{maxlikeG2} are absorbed by counterterms to the
operators $[O']$ appearing in the renormalized halo bias expansion.

In \refapp{MAPdelta}, we explicitly compute \refeq{maxlikeG2} for $O=O'=\delta$
to illustrate this reasoning quantitatively. We also show there that
the remaining residuals in \refeq{maxlikeG2} can indeed be absorbed by
counterterms, while this does not hold if one were to set $\Lin\to\infty$.

\section{Numerical implementation}
\label{sec:num}

All numerical tests presented below are based on the same set of
N-body simulations used in \cite{paperII}, which were presented in
\cite{2017MNRAS.468.3277B}. They are generated using
\textsc{GADGET-2} \cite{2005MNRAS.364.1105S} for a flat $\Lambda$CDM cosmology
with parameters $\Omega_\mathrm{m} = 0.3$, $n_\mathrm{s} = 0.967$, $h = 0.7$,
and $\se = 0.85$, a box size of $L = 2000 \, h^{-1}
\mathrm{Mpc}$, and $1536^3$ dark matter particles of mass
$M_\mathrm{part} = 1.8 \times 10^{11} \, h^{-1} M_{\odot}$.
Two realizations are available, which we refer to as ``run 1'' and ``run 2.'' 
Dark matter halos
were subsequently identified at different redshifts as spherical
overdensities \cite{1974ApJ...187..425P, 1992ApJ...399..405W,1994MNRAS.271..676L} applying the Amiga Halo Finder algorithm
\cite{2004MNRAS.351..399G, 2009ApJS..182..608K} with an
overdensity threshold of $200$ times the background matter density.
Given that the EFT approach should apply to any physical tracer, we
  expect the same conclusions for other halo definitions, e.g. friends-of-friends.
We present results for four logarithmic mass bins, each at three
redshifts.

Our lowest mass bin ranges from $10^{12.5} \Msunh$ to
$10^{13} \Msunh$. The halos in this bin contain 18 or more member particles,
while the mean mass corresponds to 30 particles. Halos with fewer than
30 particles can not necessarily be reliably identified with bound
structures at this mass, so results on the bias and stochastic parameters
for this mass bin should be taken with a grain of salt. Nevertheless,
the evolution of the collection of particles making up these low-mass halos
is still governed by local dynamics, so we expect the general bias expansion
to also describe such poorly resolved halos, and hence lead to an unbiased estimate of \se.

We employ two different approaches to generate the forward-evolved matter field to be used in the construction of the bias operators entering the likelihood.
The first is to generate particle positions using second-order Lagrangian
perturbation theory (2LPT) at the desired final redshift. We
employ the 2LPTic code \cite{2006MNRAS.373..369C, 2012PhRvD..85h3002S},
the same code used to generate the initial conditions of the full N-body
simulations at $z_{\rm ini}=99$,
and refer to this as ``2LPT density field.'' The second is to generate
2LPT particle positions at $z_{\rm ini}=99$, and perform an N-body simulation using
the same settings as the simulations described above, evolved to the desired
redshift. We refer to this as ``N-body density field.''
In both cases, we only populate modes with $k \leq \Lin$ when the initial,
linear displacement field is sampled in the 2LPTic code.
In order to generate grid representations of density fields, we assign particles to
grids of size $512^3$ using
a leading-order Fourier-Taylor expansion as described in \refapp{FT}. We choose
this assignment scheme rather than cloud-in-cell as its kernel shape is much closer
to the desired sharp-$k$ filter, avoiding the need for first assigning to a
high-resolution grid. The same assignment scheme is used for halos. 

Notice that for each value of $\Lin$, we need to generate density fields for a range of
\se. For this reason, we only generated N-body density fields for a single value
of $\Lin = 0.1\iMpch$. Specifically, the \se values are
\ba
\se &\in \{ 0.65,\	0.75,\	0.80,\	0.83,\	0.85,\	0.87,\	0.90,\	0.95,\	1.00,\	1.10,\	1.20 \} \quad\mbox{(2LPT)}\,\,, \vs
\se &\in \{ 0.78,\	0.83,\	0.85,\	0.87,\	0.92 \} \quad\mbox{(N-body)}\,\,,
\ea
where $\se = \se^{\rm fid} = 0.85$ is the value used for the simulations of \cite{2017MNRAS.468.3277B} that provide our ground truth.

At fixed $\Lin$, halo sample, and redshift, we find the profile likelihood
${-2}\ln P^\text{prof}(\se^i)$ by searching for the maximum in the
$\{b_1,\sigma_\eps,\sigma_{\eps,2}\}$ space, employing the MINUIT algorithm \cite{James:1975dr} as described in \cite{paperII}.
This procedure results in a set of values $\{ \se^i,\, -2\ln P^\text{prof}(\se^i) \}_i$
which we find is fit well by a parabola in all cases (we disregard a small number of isolated cases where the minimization failed to converge).
The best-fit value $\hse$ is given by the location of the minimum of the
best-fit parabola, while the estimated $1\sigma$ error on $\hse$ is given by the inverse square-root of the curvature of the parabolic fit.
For convenience, we phrase results in terms of
\be
\hat\alpha \equiv \frac{\hse}{\se^{\rm fid}}
\label{eq:alphadef}
\ee
below, so that $\hat\alpha=1$ corresponds to a perfectly unbiased
inference of \se. 
We emphasize that the quoted error on $\hse$ or $\hat\alpha$ does not include any residual cosmic variance,
and is essentially purely governed by the halo stochasticity which appears in
the variance of the likelihood. It is worth noting that the error reported on
  $\hse$ also includes the degeneracy between the bias parameters which are
  marginalized over, such as $b_1$, and $\hse$.

\begin{figure*}[tbp]%
	\centerline{\resizebox{\hsize}{!}{
		\includegraphics*{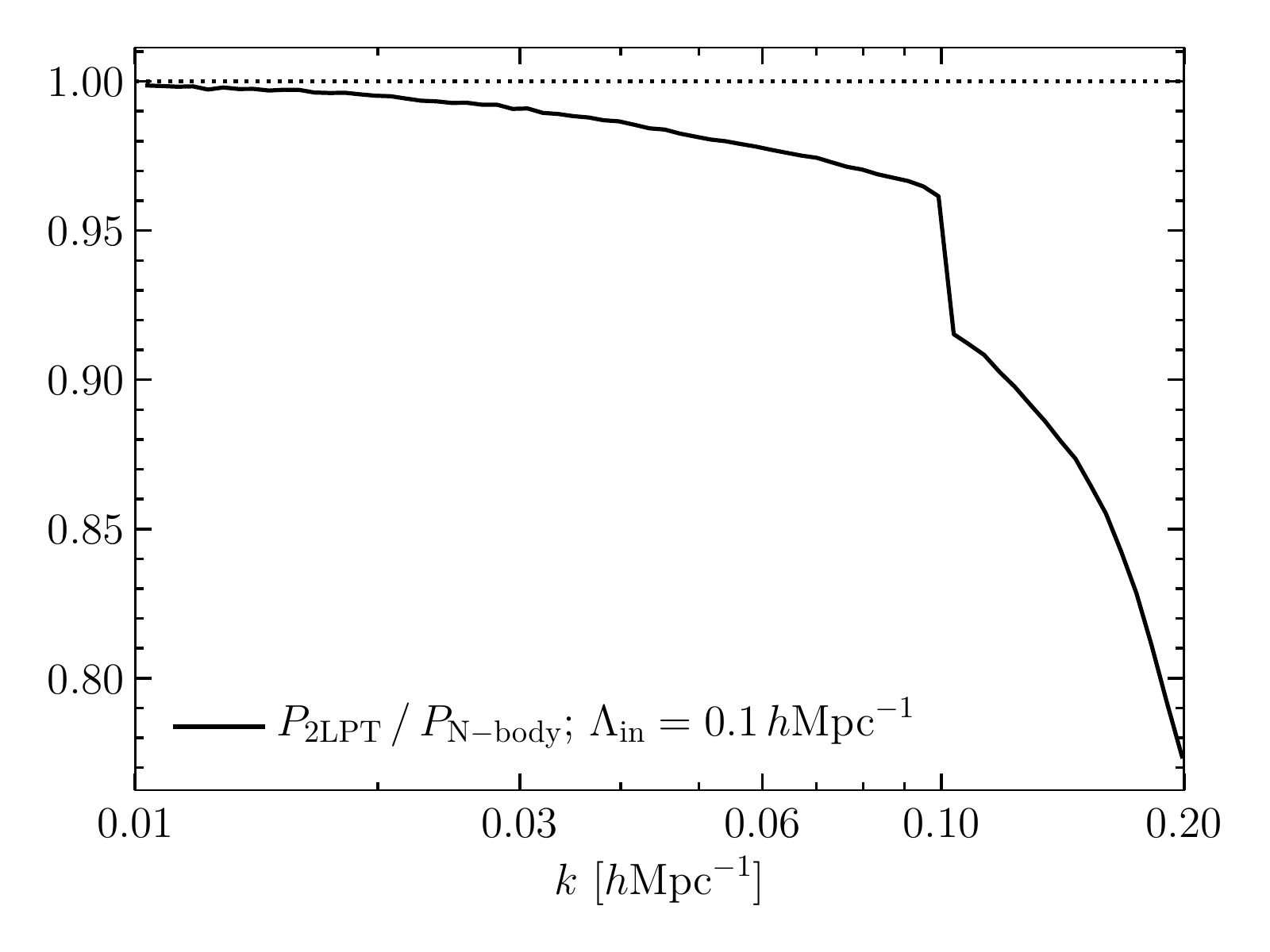}
		\includegraphics*{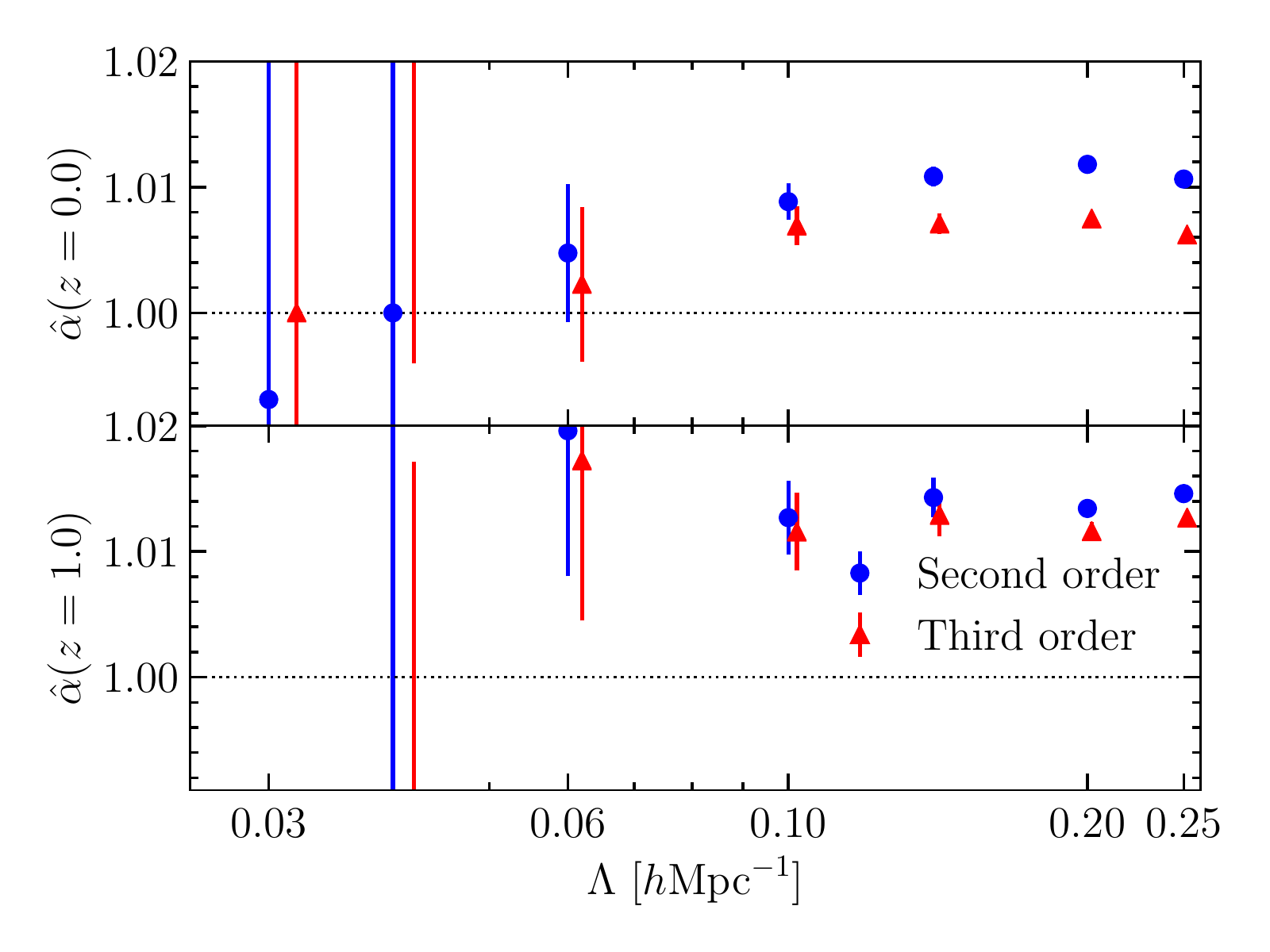}
		}}
	\cprotect\caption{\emph{Left panel:} Ratio of power spectra of the evolved matter density field at $z=0$ for 2LPT and N-body when a momentum cutoff of $\Lin=0.1\iMpch$ is employed. \emph{Right panel:} Maximum-likelihood (ML) values for $\alpha$ [\refeq{alphadef}] from the profile likelihood applied to subsampled N-body particles with mean number density $\bar n_m = 0.01 (\iMpch)^{3}$. Results are based on the 2LPT matter density field and second- or third-order bias expansions, as labeled. The top (bottom) panel shows results at $z=0$ ($z=1$).
	\label{fig:dm}}
\end{figure*}

Before moving on to the results, we justify our choice of $\kmax = \L = \Lin$.
The left panel of \reffig{dm} shows the ratio of power spectra of the evolved
matter density field using 2LPT and N-body forward evolution as described
above. In both cases, a cutoff of $\Lin=0.1\iMpch$ is employed. We see that
the disagreement between 2LPT and N-body rapidly worsens for $k > \Lin$.
This is because the modes with $k > \Lin$ are exclusively excited by nonlinear
evolution, with leading contributions that are progressively higher order as
$k/\Lin$ grows; specifically, modes with $k > n \Lin$ are only generated at $(n+1)$-th
order in perturbations. Since the 2LPT density field is only correct up to second order in perturbations, this leads
to a worse description of the density field at $k > \Lin$. For this reason,
we conservatively choose to only use modes in the evolved density fields with momenta
less or equal to $\Lin$, corresponding to setting $\kmax = \L = \Lin$.

\section{Results}
\label{sec:results}

\subsection{Test on N-body particles}
\label{sec:dm}

Before turning to halos, we begin with a simple test case of a trivially
biased tracer. Specifically, we construct a tracer density field by
randomly subsampling N-body particles to a desired number density $\bar n_{m}$.
This ``tracer'' is thus perfectly linearly biased with respect
to the full, nonlinear N-body density field, and allows us to
test the accuracy of the 2LPT matter forward model in terms of the
\se inference.

The right panel of \reffig{dm} shows the resulting maximum-likelihood (ML)
value $\hat\alpha$ as a function of $\Lin = \L = \kmax$, at redshifts
$z=0$ (top) and $z=1$ (bottom). We show results for both the second- and third-order
bias expansions where, in case of a perfect matter forward model,
one expects all bias parameters apart from $b_1$ to be consistent with zero.

Several aspects of these results are noteworthy:
$(i)$ The overall accuracy of \se for subsampled N-body particles is on the order of $1\,\%$ 
up to $\L = 0.25\iMpch$,
with a slightly better performance in case of third-order bias. This indicates
that the bias parameters absorb some of the deficiencies of the
2LPT forward model. $(ii)$ We see convergence toward $\L \to 0$ at $z=0$,
as expected. $(iii)$ The accuracy in \se is not improved toward higher redshift. This appears to indicate that the displacement contributions to the 2LPT density field that the \se inference builds on are not closer to those in the N-body density field at $z=1$ compared to $z=0$. It would be interesting to explore whether this is due to numerical reasons or an effect that is physically expected. 

One further notices from the right panel of \reffig{dm} that the error
bars on $\hat\alpha$ grow rapidly as $\L$ is reduced to below $0.06\iMpch$.
This is due to two reasons. First, at fixed simulation volume, the number
of available modes rapidly shrinks toward smaller $\L$. Second, the
shape of the matter power spectrum changes: while $\Plin(k) \propto k^{-1.5}$
at $k\simeq 0.2 \iMpch$, it gradually becomes shallower toward lower $k$,
with $\Plin(k) \sim {\rm const}$ at $k \simeq 0.02 \iMpch$.
As argued in \cite{cabass/schmidt:2019}, the EFT likelihood is based on
different scalings with wavenumber of the deterministic (contained in $\d_{h,\rm det}$)
and stochastic contributions (contained in $\sigma^2(k)$),
which are in turn controlled by the power-law index of $\Plin(k)$. 
As one reduces $\L$ to values approaching $0.02\iMpch$, the effective index of $\Plin(k)$ approaches zero
(since modes with $k \ll \L$ contribute fairly little due to their small number),
at which point all contributions become degenerate. In order to counteract
this effect, one would require either enormous simulation volumes, or
simulations with non-$\Lambda$CDM forms of $\Plin(k)$ that do not have a turnover,
such that the power spectrum index remains significantly
negative for small $k$. The latter would indeed provide interesting
possibilities to investigate the convergence properties rigorously.

We conclude from the test on subsampled N-body particles that we do not expect \se inference from halo
catalogs based on the 2LPT density field to be more accurate than $\sim 1\,\%$.

\begin{figure*}[tbp]%
	\centerline{\resizebox{\hsize}{!}{
		\includegraphics*{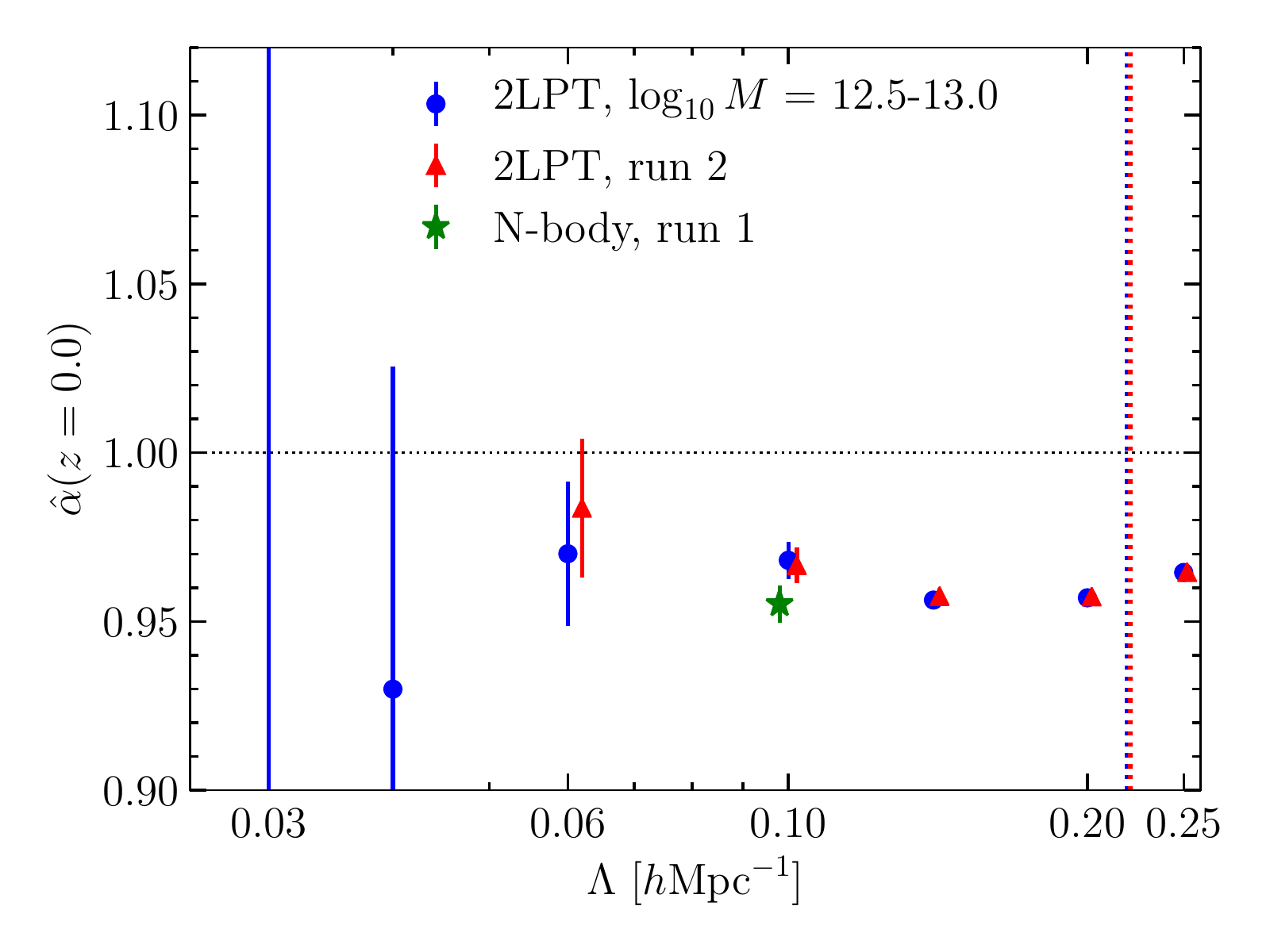}
		\includegraphics*{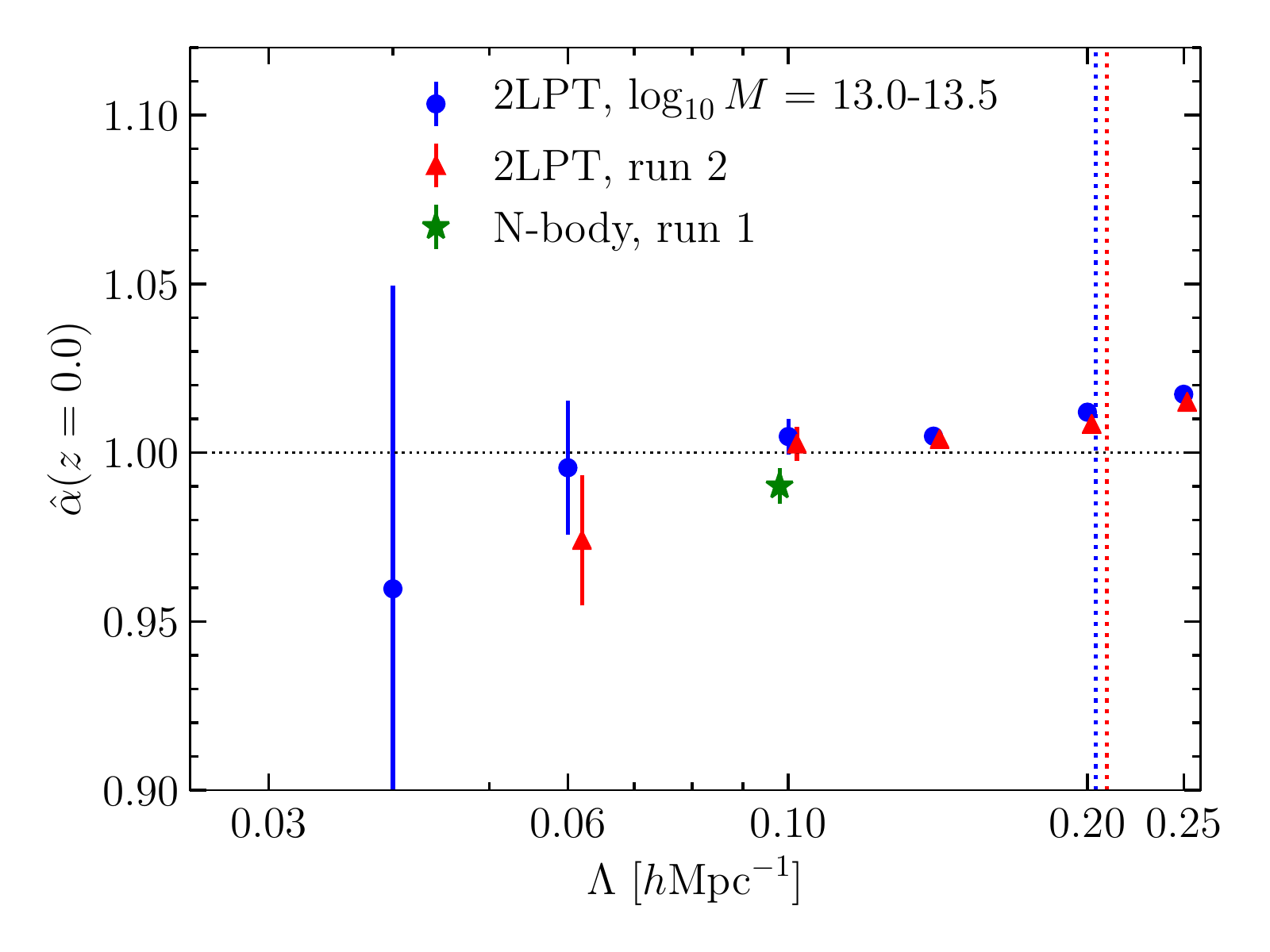}
		}}
	\centerline{\resizebox{\hsize}{!}{
		\includegraphics*{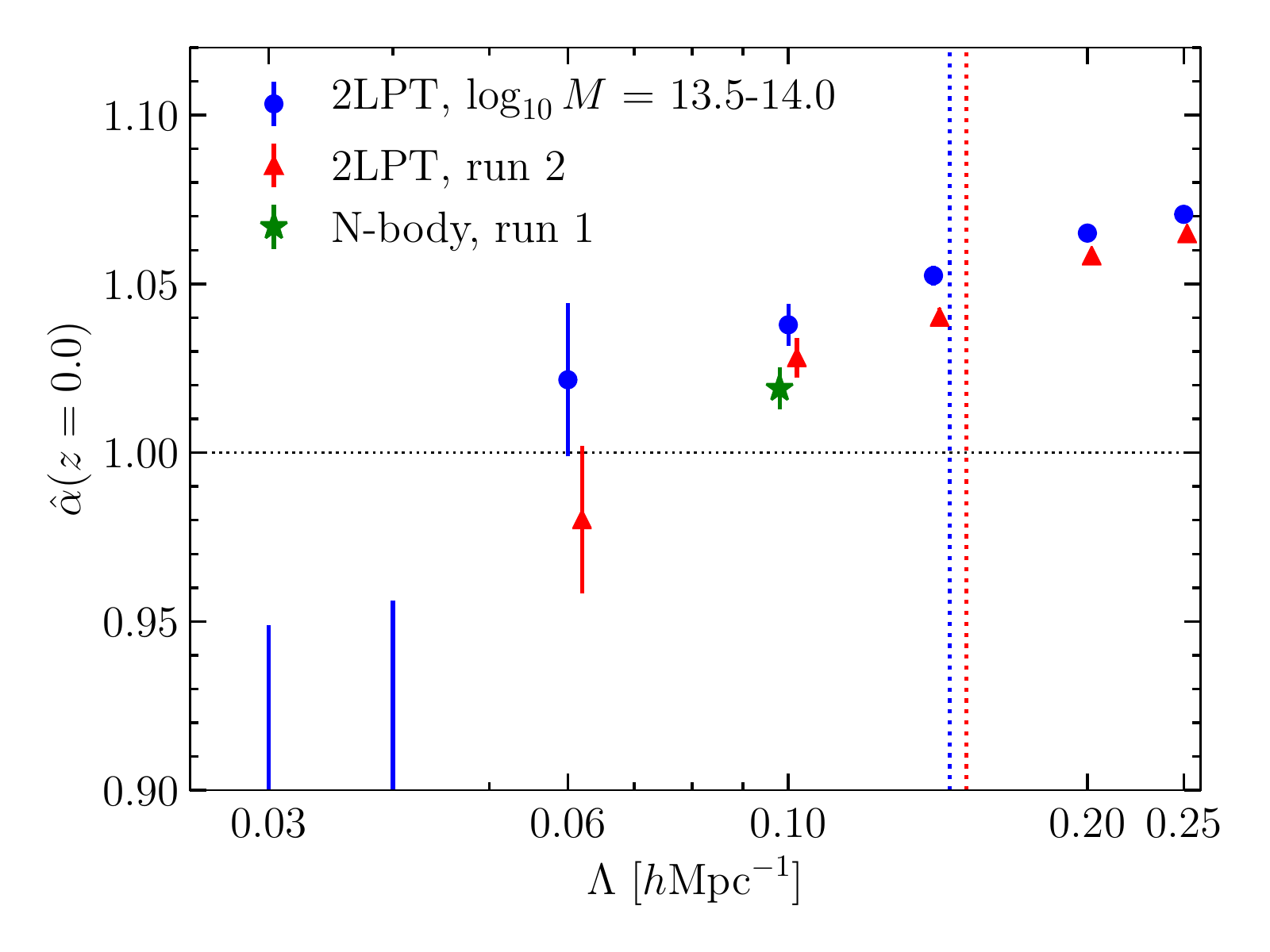}
		\includegraphics*{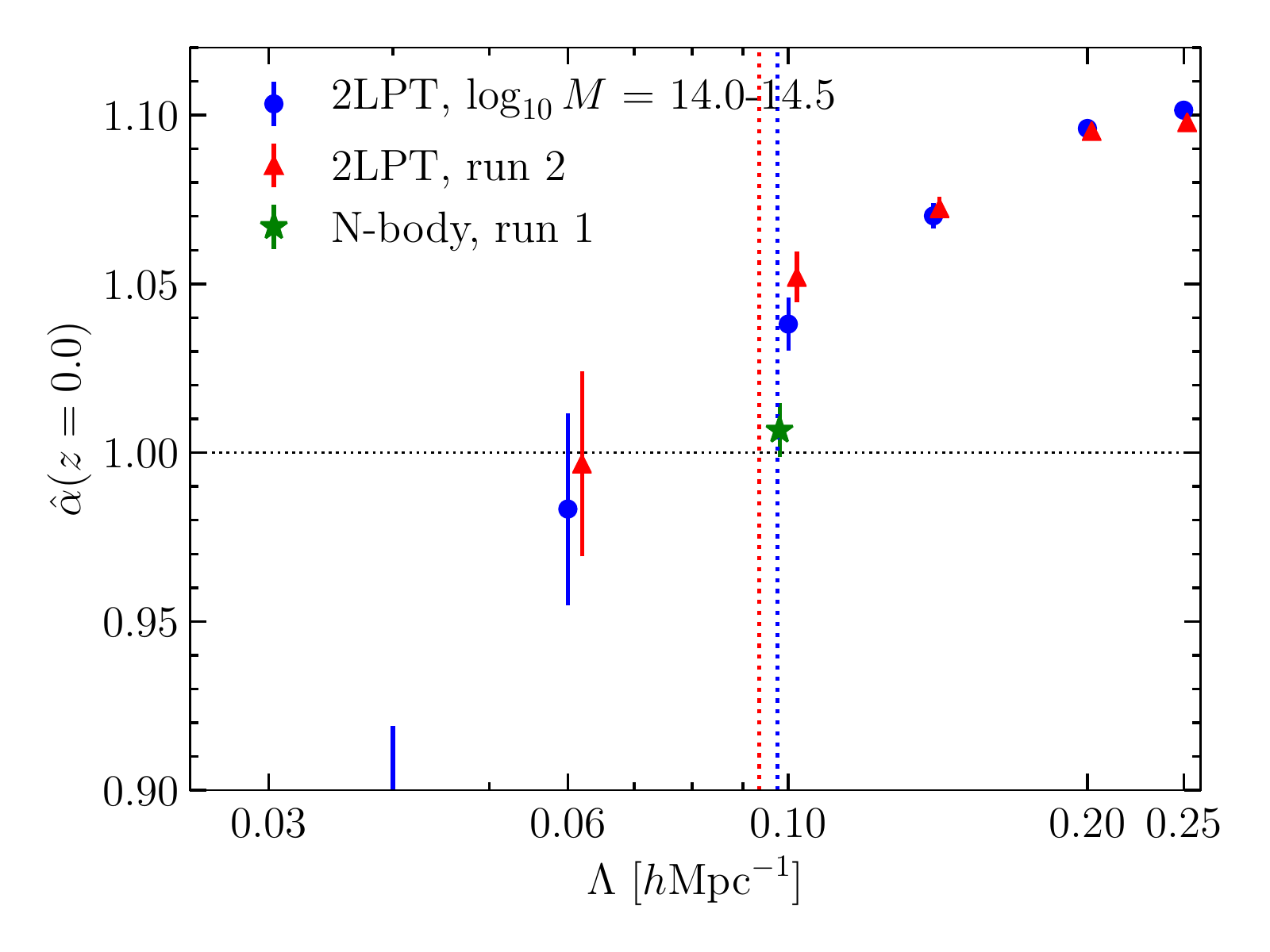}
		}}
	\cprotect\caption{ML values for $\alpha$ using the second-order bias expansion as a function of $\L = k_{\rm max} = \Lin$ at $z=0$. Shown here are different mass bins each for run 1 and 2 (in the legend of these and the following plots we drop the units on $M$, which is always understood to be in units of $\Msunh$). Values $\L < 0.06\iMpch$ are only available for run 1, while the results using an N-body forward-evolved matter density field are only available for $\L=0.1\iMpch$. The vertical dotted lines indicate the scale $\L = k_{\rm st}$ where $\nbarh b_1^2 \Plin(k_{\rm st})=1$ in each case; no significant further information is expected on scales much smaller than this.
	\label{fig:Linrun12}}
\end{figure*}%

\subsection{Second-order bias}

We now turn to the application to halo catalogs, beginning with the
second-order bias expansion. \reffig{Linrun12} shows results at $z=0$
for different halo mass bins, and the two simulation realizations
in each case. These give a rough indication of the expected cosmic
variance error.
We again see the expected convergence behavior as $\L \to 0$.
At $\L=0.1\iMpch$, $\se$ is recovered to within $7\,\%$ for all mass bins.
The quantitative results are summarized in \reftab{results}. Notice that the
impact of the higher-derivative contribution $\sigma_{\eps,2}$ to the noise
is numerically very small. We have found that fixing $\sigma_{\eps,2}=0$ in the
minimization leads to negligible shifts in the maximum-likelihood values
for \se. This is in keeping with the theoretical expectation that the higher-derivative
stochasticity is less relevant than higher-order bias terms \cite{cabass/schmidt:2019}.

\begin{figure*}[tbp]%
	\centerline{\resizebox{\hsize}{!}{
		\includegraphics*{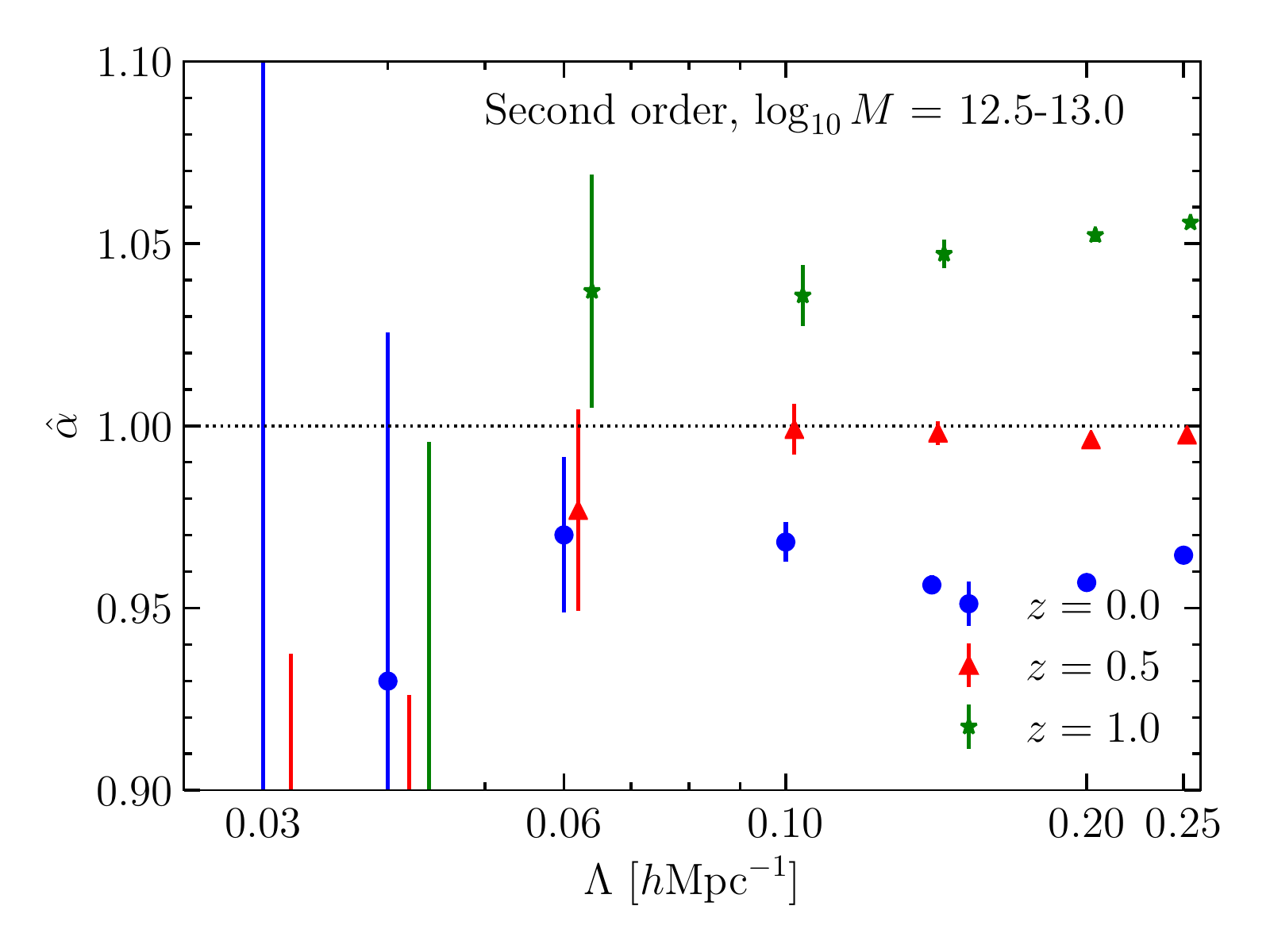}
		\includegraphics*{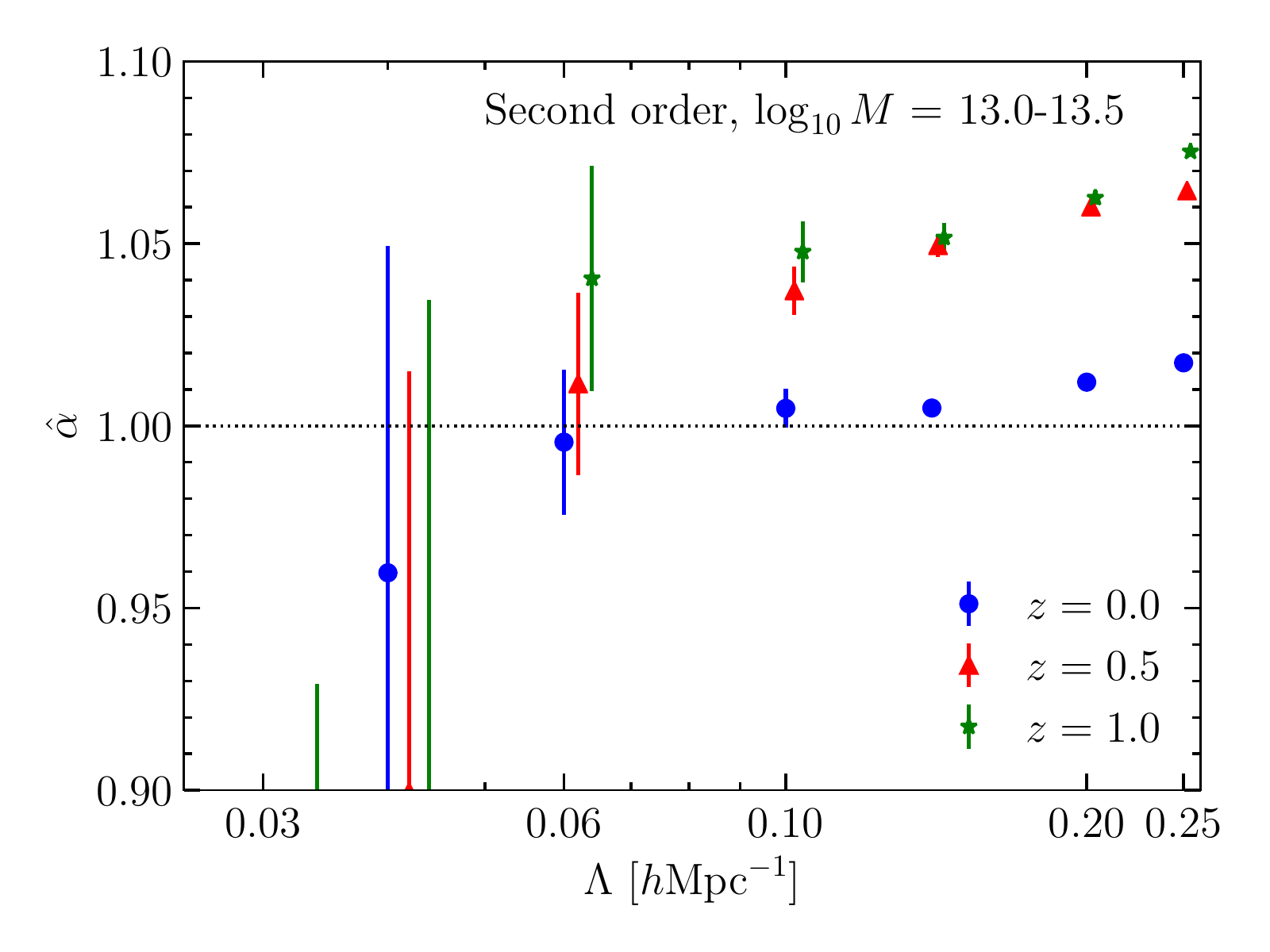}
		}}
	\centerline{\resizebox{\hsize}{!}{
		\includegraphics*{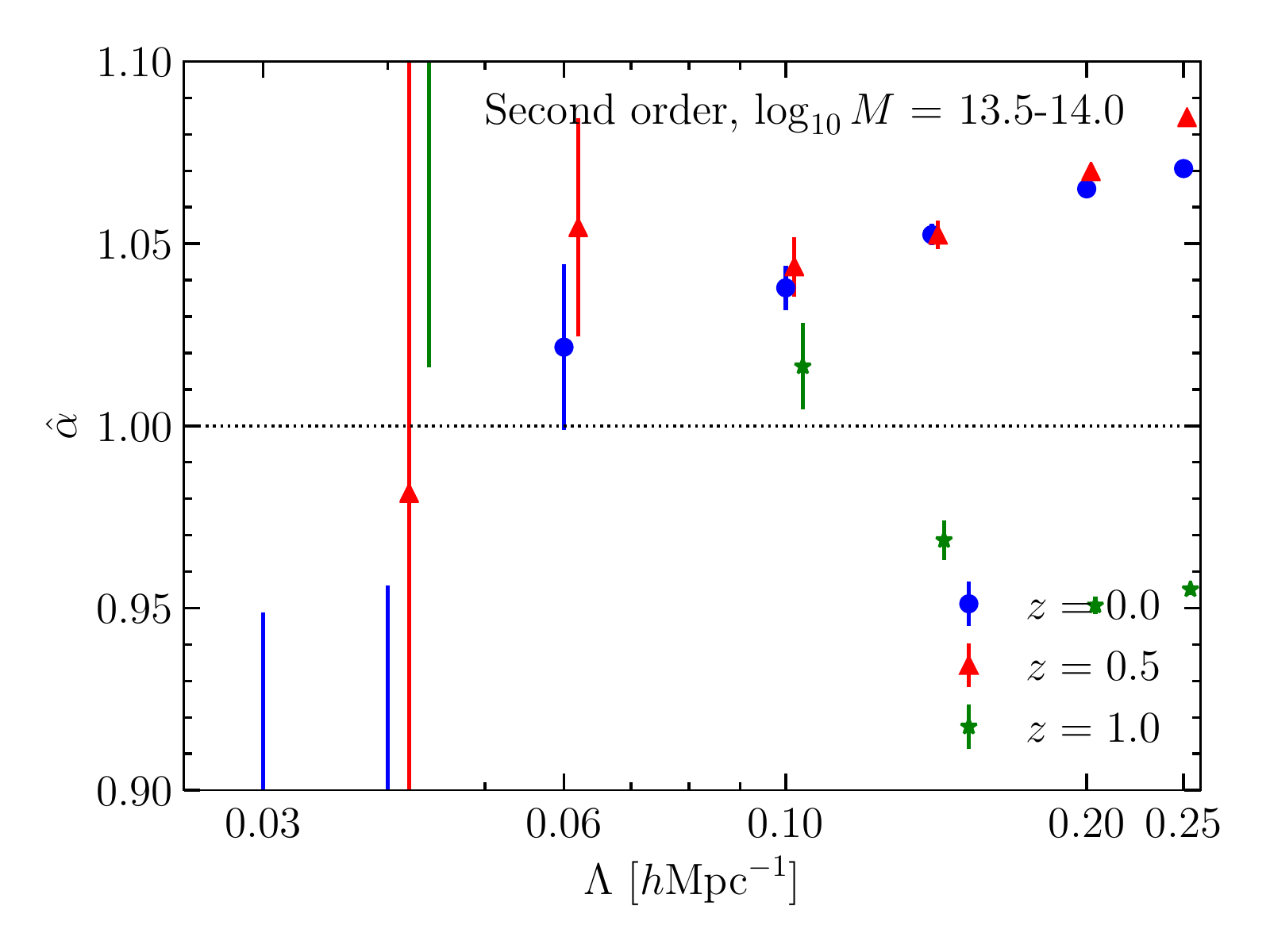}
		\includegraphics*{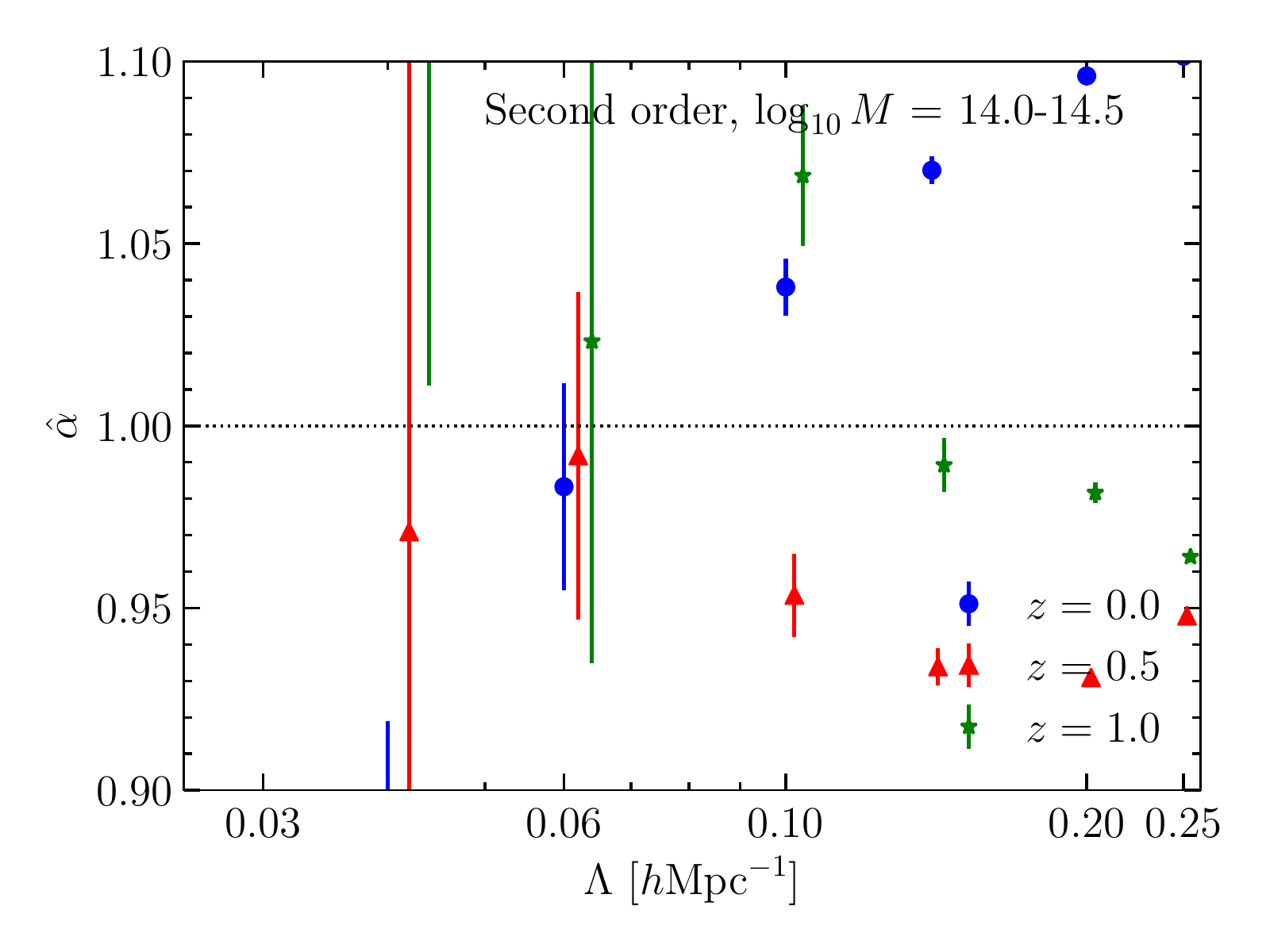}
	}}
	\cprotect\caption{ML values for $\alpha$ using second-order bias expansion as a function of $\L = k_{\rm max} = \Lin$. Different panels show the four mass bins, each at different redshifts for run 1. 
	\label{fig:Linvsz}}
\end{figure*}%

We have already discussed the rapidly growing error bars for $\L < 0.06\iMpch$,
which are due to the limited information available on those scales for
a $\Lambda$CDM power spectrum. In case of halos, there is also a limit
on the information on small scales: due to their finite number density,
there is a cross-over scale $k_{\rm st}$ where $b_1^2 \Plin(k_{\rm st}) = 1/\nbarh$, 
with $\nbarh$ being the mean number density of halos. 
Modes with $k > k_{\rm st}$ are dominated by the stochasticity of halos,
and we do not expect significant additional information from including
such modes \cite{cabass/schmidt:2019} (in fact, one expects additional
stochastic corrections to the likelihood to become relevant on those
scales, potentially leading to a bias in \se). The scale $k_{\rm st}$ is marked by
the vertical lines for each halo sample in \reffig{Linrun12}.

\reffig{Linrun12} also shows the result when using the matter density field from N-body rather than 2LPT, for $\L=0.1\iMpch$. 
The results show that the ML value of \se is shifted to lower values, reducing the bias in \se relative to the 2LPT one in almost all cases. Again, this is as expected physically.

We next turn to the evolution with redshift, shown in \reffig{Linvsz} for the same mass bins and run 1. For most mass bins, the deviation of $\hat\alpha$ from 1
grows toward higher redshift. This is in contrast to the reach of
perturbation theory, which is expected to extend to higher wavenumbers at
higher redshifts. To understand this result, recall
that the discrepancy in \se is dominated by higher-order bias contributions
(since we have seen that the 2LPT matter forward model shifts $\hat\alpha$
only at the $1\,\%$ level). While the higher-order bias operators themselves
are relatively suppressed by powers of $D_{\rm norm}(z)$ at higher redshifts,
where $D_{\rm norm}(z) = D(z)/D(0)$ and $D(z)$ is the linear growth factor,
the increase in their coefficients, i.e.~the higher-order bias parameters,
with redshift might in fact more
than compensate for this suppression when considering halos within a fixed
mass range. To investigate this, we can make the very rough approximation
that higher-order bias parameters are proportional to $b_1^L \equiv b_1-1$. This
approximation is motivated by the ``Lagrangian local-in-matter-density''
assumption, coupled with thresholding or excursion-set pictures
(see Sec.~2.1--2.2 in \cite{biasreview}). Under these assumptions,
all higher-order bias parameters are controlled by powers of $b_1^L$.

\begin{figure*}[tbp]%
	\centerline{\resizebox{0.5\hsize}{!}{
		\includegraphics*{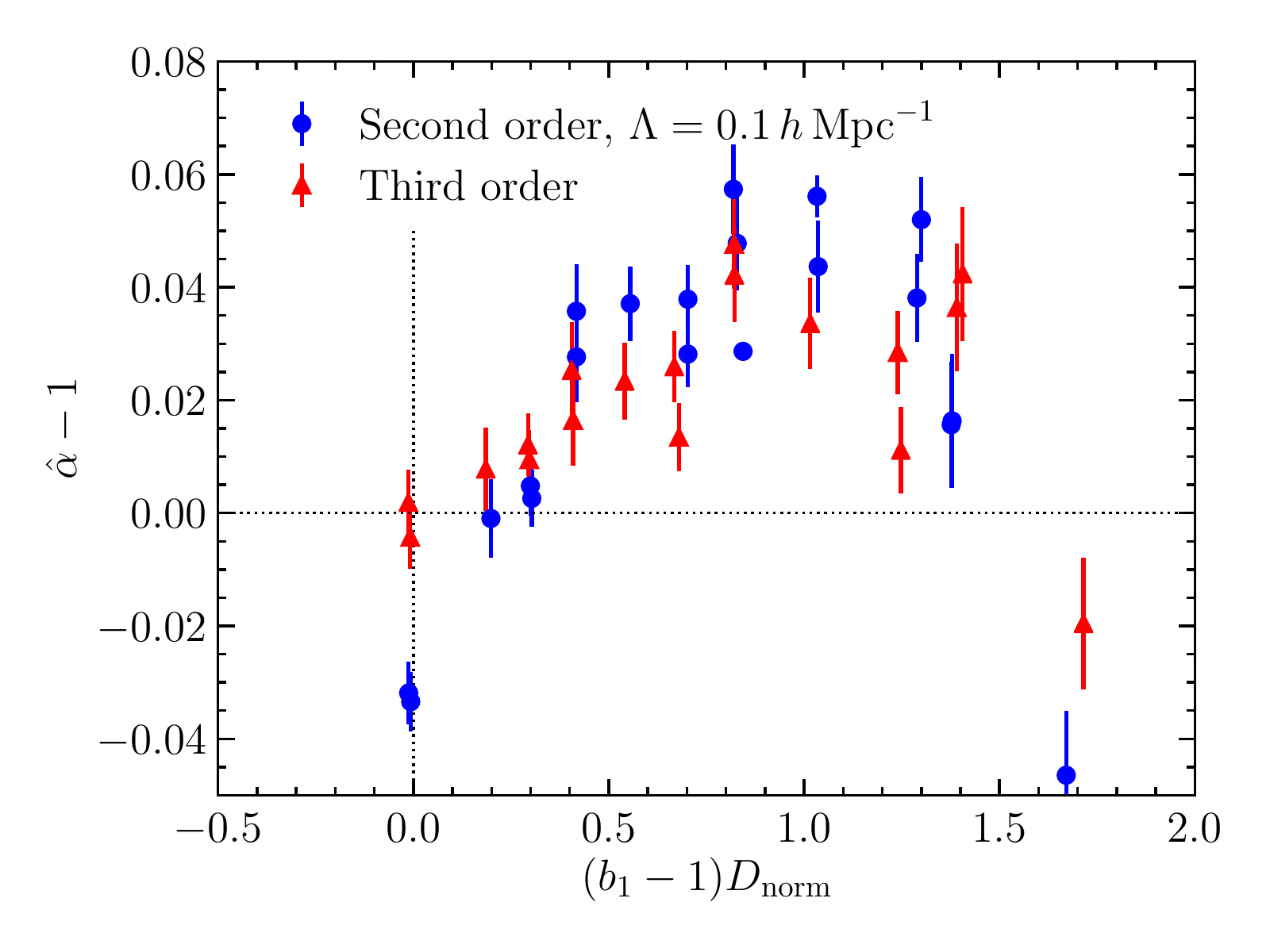}
		}}
	\cprotect\caption{Fractional systematic error in \se, i.e.~ML values for $\alpha-1$ for all halo mass bins and redshifts, including those shown in \reffig{Linvsz} and \reffig{Lincubicvsz}, respectively, plotted against $(b_1-1)D_{\rm norm}$, where $D_{\rm norm}(z) = D(z)/D(0)$. The latter quantity is a very rough estimate of higher-order bias corrections (see discussion in text), and shows a clear correlation with the relative bias in \se, $\hat\alpha-1$.
	\label{fig:alphavsX}}
\end{figure*}%

While very rough, and not expected to be accurate for actual halos
\cite{Baldauf:2012hs,Sheth:2012fc,Modi:2016dah,Lazeyras:2017hxw,Abidi:2018eyd},
this approximation gives us a means of testing the source
of the increasing deviation in \se with redshift.
\reffig{alphavsX} shows a scatter plot of $\hat\alpha$ against the
combination expected to control the higher-order bias contributions,
namely $(b_1(z)-1) D_{\rm norm}(z)$. Here, $b_1$ is taken as the maximum-likelihood value
for the true \se for the corresponding halo sample and redshift, while $D_{\rm norm}$ is computed in the simulation cosmology. 
We show results for second-order (blue)
and third-order bias expansions (red, discussed in the next section)
for all mass bins and redshifts, but at fixed $\L=0.1\iMpch$. The correlation
of $\hat\alpha-1$ with $(b_1-1) D_{\rm norm}$ is clearly visible.
This lends strong support to the conjecture that the residual bias in \se
is due to higher-order bias contributions which scale nontrivially with mass
and redshift. The results from third-order bias in the next section will
provide further independent evidence for this conjecture.

\begin{figure*}[tbp]%
	\centerline{\resizebox{\hsize}{!}{
		\includegraphics*{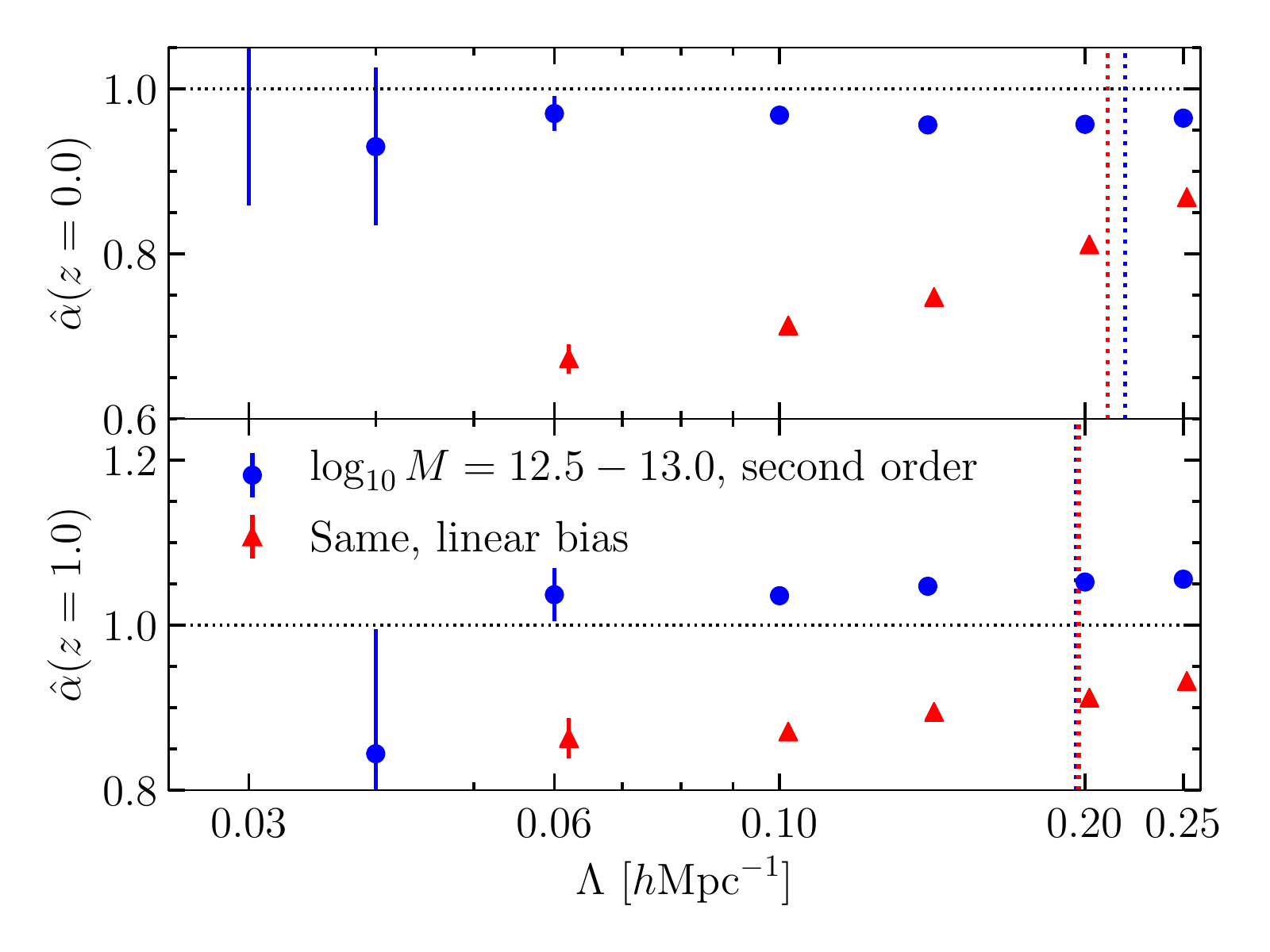}
		\includegraphics*{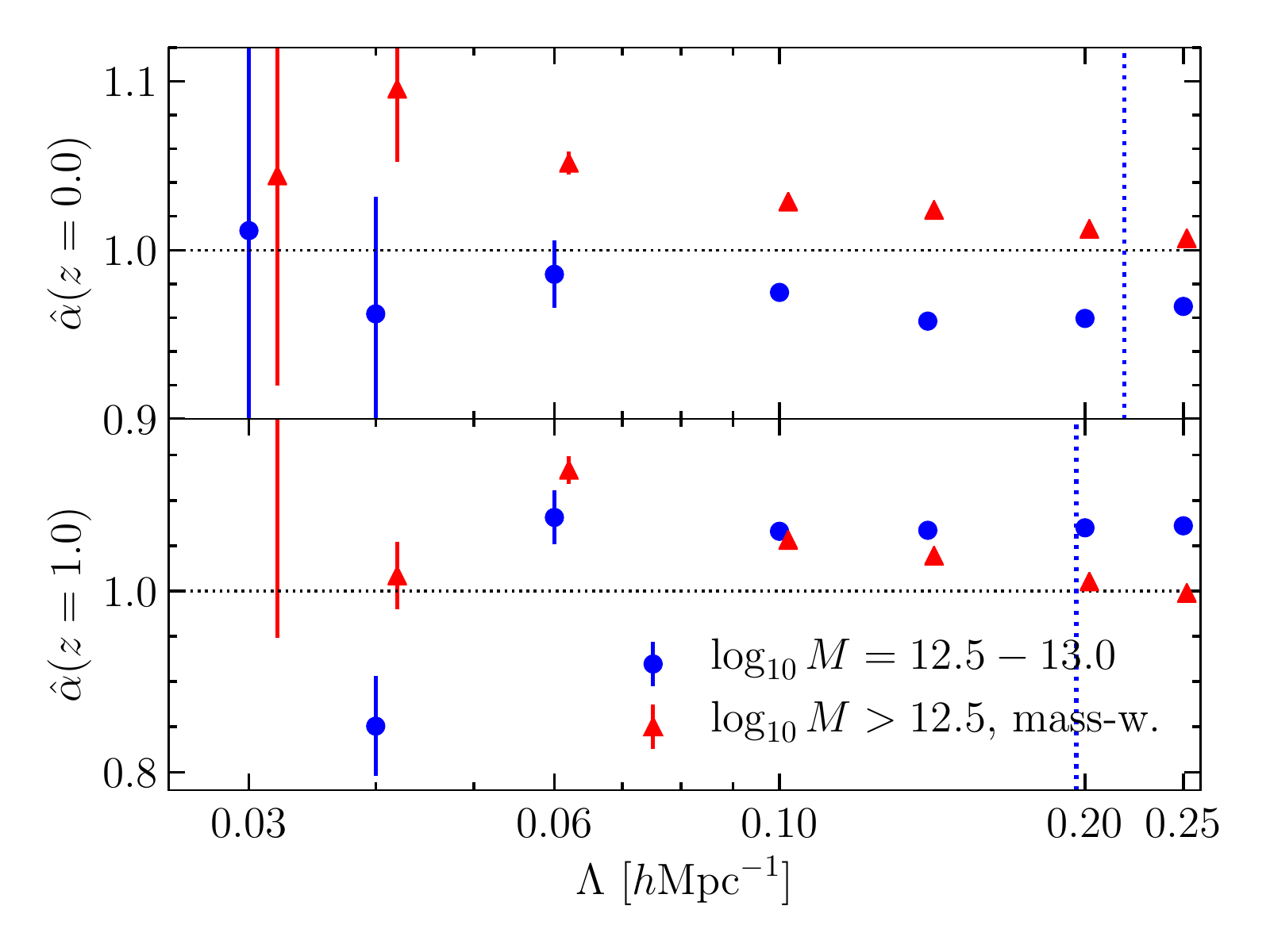}
		}}
	\cprotect\caption{\emph{Left panel:} ML values for $\alpha$ for linear and second-order bias expansions. The linear bias expansion leads to a significantly larger discrepancy in the inferred \se value. 
	\emph{Right panel:} results for mass-weighted halos with $\log_{10}(M/h^{-1} M_\odot) > 12.5$ and using the second-order bias expansion.
	\label{fig:Lin2}}
\end{figure*}

Before we move on, two further tests are shown in \reffig{Lin2}: the left panel compares the result of linear bias with second-order bias for the lowest-mass bin, $\log_{10}(M/h^{-1} M_\odot) \in [12.5,13.0]$. Clearly, linear bias performs much worse, which is as expected. The right panel shows the result for mass-weighting all halos with $\log_{10}(M/h^{-1} M_\odot) > 12.5$. The error bars shrink significantly due to the reduced stochasticity, as has been found previously 
\cite{seljak/hamaus/desjacques:2009,hamaus/seljak/etal:2010}
and is expected within the halo model \cite{schmidt:2016a}. The bias in the central $\se$ value is comparable to the case of unweighted halos in mass bins. We conclude that, while mass weighting can significantly reduce halo stochasticity, it does not by itself substantially reduce the effect of higher-order bias corrections. Note however that these conclusions might depend on the lower mass cut used.

\begin{table}[b]
\centering
\begin{tabular}{c c c c c c c}
\hline
\hline
Redshift & \specialcell{Mass range\\$\log_{10} M [\Msunh]$}
& \specialcell{$100(\hat{\alpha}-1)$\\(run 1)}
& \specialcell{$100(\hat{\alpha}-1)$\\(run 2)}
& $b_1$
& \specialcell{$\sigma_\eps^2$\\$[$Poisson$]$}
& \specialcell{$\sigma_{\eps,2}/\sigma_\eps$\\ $[(\Mpch)^2]$} \\
\hline
0 & [12.5-13.0] & $-3.2 \pm 0.6$ & $-3.3 \pm 0.5$ & $0.99$ & $1.18$ & $7.6$ \\
0.5 & [12.5-13.0] & $-0.1 \pm 0.7$ & --- & $1.26$ & $1.16$ & $-0.3$ \\
1 & [12.5-13.0] & $3.6 \pm 0.8$ & $2.8 \pm 0.8$ & $1.68$ & $1.03$ & $1.5$ \\
\hline
0 & [13.0-13.5] & $0.5 \pm 0.5$ & $0.3 \pm 0.5$ & $1.30$ & $1.05$ & $8.8$ \\
0.5 & [13.0-13.5] & $3.7 \pm 0.7$ & --- & $1.72$ & $0.99$ & $4.5$ \\
1 & [13.0-13.5] & $4.8 \pm 0.8$ & $5.7 \pm 0.8$ & $2.35$ & $0.92$ & $3.7$ \\
\hline
0 & [13.5-14.0] & $3.8 \pm 0.6$ & $2.8 \pm 0.6$ & $1.70$ & $0.93$ & $7.3$ \\
0.5 & [13.5-14.0] & $4.4 \pm 0.8$ & --- & $2.34$ & $0.90$ & $3.7$ \\
1 & [13.5-14.0] & $1.6 \pm 1.2$ & $1.6 \pm 1.1$ & $3.25$ & $0.90$ & $4.1$ \\
\hline
0 & [14.0-14.5] & $3.8 \pm 0.8$ & $5.2 \pm 0.8$ & $2.29$ & $0.87$ & $4.1$ \\
0.5 & [14.0-14.5] & $-4.6 \pm 1.1$ & --- & $3.16$ & $0.88$ & $4.0$ \\
1 & [14.0-14.5] & $6.9 \pm 1.9$ & $4.3 \pm 1.8$ & $4.37$ & $0.93$ & $3.8$ \\
\hline
\hline
\end{tabular}
\caption{Summary of results for the second-order bias expansion and $\kmax = \Lambda = \Lin = 0.1\iMpch$ for different mass bins and redshifts; for $z=0.5$, only results for run 1 are available. The fractional deviation of the maximum-likelihood \se, $\hat\alpha-1$, is quoted in percent; results from run 1 and run 2 are shown individually with estimated $68\,\%$ confidence-level error bars. $b_1$ and stochastic amplitudes are reported for the fiducial $\se=\se^\text{fid}$ and averaged over both runs. The stochastic variance $\sigma_\eps^2$ is scaled to the Poisson expectation for the given halo sample, as described in App.~A of \cite{paperII}. The last column shows the ratio of the higher-derivative stochastic amplitude to the leading one, indicating the scale associated with the expansion of $\s^2(k)$ in $k$.}
\label{tab:results}
\end{table}

\clearpage
\subsection{Third-order bias}

\begin{figure*}[tbp]%
	\centerline{\resizebox{\hsize}{!}{
		\includegraphics*{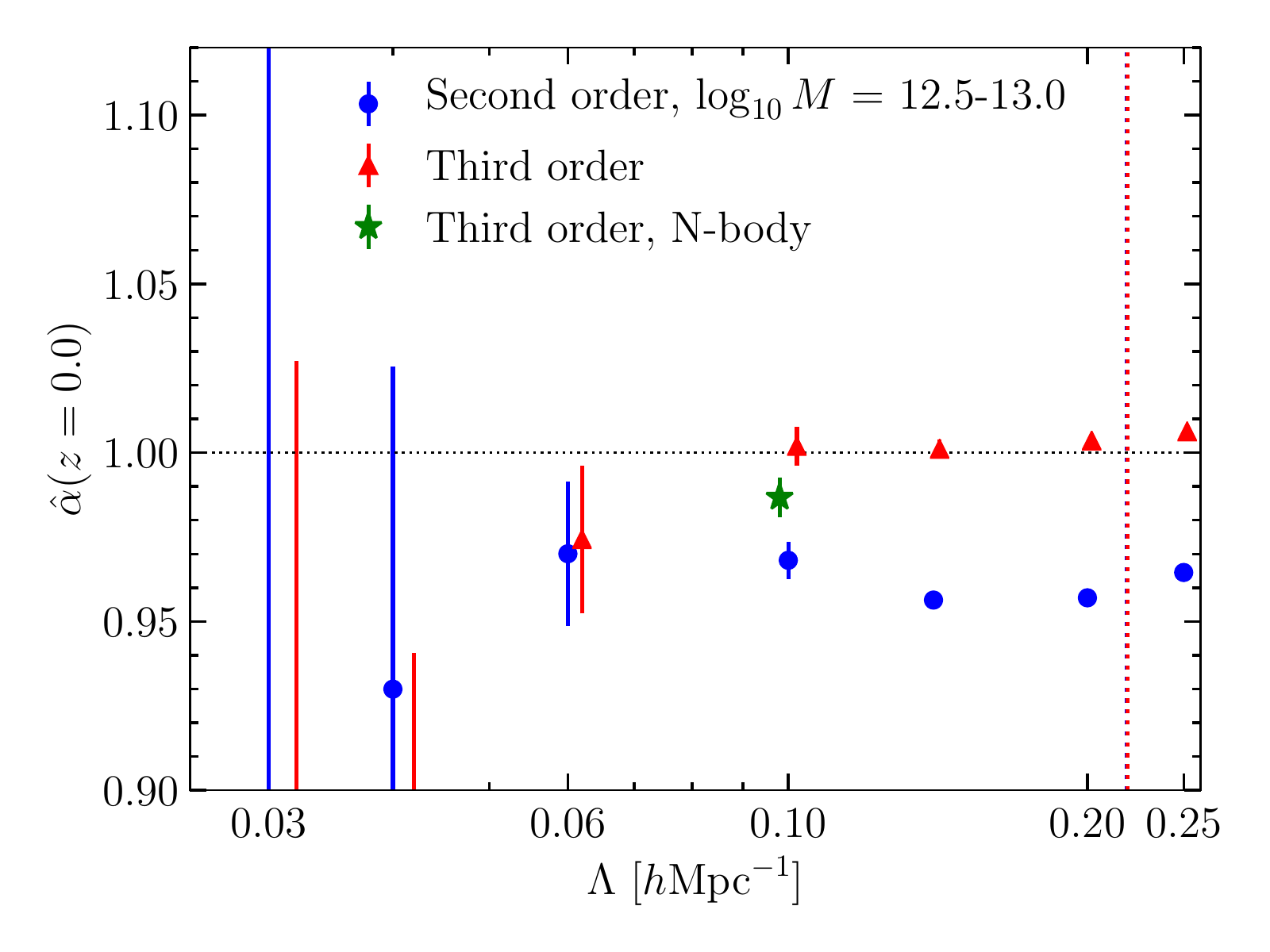}
		\includegraphics*{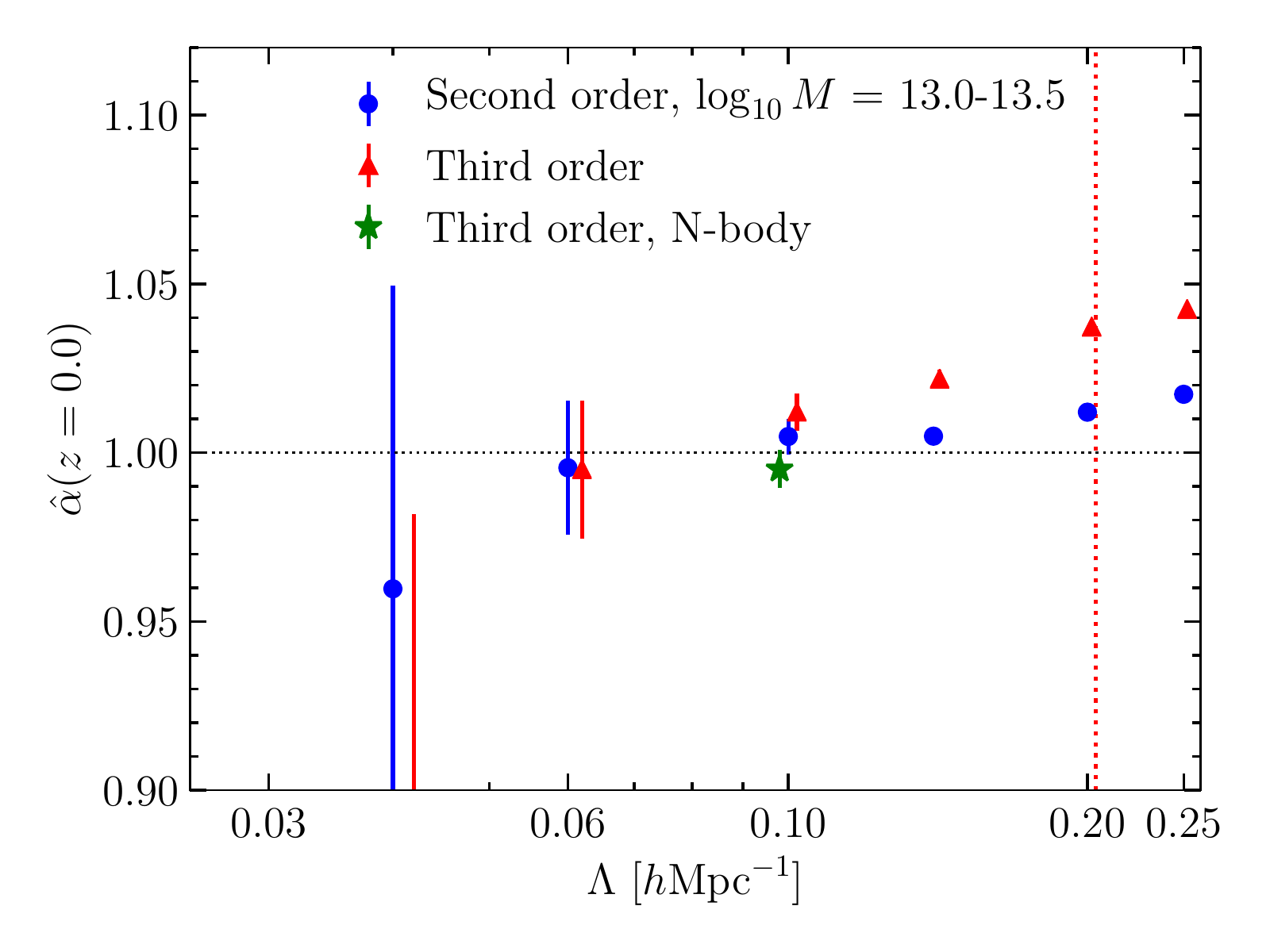}
		}}
	\centerline{\resizebox{\hsize}{!}{
		\includegraphics*{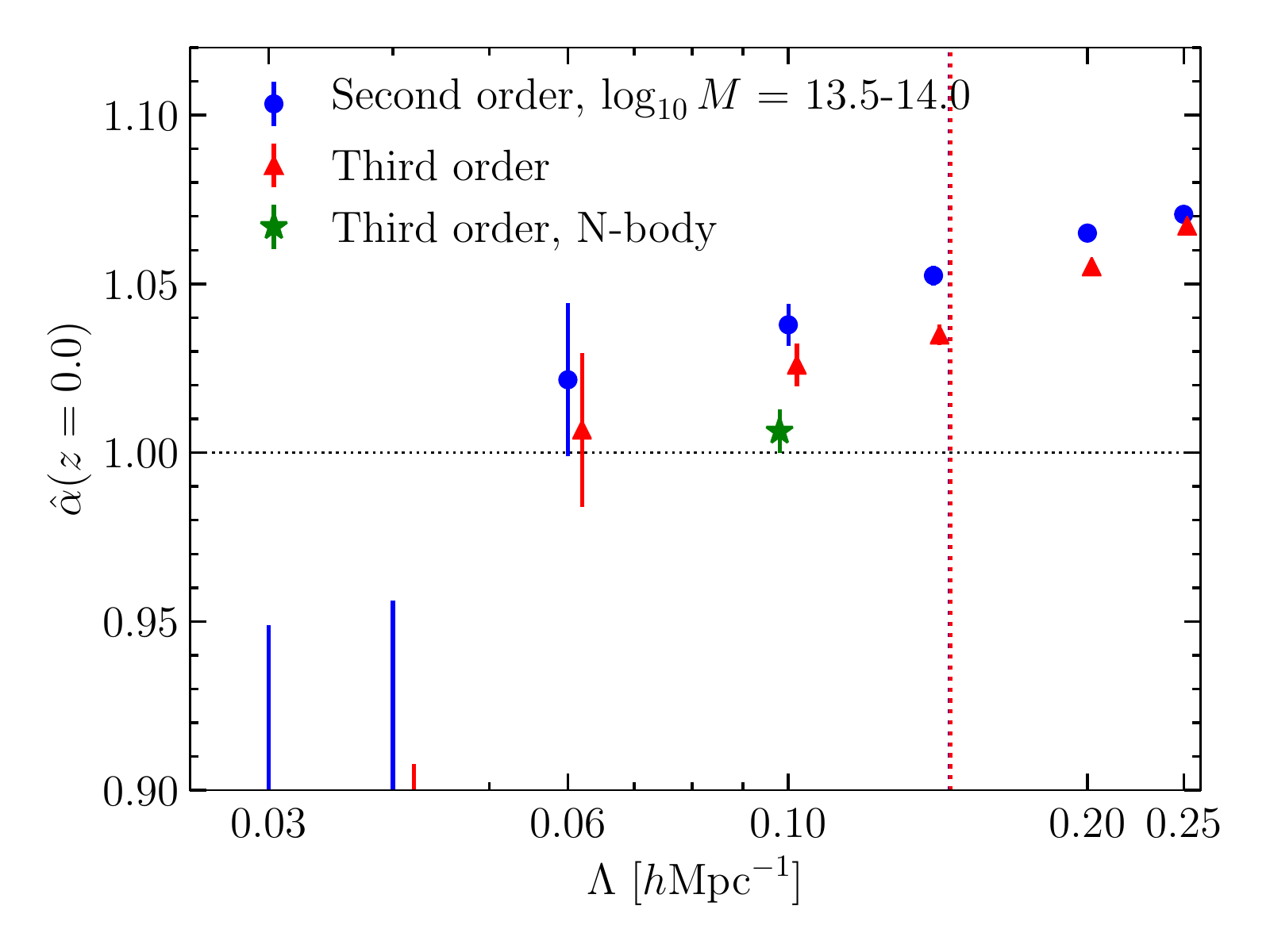}
		\includegraphics*{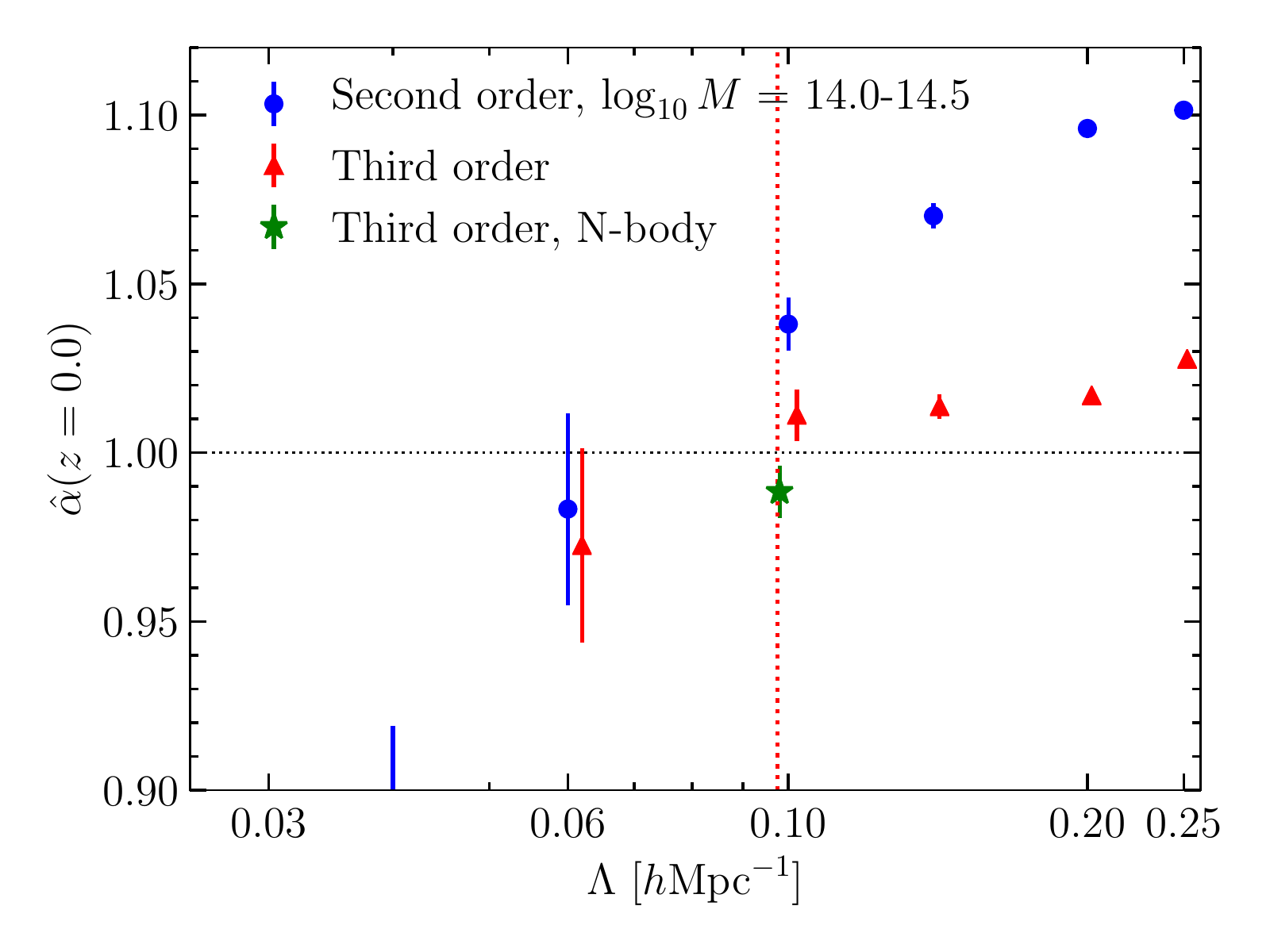}
		}}
	\cprotect\caption{ML values for $\alpha$ as a function of $\L = k_{\rm max} = \Lin$ for run 1 at $z=0$. Shown here are different mass bins each for second- and third-order bias expansions (the former being the same results as shown in \reffig{Linrun12}). 
	Also shown is the third-order result when using the N-body instead of 2LPT density field (only $\L=0.1\iMpch$). 
	\label{fig:Lincubic}}
\end{figure*}

We next turn to results using the third-order bias expansion, with the list of operators given in \refeq{Ocubic}. The results are summarized in \reftab{resultscubic} (for $\L=0.1\iMpch$), and are shown as function of $\L$ in \reffig{Lincubic}, comparing to the second-order bias case for each mass bin (at $z=0$ and for run 1 in all cases). We find that the bias in \se is reduced for all mass bins and redshifts, in some cases substantially. For many samples, the bias is under $2\,\%$ and hence approaching the bias found when subsampling DM particles (\reffig{dm}). That is, a significant part of the residual misestimation of \se might be explained by the deficiency of our matter forward model (i.e.~2LPT);
we also show results at $\L=0.1\iMpch$ using the N-body density field, which
indeed moves $\hat\alpha$ even closer to unity.

\begin{figure*}[tbp]%
	\centerline{\resizebox{\hsize}{!}{
		\includegraphics*{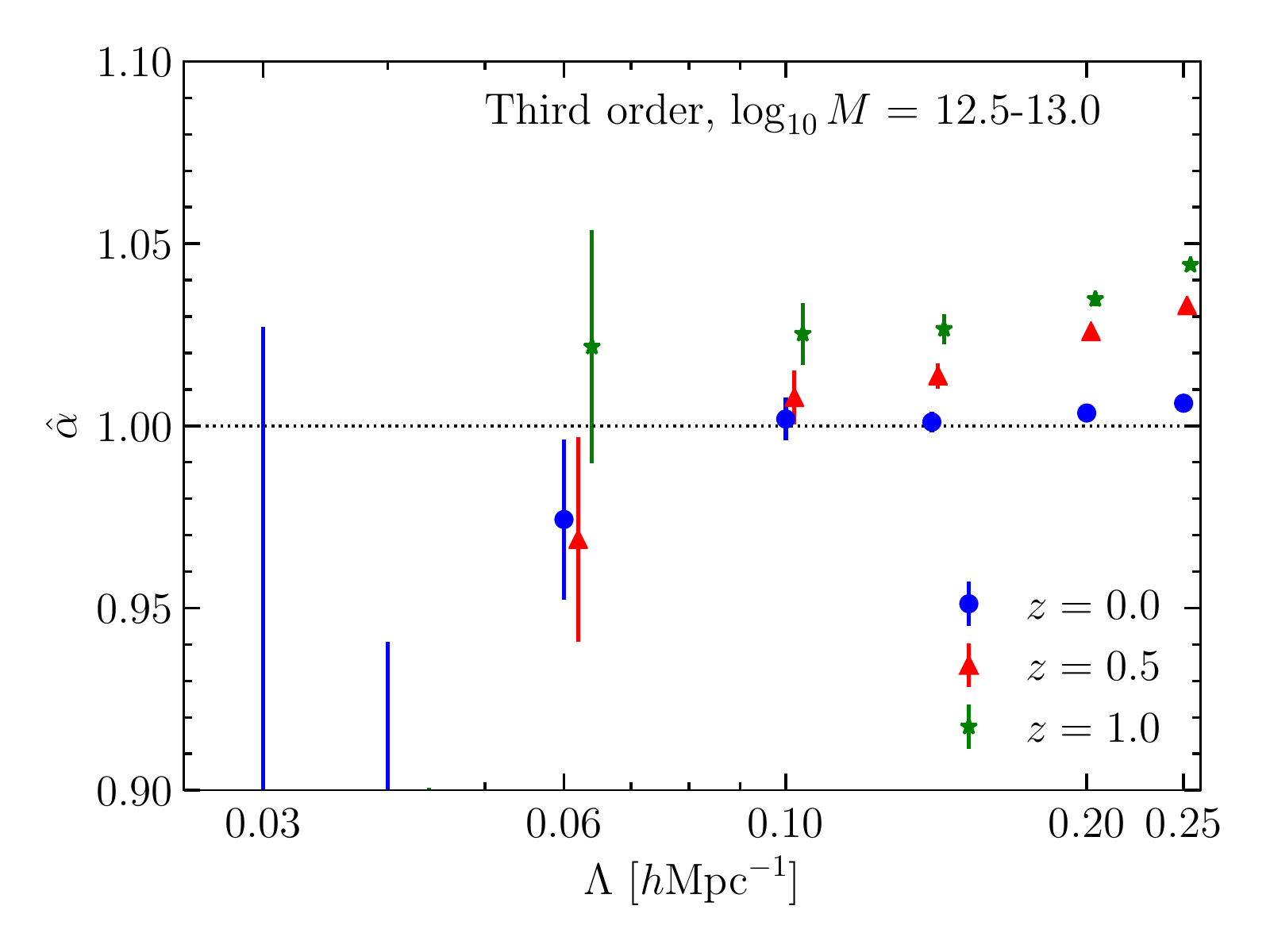}
		\includegraphics*{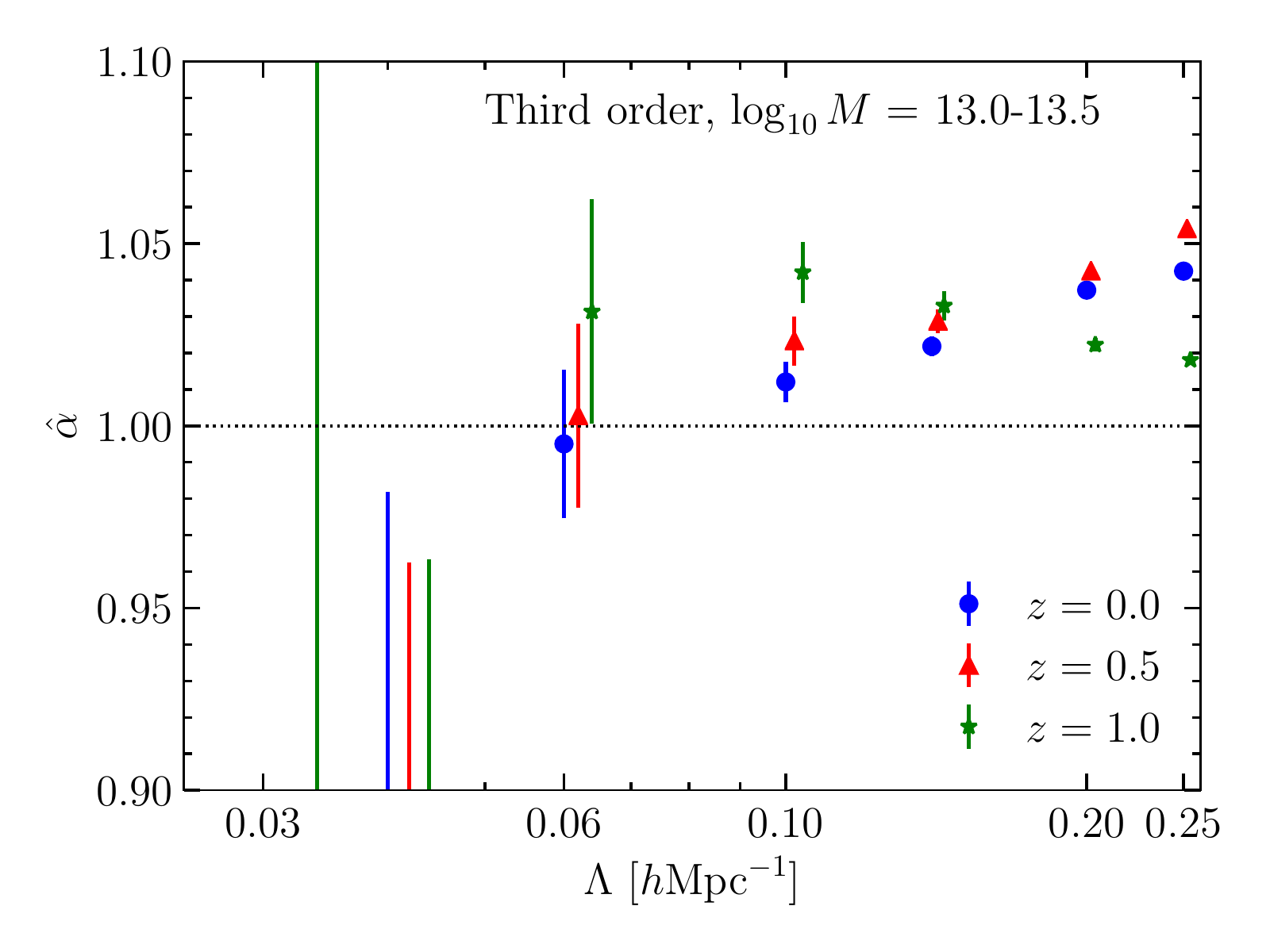}
		}}
	\centerline{\resizebox{\hsize}{!}{
		\includegraphics*{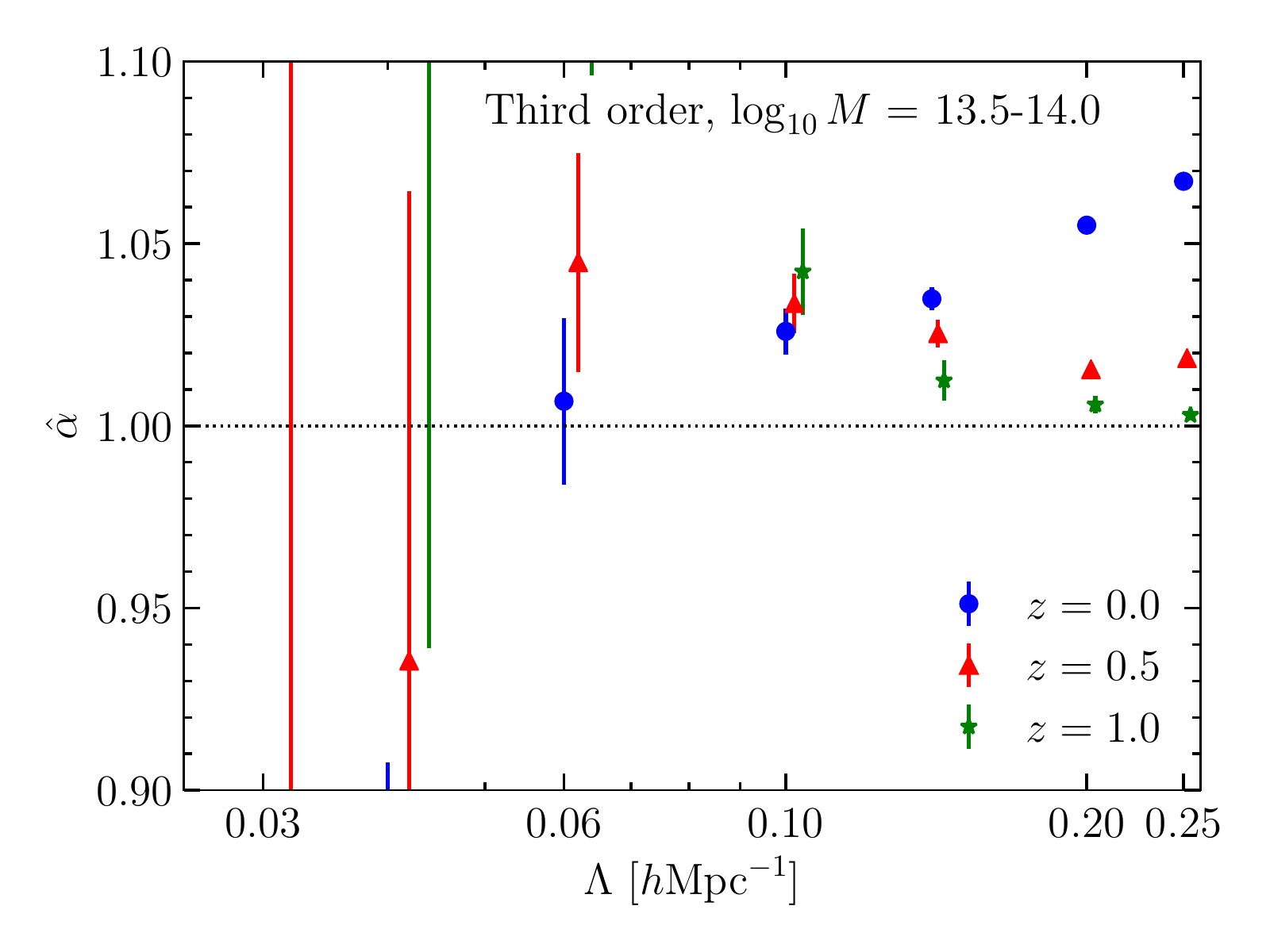}
		\includegraphics*{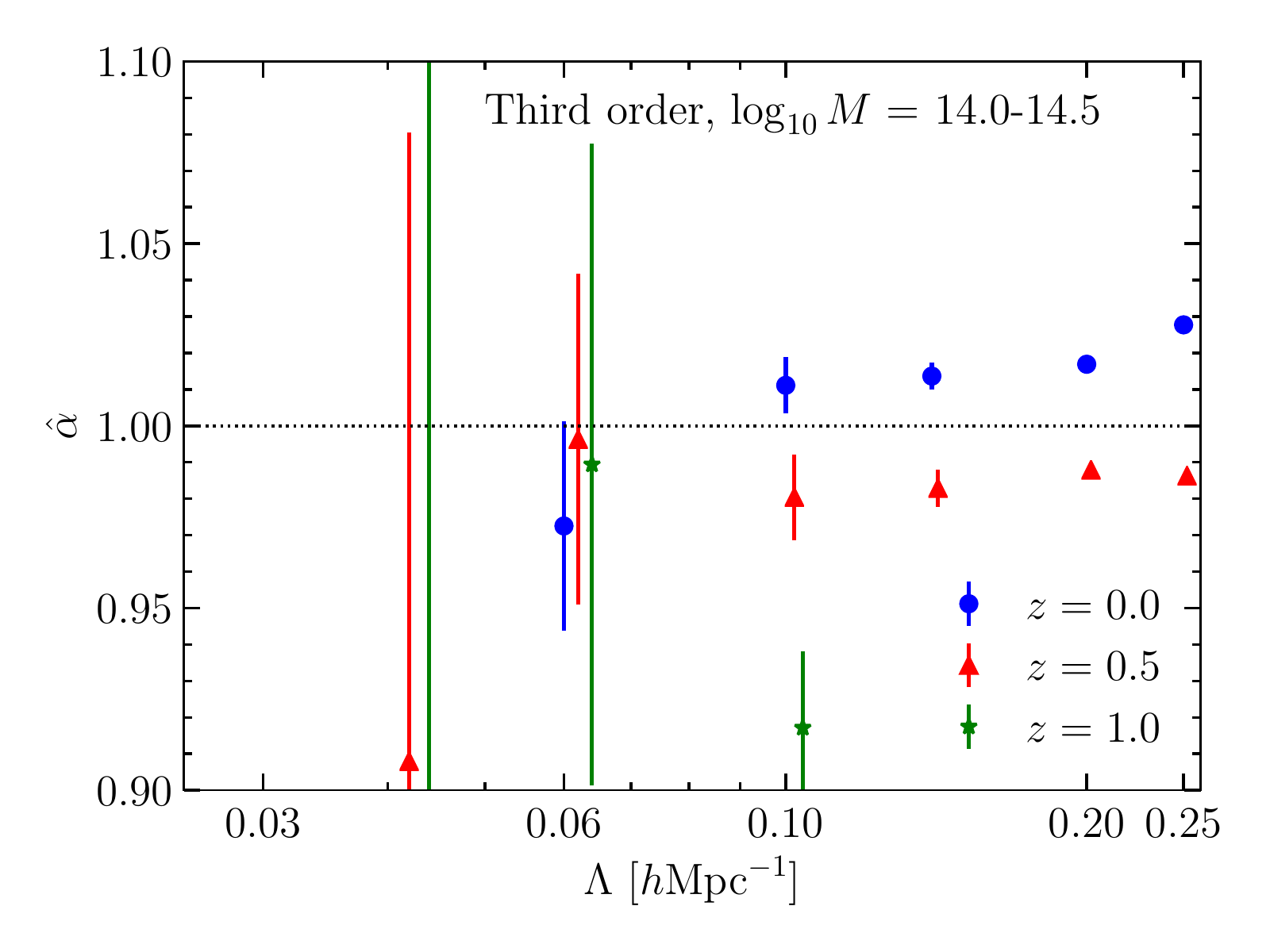}
		}}
	\cprotect\caption{ML values for $\alpha$ using the third-order bias model as a function of $\L = k_{\rm max} = \Lin$. Different panels show the four mass bins, each at different redshifts for run 1. 
	\label{fig:Lincubicvsz}}
\end{figure*}

\reffig{Lincubicvsz} shows the result (third-order bias only) as a function of redshift. We generally see the same trend as in the second-order case, namely that the discrepancy in the estimated $\alpha$ grows toward higher redshift. This indicates that the residual discrepancy in \se is indeed due to 
higher-order bias terms, at least in the case of highly biased tracers, and is
another very important test of theoretical consistency. In \reffig{alphavsX},
we also show the fractional deviation in \se, $\hat\alpha-1$, obtained from the third-order bias expansion at $\L=0.1\iMpch$
as a function of $(b_1-1) D_{\rm norm}$. A clear correlation can again be seen, with
the overall deviation being smaller than in the second-order case.

\begin{table}[b]
\centering
\begin{tabular}{c c c c c c c}
\hline
\hline
Redshift & \specialcell{Mass range\\$\log_{10} M [\Msunh]$}
& \specialcell{$100(\hat{\alpha}-1)$\\(run 1)}
& \specialcell{$100(\hat{\alpha}-1)$\\(run 2)}
& $b_1$
& \specialcell{$\sigma_\eps^2$\\$[$Poisson$]$}
& \specialcell{$\sigma_{\eps,2}/\sigma_\eps$\\ $[(\Mpch)^2]$} \\
\hline
0 & [12.5-13.0] & $0.2 \pm 0.6$ & $-0.4 \pm 0.6$ & $0.99$ & $1.17$ & $8.2$ \\
0.5 & [12.5-13.0] & $0.8 \pm 0.7$ & --- & $1.24$ & $1.16$ & $-0.2$ \\
1 & [12.5-13.0] & $2.5 \pm 0.8$ & $1.6 \pm 0.8$ & $1.67$ & $1.03$ & $1.4$ \\
\hline
0 & [13.0-13.5] & $1.2 \pm 0.6$ & $0.9 \pm 0.5$ & $1.29$ & $1.05$ & $8.8$ \\
0.5 & [13.0-13.5] & $2.3 \pm 0.7$ & --- & $1.70$ & $0.99$ & $4.4$ \\
1 & [13.0-13.5] & $4.2 \pm 0.8$ & $4.8 \pm 0.8$ & $2.34$ & $0.91$ & $4.0$ \\
\hline
0 & [13.5-14.0] & $2.6 \pm 0.6$ & $1.3 \pm 0.6$ & $1.67$ & $0.93$ & $7.2$ \\
0.5 & [13.5-14.0] & $3.4 \pm 0.8$ & --- & $2.31$ & $0.89$ & $3.8$ \\
1 & [13.5-14.0] & $4.2 \pm 1.2$ & $3.6 \pm 1.1$ & $3.29$ & $0.90$ & $4.5$ \\
\hline
0 & [14.0-14.5] & $1.1 \pm 0.8$ & $2.8 \pm 0.7$ & $2.24$ & $0.86$ & $4.7$ \\
0.5 & [14.0-14.5] & $-2.0 \pm 1.2$ & --- & $3.22$ & $0.88$ & $4.5$ \\
1 & [14.0-14.5] & $-8.3 \pm 2.1$ & $-10.4 \pm 2.1$ & $4.49$ & $0.93$ & $3.6$ \\
\hline
\hline
\end{tabular}
\caption{Summary of results for $\Lambda = 0.1\iMpch$ and third-order bias. The columns are the same as in \reftab{results}.}
\label{tab:resultscubic}
\end{table}

\clearpage
\subsection{Halo density field}
\label{sec:rk}

After having presented the inference of \se from the EFT likelihood, we now
present some results on the predicted, deterministic halo density field
$\d_{h,\rm det}$, cf.~\refeq{dhdet}, specifically its power spectrum and cross-correlation
coefficient with the actual halo density field $\d_h(\vk)$. This is
the deterministic halo field obtained using our forward model, which consists
of the matter forward model plus bias fields, after inserting the maximum-likelihood
bias coefficients at the fiducial value of \se.

We start from the ansatz
\be
\d_h(\vk) = \d_{h,\rm det}(\vk) + \eps(\vk)\,\,.
\label{eq:dhdhdet}
\ee
We now assume that the noise field $\eps(\vk)$ is a Gaussian random field with
zero mean and power spectrum
\be
\< \eps(\vk) \eps(\vk') \>' = P_\eps(k) = P_\eps^{\{0\}} + P_\eps^{\{2\}} k^2\,\,.
\ee
When integrating out the noise field with this power spectrum, one
obtains the conditional likelihood of \refeq{PcondGFT}, as shown in
\cite{paperI}.
Further, the constants $P_\eps^{\{0,\,2\}}$ can be obtained directly from the constants $\sigma_\eps, \sigma_{\eps,2}$ of the likelihood \cite{paperII}:
\be
P_\eps^{\{0\}} = \frac{L_{\rm box}^3}{N_g^6} \sigma_\eps^2\,\,; \quad
P_\eps^{\{2\}} = 2 \frac{L_{\rm box}^3}{N_g^6} \sigma_\eps \sigma_{\eps,2}\,\,. 
\ee
At the order we work in throughout this paper, these assumptions on $\eps(\vk)$ are consistent
with the EFT likelihood \cite{cabass/schmidt:2019}. 

\begin{figure*}[tbp]%
	\centerline{\resizebox{\hsize}{!}{
		\includegraphics*{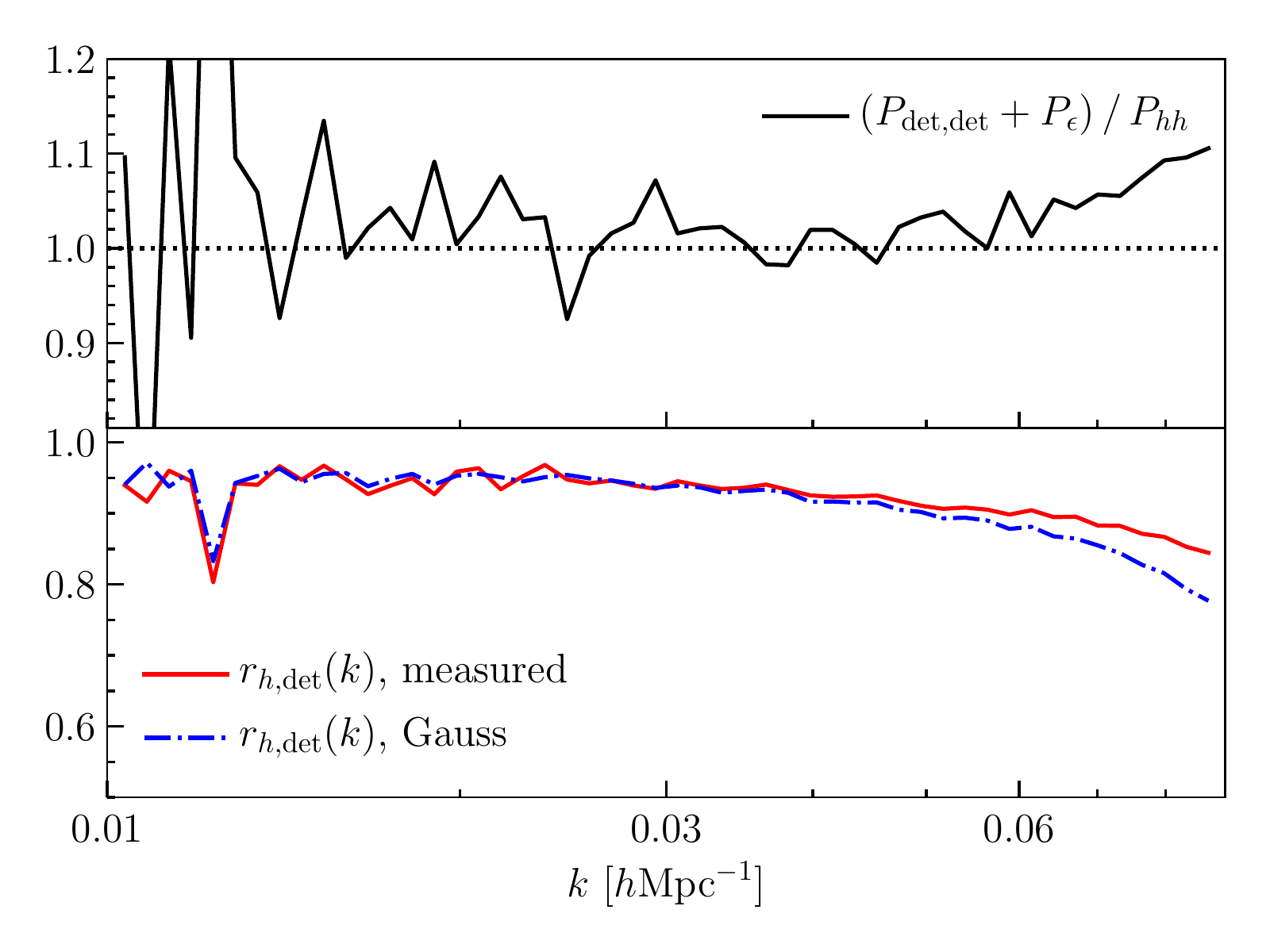}
		\includegraphics*{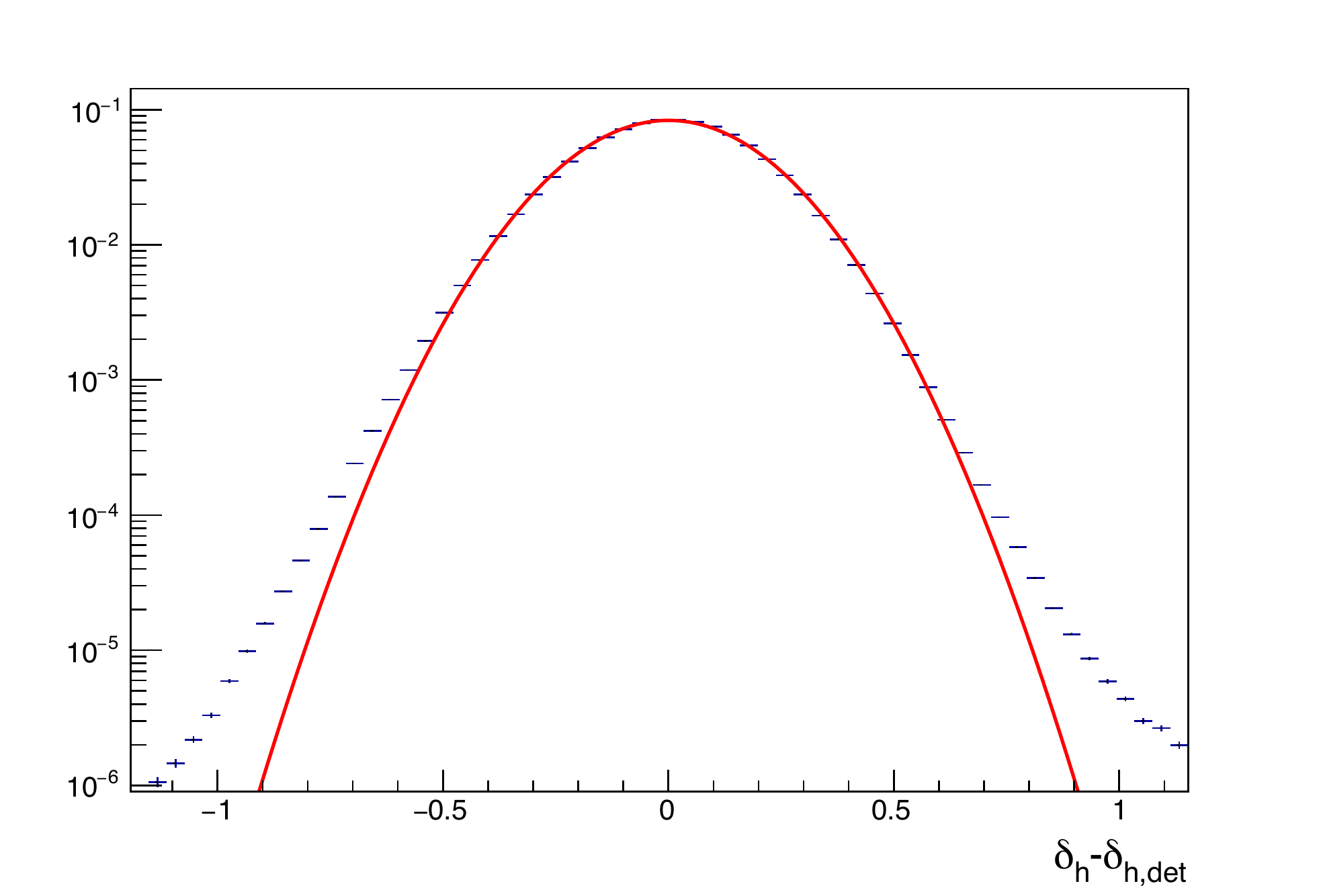}
		}}
	\cprotect\caption{	
          Properties of the noise $\d_h-\d_{h,\rm det}$. We show results for the reference cutoff value employed in the analysis, $\L=0.1\iMpch$. \textit{Left panel:} accuracy of the Gaussian-noise assumption, cf.~\refeq{Phrk_gauss}, 
	in the halo power spectrum (top panel) and the cross-correlation coefficient (bottom panel), both for halos with $\log_{10}(M/h^{-1} M_\odot) \in [12.5-13]$ at $z=0$. Clearly, on the range of scales probed, the halo power spectrum and cross-correlation coefficient between the halo density field $\d_h$ and the deterministic prediction $\d_{h,\rm det}$ are well described by \refeq{Phrk_gauss}. 
          \textit{Right panel:} normalized histogram of $\d_h-\d_{h,\rm det}$ for the same halo sample and deterministic model. The solid line shows a Gaussian fit. A logarithmic scale is chosen in order to increase the visibility of the non-Gausian tails. The Gaussian approximation is excellent. 
	\label{fig:Gauss}}
\end{figure*}

\refeq{dhdhdet} and the Gaussianity of $\eps$ then allow us to derive the
relation between the power spectra of $\d_h$ and $\d_{h,\rm det}$,
denoted as $P_{hh}(k)$ and $P_{\rm det,det}(k)$, respectively, as well as their
cross-correlation coefficient, $r_{h,\rm det}(k)$:
\ba
P_{hh}(k)\Big|_\text{Gaussian noise} &= P_{\rm det,det}(k) + P_\eps(k) \vs
r_{h,\rm det}(k)\Big|_\text{Gaussian noise} &= \left(1 - \frac{P_\eps(k)}{P_{hh}(k)}\right)^{1/2}\,\,.
\label{eq:Phrk_gauss}
\ea
The left panel of \reffig{Gauss} compares both of these quantities with the measured
halo power spectrum (top) and correlation coefficient with $\d_{h,\rm det}$ (bottom;
this is inferred via $r_{h,\rm det}^\text{meas} = P_{h,\rm \det}/\sqrt{P_{hh} P_{\rm det,det}}$);
in case of the power spectrum, the ratio is shown.

In each case, we choose the lowest-mass halo sample at $z=0$, which is the least noisy
sample, and our reference cutoff value $\L = 0.1\iMpch$.
Clearly, \refeq{Phrk_gauss} matches the measurements quite well.
The right panel of \reffig{Gauss} shows a histogram of the effective noise,
$\d_h(\vx)-\d_{h,\rm det}(\vx)$, for the same two fields. This is computed in real space
on the $512^3$ grid where all fields are represented, after both fields have been
sharp-$k$ filtered at the cutoff $\L$. In the limit of infinite volume, this histogram
shows the 1-point probability distribution for the residual field $\d_h(\vx)-\d_{h,\rm det}(\vx)$. 
Both results show that the Gaussian noise assumption is an excellent approximation on the scales probed.

\begin{figure*}[tbp]%
	\centerline{\resizebox{0.5\hsize}{!}{
		\includegraphics*{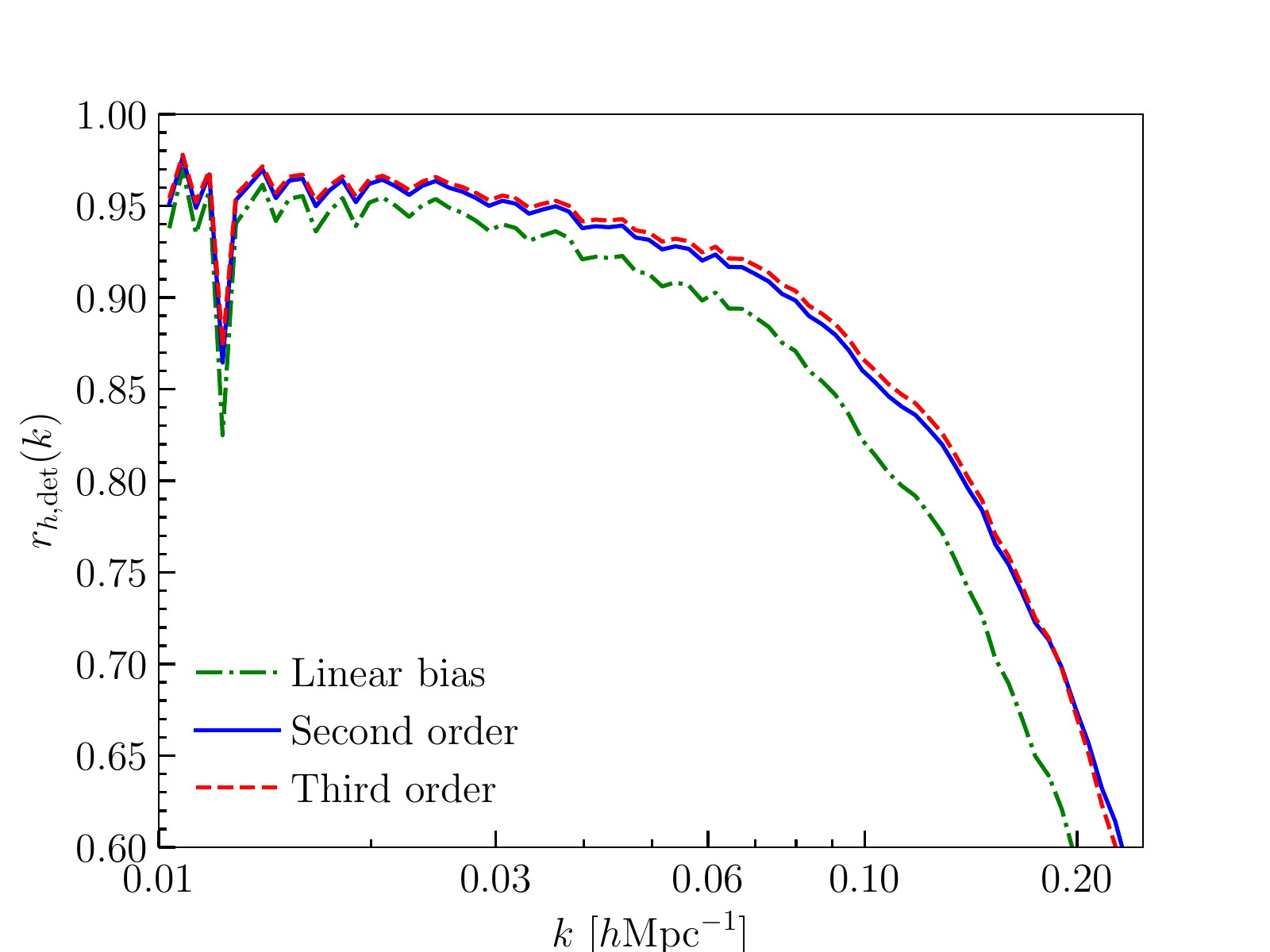}
		}}
	\cprotect\caption{Cross-correlation coefficient between $\d_h$ and $\d_{h,\rm det}$ as inferred using \refeq{Phrk_gauss} for linear (including higher-derivative), second-order, and third-order bias expansions. Here, results are shown for the highest cutoff $\L=0.25\iMpch$. The cross-correlation is improved significantly when going from linear to second-order bias, after which the improvement is small.
	\label{fig:rk}}
\end{figure*}

We next turn to the cross-correlation coefficient $r_{h,\rm det}(k)$ for different bias expansions. This was studied in detail in \cite{schmittfull/etal:2018}. One expects that adding additional bias terms should improve the correlation between $\d_{h,\rm det}$ and the
actual halo field $\d_h$. We can determine $r_{h,\rm det}(k)$ simply by using the
measured $P_{hh}(k)$ and the inferred
maximum-likelihood values of $\sigma_\eps,\,\sigma_{\eps,2}$. The result is shown
in \reffig{rk} for different bias expansions. We now choose the maximum
  cutoff value considered in this paper, $\L=0.25\iMpch$, in order to test how
  much the higher-order bias contributions improve the cross-correlation on smaller scales.
The linear-bias expansion here includes the leading higher-derivative term. Clearly, including second-order
bias terms significantly increases the cross-correlation of $\d_{h,\rm det}$ with $\d_h$,
in agreement with the findings of \cite{schmittfull/etal:2018} based on a slightly different
forward model. On the other hand, going to third order in the bias expansion only
mildly improves the cross-correlation. In the previous section, we saw however that including
third-order bias terms substantially improves the cosmological parameter estimates
(in this case, \se or $\AS$). Thus, we conclude that the cross-correlation coefficient
between data and model is not necessarily indicative of the quality of inferred
cosmological parameters. As an example for how this can arise, consider an overall rescaling of the model prediction by a function $f(k)$. This does not change the correlation coefficient with the data or the truth, but it clearly impacts cosmological parameters inferred from the model by changing the amplitude and shape of the linear power spectrum that leads to the apparent best match with the data.

\section{Conclusions}
\label{sec:concl}

We have presented results on the inference of \se from a rest-frame halo catalog using the EFT likelihood. This inference is completely based on nonlinear information that is protected by the equivalence principle and that is necessary to break the bias--$\se$ degeneracy. 
The EFT likelihood isolates precisely those parts of the likelihood $P(\d_h | \{ b_O, \sigma_a \}, \se, \cdots)$ of a biased tracer density field $\d_h$ that are uncertain, by making use of a full nonlinear matter forward model. That is, no perturbative expansion of the matter density field is necessary.

In contrast to previous work, we impose a cutoff on the momenta of the initial density perturbations, that is, before the perturbations are evolved forward under gravity. We show that this is crucial to regularize certain loop contributions that are otherwise not under control (\refsec{MAP}), which we conjecture to be responsible for the biased \se estimate reported in \cite{paperII}.

Our numerical results, presented in \refsec{results}, show the expected convergence behavior as a function of scale, both for second-order and third-order bias. At the reference scale of $\L=0.1\iMpch$, and excluding the most highly biased sample (i.e. $\log_{10} M > 10^{14}\Msunh$ at $z=1$, with $b_1\simeq 4.4$), \se is recovered to within $6\,\%$ for the second-order bias expansion, and $4\,\%$ for the third-order bias expansion. To our knowledge, this is the \emph{most precise inference of cosmological parameters from purely nonlinear information in biased tracers of large-scale structure} demonstrated to date---albeit fixing the phases and other cosmological parameters.

We present strong indications that the residual bias in \se is due to higher-order bias terms, which is expected in the context of the EFT likelihood. Interestingly, due to the scaling of the bias parameters, these corrections do not necessarily decrease toward higher redshifts. This fact applies much more broadly to cosmology inference from biased tracers, such as via the power spectrum and bispectrum, and suggests that systematic errors from higher-order bias terms
are roughly controlled by $b_1-1$, at least for halos.

Given that the expected convergence behavior for \se is seen, there are several
immediate applications of the EFT likelihood, which we leave to future work:
\begin{itemize}[leftmargin=*]
\item Continue the bias expansion and likelihood to higher order in perturbations by including the field-dependent noise covariance \cite{cabass/schmidt:2020};
\item Implement the likelihood in a forward-modeling framework which samples the phases of the initial conditions;
\item Apply the likelihood to measure bias parameters of halos or other tracers in simulations where phases are known, leading to the optimal cancelation of cosmic variance in bias measurements.
\end{itemize}

Finally, the connection to observations will require the incorporation of redshift-space distortions. As shown recently by Ref.~\cite{Cabass:2020jqo}, this is in principle straightforward within the EFT likelihood framework, by transforming the deterministic field $\d_{h,\rm det}$ to redshift space. We leave an implementation of this to future work as well.

\acknowledgments

We thank Alexandre Barreira, Andrija Kosti{\'c}, Minh Nguyen, Marcel Schmittfull and Marko Simonovi{\'c} for helpful discussions, and Franz Elsner and Minh Nguyen for collaboration on previous work leading up to these results. We further thank Titouan Lazeyras for supplying us with the N-body and 2LPT codes used to generate initial conditions for the simulations our results are based on. FS thanks Oliver Hahn, Cornelius Rampf and the OCA, Nice, for hospitality while this work was being completed.

GC and FS acknowledge support from the Starting Grant (ERC-2015-STG 678652) ``GrInflaGal'' of the European Research Council.
GL acknowledges financial support from the ILP LABEX (under reference ANR-10-LABX-63) which is financed by French state funds managed by the ANR within the Investissements d'Avenir programme under reference ANR-11-IDEX-0004-02. This work was supported by the ANR BIG4 project, grant ANR-16-CE23-0002 of the French Agence Nationale de la Recherche.
This work is done within the Aquila Consortium.\footnote{\url{https://aquila-consortium.org}}

\clearpage
\appendix

\section{Deriving the maximum-a-posteriori relation}
\label{app:MAP_review}

\noindent In this appendix we review how to obtain the maximum-a-posteriori relation of \refeqs{maxlikeG}{maxlikeG2}. 
The logarithm of the likelihood is given in \refeq{PcondGFT}. The ``data'' $\delta_h(\vk)$ are obtained for 
a given set of initial conditions $\hat{\delta}_{\rm in} = \hat{\delta}_{{\rm in},\infty}$ (i.e., as discussed in Sec.~\ref{sec:MAP}, 
the initial conditions are not cut at $\Lambda_{\rm in}$).
In the perturbative description, $\delta_h$ is then a linear combination of
renormalized operators $[O]$ constructed from $\hat{\delta}_{{\rm in},\infty}$, 
and a set of bias parameters $\bt_O$. The ``model'' is instead given by Eqs.~\eqref{eq:dhdet}, \eqref{eq:d_fwd_def}, i.e.~ 
\begin{equation}
\label{eq:app-1-a}
\d_{h,\rm det} = \sum_Ob_OO[\delta_\Lambda]\,\,, 
\end{equation}
where 
\begin{equation}
\label{eq:app-1-b}
\d_\L(\vk) = W_\L(\vk) \d_{\rm fwd}\left[ \hat{\d}_{{\rm in},\Lin} \right](\vk)\,\,\quad\mbox{and}\quad 
\hat{\d}_{{\rm in},\Lin} = W_{\Lin}(\vk) \hat{\d}_{\rm in}\,\,. 
\end{equation}
That is, while in the application to real data one must perform an inference of the initial conditions as well, we here instead fix $\d_{\rm in}$ in \refeqs{app-1-a}{app-1-b} to be equal to the set of initial conditions $\hat{\delta}_{\rm in}$ 
we have used to generate the data. Our likelihood then becomes only a function 
of cosmological and bias parameters. Since the bias parameters $b_O$ appear only quadratically, it is straightforward to take the derivative of
the logarithm of the likelihood with respect to them and look for the maximum. The relation one obtains is 
\ba
\fsum{\vk} \frac1{\s^2(k)} \d_h(\vk) O(-\vk) = \fsum{\vk} \frac1{\s^2(k)} \sum_{O'} b_{O'} O'(\vk) O(-\vk) 
\qquad\forall\ O\,\,. 
\label{eq:app-1-c}
\ea 
Importantly, $\fsum{\vk}1/\sigma^2{(k)}$ defines a scalar product given that $\sigma^2(k)$ is strictly nonnegative. 
Together with the fact that Eq.~\eqref{eq:app-1-c} must hold for any shape of the linear power spectrum, this allows us to write the maximum-a-posteriori relation at a fixed $\vk$, i.e.~ 
\ba
\d_h(\vk) O(-\vk) = \sum_{O'} b_{O'} O'(\vk) O(-\vk) 
\qquad\forall\ O\,\,. 
\label{eq:app-1-d}
\ea 
We can then multiply this equality on both sides by the Gaussian prior on the initial conditions and 
functionally integrate over $\hat\delta_{\rm in}$. Using translational invariance, we arrive at Eq.~\eqref{eq:maxlikeG}.

\section{Maximum-a-posteriori relation for \texorpdfstring{$\bm{O=O'=\delta}$}{O=O'=\textbackslash delta}}
\label{app:MAPdelta}

In this appendix, we explicitly compute the left- and right-hand sides of
\refeq{maxlikeG2} for $O=\d$:
\ba
\< \d_\L(\vk) \d_h(\vk') \> = \sum_{O'} b_{O'} \< \d_\L(\vk) O'[\d_\L](\vk')\>\,\,.
\label{eq:MAPdelta}
\ea
As in App.~\ref{app:MAP_review}, we will denote the halo bias parameters on the left-hand side with $\bt_O$, while the
bias parameters in the likelihood on the right-hand side will be denoted as $b_O$. 
In the following, we explicitly include both the cutoff $\L$ on the final density field as well as that imposed on the initial density field, $\Lin$. 
Throughout, we assume $k < {\rm min}(\Lin,\L)$ so that we can set factors of $W_\L(\vk)$, $W_{\Lin}(\vk)$ to unity.

\noindent\textbf{Left-hand side:} 
Using the results from Sec.~4.1 of \cite{biasreview}, and paying attention
to the cutoff in the initial conditions following \refsec{MAP}, we can write
\ba
\< \d_\L(\vk) \d_h(\vk') \>' &= \bt_1 \left[\Plin(k) + P_{mm}^\NLO(k;\Lin,\infty) \right]
\vs
&\ + 2\!\sum_{O'^{[2]}} \bt_{O'}\!\!\int_{\vp} S_{O'}(\vp,\vk-\vp) F_2(\vp,\vk-\vp) \Plin(p) \Plin(|\vk-\vp|) W_{\Lin}(\vp) W_{\Lin}(\vk-\vp)
\vs
&\ 
	+ 4\!\sum_{O'^{[2]}} \bt_{O'}\!\!\int_{\vp} S_{O'}(\vp,\vk-\vp) F_2(\vk,-\vp)
	\Plin(p)\,\Plin(k) \vs
	&\ + \frac25 \bt_{\otd} f_\NLO(k) \Plin(k) - \bt_{\lapl\d} k^2 \Plin(k) \vs
	&\ + \mbox{counterterms}\,\,,
\label{eq:MAPlhs}
\ea
where $\sum_{O'^{[2]}}$ denotes a sum over all bias operators that start at second order, and 
\ba
P_{mm}^\NLO(k;\Lin,\Lin') &\stackrel{\Lin \leq \Lin'}{=} 2 \int_{\vp} \left[F_2(\vp,\vk-\vp)\right]^2 \Plin(p) \Plin(|\vk-\vp|) W_{\Lin}(\vp) W_{\Lin}(\vk-\vp) \vs
&\qquad + 3 \Plin(k) \int_{\vp} F_3(\vp,-\vp,\vk) \Plin(p) \left[W_{\Lin}(\vp) + W_{\Lin'}(\vp)\right]
\ea
is the NLO ($1$-loop) contribution to the cross-correlation between forward-evolved matter density fields with two different initial cutoffs $\Lin,\Lin'$
(and we have assumed $\Lin \leq \Lin'$ without loss of generality).
Thus, $P_{mm}^{\NLO}(k;\Lin,\infty)$ is the NLO cross-correlation between
forward-evolved matter density fields \emph{with} and \emph{without} cutoff in the initial conditions, 
corresponding to the contributions shown in Eq.~\eqref{eq:delta_delta}. 
Further, $f_\NLO(k)$ is defined as 
\ba
f_\NLO(k) =\:& 4 \int_{\vp} \left[\frac{[\vp\cdot(\vk-\vp)]^2}{p^2 |\vk-\vp|^2}-1\right] F_2(\vk,-\vp) \Plin(p)\,\,.
\label{eq:fNLO}
\ea
The terms in the second (third) line of \refeq{MAPlhs} can be identified with
the first (second) term in Eq.~\eqref{eq:D3}. The first term in the fourth line corresponds to Eq.~\eqref{eq:D4}. 

\noindent\textbf{Right-hand side:} extending App.~C of \cite{paperI} by the cutoff in the initial conditions, we straightforwardly obtain
\ba
\sum_{O'} b_{O'} \< \d_\L(\vk) O'[\d_\L](\vk')\> &=
\left(b_1 - \blapl k^2 \right) \left[\Plin(k) + P_{mm}^\NLO(k;\Lin,\Lin) \right]
\label{eq:MAPrhs} \\
&\ + 2 \sum_{O'^{[2]}} b_{O'} \int_{\vp} S_{O'}(\vp,\vk-\vp) W_\L(\vp) W_\L(\vk-\vp) \vs
&\quad \times \Big[F_2(\vp,\vk-\vp) \Plin(p) \Plin(|\vk-\vp|) \vs
&\quad\quad + \{ F_2(\vp,-\vk)\Plin(p) + F_2(\vk-\vp,-\vk) \Plin(|\vk-\vp|) \} \Plin(k) \Big]\,\,.
\nonumber
\ea
The two terms in braces on the last line yield the same result, as can
be see by shifting integration variables $\vp \to \vp' =\vk-\vp$.
Notice that all loop integrals here are regularized, i.e. no modes with
momenta greater than $2\L$ or $2\Lin$ appear.

\noindent\textbf{Residual:} taking the difference of left- and right-hand
sides, and setting $\bt_O = b_O$ (we will see that this applies to all bias parameters except for $\blapl$, which absorbs additional contributions), we have
\ba
\mbox{\refeq{MAPlhs}--\refeq{MAPrhs}} &= b_1 \left[ P_{mm}^\NLO(k;\Lin,\infty) - P_{mm}^\NLO(k;\Lin,\Lin) \right] \vs
&\ + 2 \sum_{O'^{[2]}} b_{O'} \int_{\vp} S_{O'}(\vp,\vk-\vp) F_2(\vp,\vk-\vp) \Plin(p) \Plin(|\vk-\vp|) \vs
&\hspace*{3cm}\times\left[W_{\Lin}(\vp) W_{\Lin}(\vk-\vp) - W_\L(\vp) W_\L(\vk-\vp) \right] \vs
& \
	+ 4 \Plin(k) \sum_{O'^{[2]}} b_{O'} \int_{\vp} S_{O'}(\vp,\vk-\vp) F_2(\vk,-\vp)
	\Plin(p) \vs
	&\hspace*{3cm} \times \left[ 1 - W_\L(\vp) W_\L(\vk-\vp) \right] \vs
	&\ + \frac25 b_{\otd} f_\NLO(k) \Plin(k) \vs
	&\ + \mbox{counterterms}\,\,.
\label{eq:MAPdiff}
\ea
Let us go through the residuals line by line.

\noindent \textbf{(1)} The residual in the first line of \refeq{MAPdiff} scales as
\ba
\mbox{$(1):$}\quad 
\Plin(k)& \int_{\vp} F_3(\vp,-\vp,\vk) \Plin(p) \left[1 - W_{\Lin}(\vp)\right] \vs
&= \int_{|\vp|> \Lin} F_3(\vp,-\vp,\vk) \Plin(p)
\sim \frac{k^2}{\knl^2} \Plin(k)\,\,.
\ea
That is, only modes $|\vp| > \Lin$ contribute. In the last, approximate scaling we have assumed that $k \ll \knl$, which is the wavenumber around which
the integrand peaks, and used the fact that
$F_3(\vp,-\vp,\vk) \propto k^2/p^2$ in the regime where $k \ll p \sim \knl$.
In \refsec{MAP}, we showed that this residual is absorbed by the counterterm
in \refeq{ct1}. The finite contribution $\propto C_s^2$ can be absorbed by
the higher-derivative bias $\blapl$, which at this order is perfectly
degenerate with the $C_s^2$ contribution. Notice that this particular residual
is absent when \emph{not} cutting on the initial modes, but is precisely
absorbed by a counterterm when the cutoff $\Lin$ is employed.

\begin{figure*}[tbp]%
	\centerline{\resizebox{\hsize}{!}{
		\includegraphics*{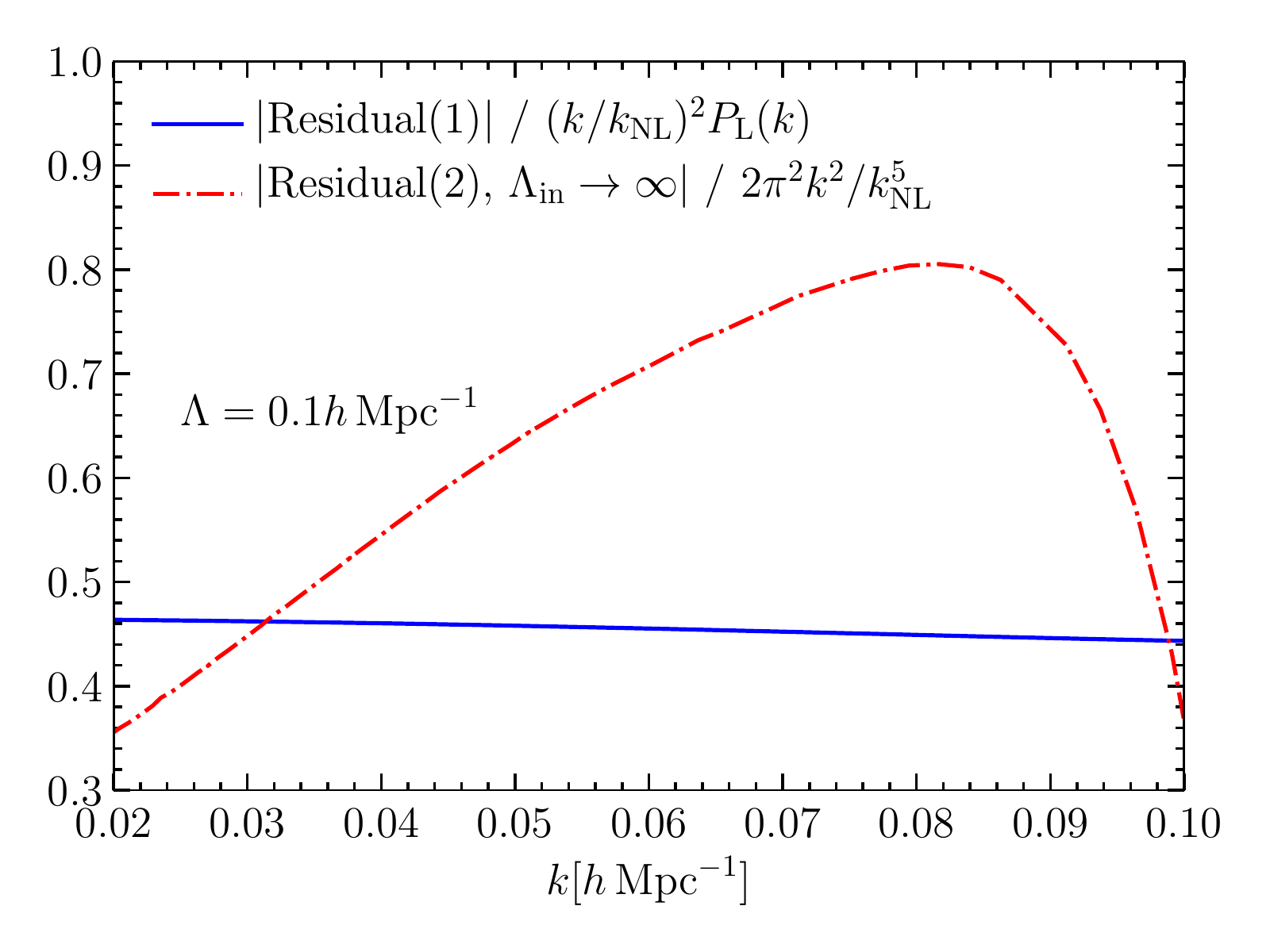}
		\includegraphics*{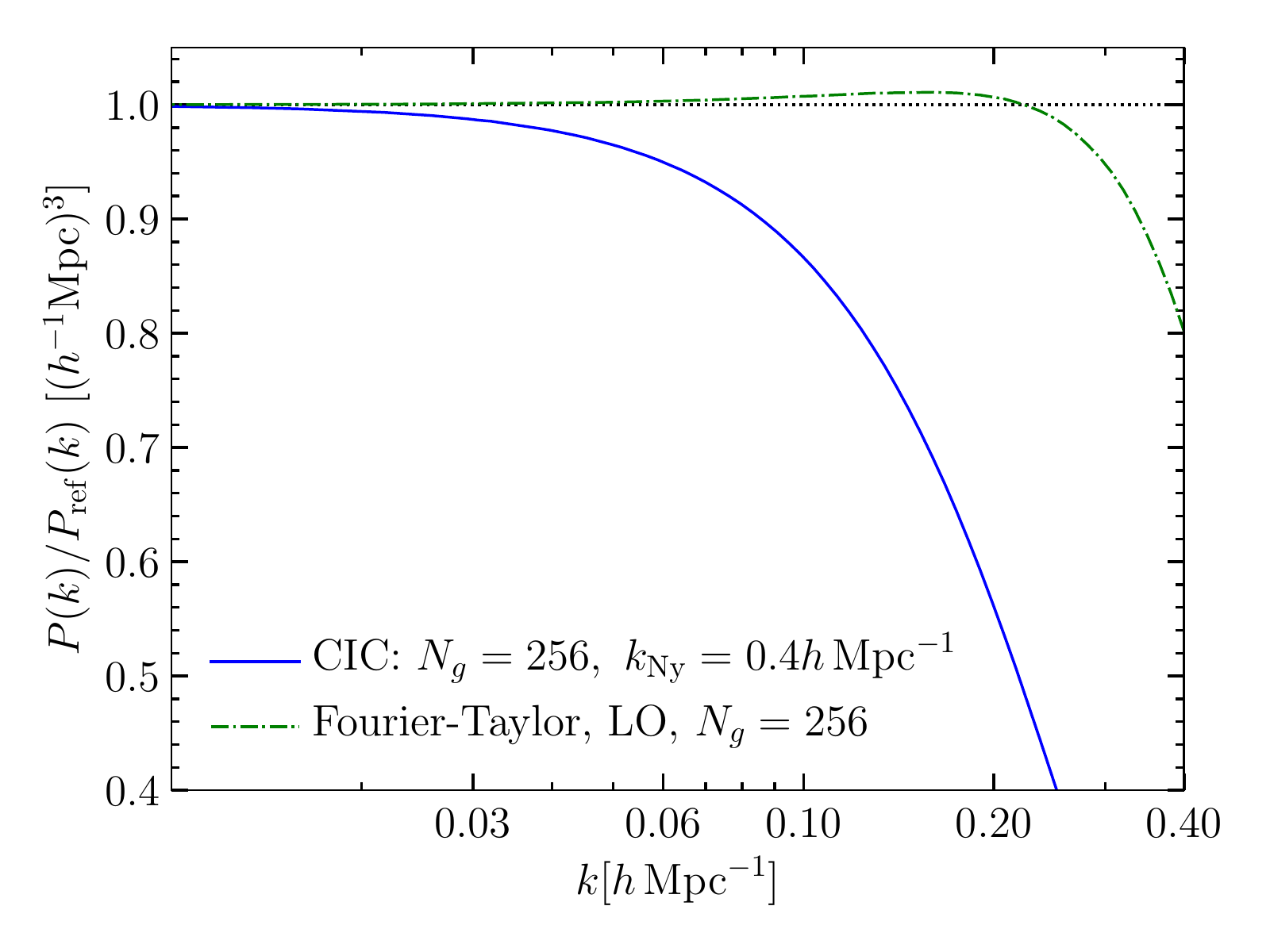}
		}}
	\cprotect\caption{\emph{Left panel:} Residuals in the MAP relation for $O=\d$, \refeq{MAPdelta}. Shown are ratios of residuals (1), for $\Lin=\L$, and (3), for $\Lin=\infty$, to their expected scalings. While residual (3) is smaller numerically, it shows a nontrivial scaling which cannot be absorbed by other counterterms. On the other hand, residual (1) scales as $k^2\Plin(k)$ to within a few percent, and can thus be absorbed by $\blapl$.
	\emph{Right panel:} Effective assignment kernels squared for CIC (solid) and leading-order Fourier-Taylor (dot-dashed), determined by measuring the ratio of power spectra to those from a high-resolution grid. See \refapp{FT} for details.
	\label{fig:residual}}
\end{figure*}

\noindent\textbf{(2)} The next contribution (second and third lines of \refeq{MAPdiff}) is
\ba
\mbox{$(2):$}\quad &
2 \sum_{O'^{[2]}} b_{O'} \int_{\vp} S_{O'}(\vp,\vk-\vp) F_2(\vp,\vk-\vp) \Plin(p) \Plin(|\vk-\vp|) \vs
&\hspace*{3cm}\times\left[W_{\Lin}(\vp) W_{\Lin}(\vk-\vp) - W_\L(\vp) W_\L(\vk-\vp) \right]
\,\,.
\ea
Clearly, this residual vanishes if $\Lin=\L$. On the other hand, if we set $\Lin\to\infty$, corresponding to no cutoff in the initial conditions, then modes with $|\vp| > \L$ or $|\vk-\vp|>\L$ contribute to the residual. Following similar reasoning as above, one roughly expects the result to scale as 
\ba
\mbox{$(2),\ \Lin\to\infty:$}\quad & \int_{|\vp|>\L} \frac{k^2}{p^2} \left[\Plin(p)\right]^2 \sim 2\pi^2 \frac{k^2}{\knl^5}\,\,,
\ea
although this scaling is not actually attained in practice (\reffig{residual}).

\noindent\textbf{(3)} We next consider the third contribution (fourth and fifth lines of \refeq{MAPdiff}),
\ba
\mbox{$(3):$}\quad&
4 \Plin(k) \sum_{O'^{[2]}} b_{O'} \int_{\vp} S_{O'}(\vp,\vk-\vp) F_2(\vk,-\vp)
\Plin(p) \vs
&\hspace*{3cm} \times \left[ 1 - W_\L(\vp) W_\L(\vk-\vp) \right]\,\,.
\ea
This contribution corresponds to the counterterm in \refeq{ct2}.
For $O'=\d^2$, it yields a contribution of $\Plin(k)$ multiplied by a
formally divergent constant, while for $O'=K^2$ there is an additional
term scaling as $k^2 \Plin(k)$ for $k\ll p$, to which the reasoning made
for residual (1) applies. 

\noindent\textbf{(4)} Finally, we have the sixth line of \refeq{MAPdiff},
\be
\mbox{$(4):$}\quad \frac25 b_{\otd} f_\NLO(k) \Plin(k)\,\,.
\ee
This residual remains due to the second-order bias expansion adopted in
this derivation. As argued in \cite{paperI}, this contribution fairly
accurately follows a scaling of $k^2 \Plin(k)$, allowing it to be approximately
absorbed by the higher-derivative bias term $\propto \blapl$ as well. 
Notice that residual (4) is taken into account by the operator $O_\otd$ when going to a
third-order bias expansion, which could partially explain the significantly
improved results obtained for the third-order bias.

To summarize, choosing a cutoff $\Lin$ in the initial conditions introduces
an additional residual, (1), at the MAP level, while it removes the residual (2). Both in terms
of scaling with $k$ as well as absolute size, one might expect residual (1)
($\propto k^2\Plin(k)$) to be more important than residual (3) ($\propto k^2$).
However, the scaling of residual (1) is precisely of the form expected
for the counterterm,
to within a few percent, and can hence be absorbed in $\blapl$ to the same
precision. On the other hand, residual (2) scales nontrivially with
$k$, and cannot be absorbed by a counterterm, as is illustrated in the left panel of \reffig{residual}.
This is due to the different support of the integrand of residual (2) as
compared to the other residuals, as well as cancelations
between modes with $p> \L$ and $p < \L$ which no longer happen when a cut
on $|\vk-\vp|$ is present. These results illustrate the
conclusions of \refsec{MAP}: all UV-sensitive loop integrals that remain
after a cutoff $\Lin$ in the initial conditions is imposed can be absorbed by
counterterms in the bias expansion, while this does not hold in the absence of
the cutoff $\Lin$.

\section{Fourier-Taylor density assignment}
\label{app:FT}

In order to obtain halo and matter density representations from a set of discrete
halo or particle positions $\{\vx_p\}$, a density assignment has to be used.
Since we are interested in obtaining fields that are sharp-$k$ filtered, the
density assignment kernel should be as close as possible to a sharp-$k$ filter, defined
as being precisely unity for modes below the cutoff, and zero otherwise. 
One possibility to achieve this is to first perform an assignment on a high-resolution
grid using a standard kernel such as cloud-in-cell (CIC), and to then apply a
sharp-$k$ filter on the density grid in Fourier space. This is the approach followed
for the results in \cite{paperII}. An alternative, more memory-efficient method
is the Fourier-Taylor assignment scheme presented in \cite{anderson/daleh} (see also \cite{colombi/etal:2009}) which we employ for the results presented in
this paper. 

The exact representation of the Fourier-space density field of a set of point-like
particles is given by
\ba
\d(\vk) &\stackrel{\vk\neq\v{0}}{=} m \sum_p e^{-i \vx_p\cdot\vk}
= m \sum_p e^{-i [\vx_{g,p} + \v{s}_p]\cdot\vk}\,\,,
\label{eq:FT1}
\ea
where $m=N_g^3/N_p$ is the grid mass, given by the number of grid cells $N_g^3$ divided by the number of particles $N_p$, and we only consider nonzero wavenumbers as $\d(\vk=0)=0$. 
In the second equality, we have introduced a finite grid and separated each particle position $\vx_p$ into the
position of the center of the grid cell $\vx_{g,p}$ containing the particle,
and the displacement $\v{s}_p$ of the particle from the cell center.
Since $|\v{s}_p| < r_{\rm cell}/2$, the quantity $\v{s}_p\cdot\vk$ is at most
of order $k/k_{\rm Ny}$, where $k_{\rm Ny}$ is the Nyquist frequency of the grid.
Since, in our application, all Fourier modes above some cutoff $\L < k_{\rm Ny}$
are set to zero (usually we choose $\L$ to be at most $1/3$ to $1/2$ of $k_{\rm Ny}$),
$\v{s}_p\cdot\vk$ is a small quantity for all Fourier modes that are kept.

Hence, we can expand \refeq{FT1} in $\v{s}_p\cdot\vk$ to obtain
\ba
\d(\vk) &\stackrel{\vk\neq\v{0}}{=} m \sum_p e^{-i \vx_{g,p}\cdot\vk} \left[1 - i \v{s}_p\cdot\vk + \frac12 \left(\v{s}_p\cdot\vk\right)^2 + \cdots \right] \vs
&= m \left[\sum_p e^{-i \vx_{g,p}\cdot\vk} - i k_j \cdot \sum_p e^{-i \vx_{g,p}\cdot\vk} s_p^j
+ \frac12 k_j k_l \sum_p e^{-i \vx_{g,p}\cdot\vk} s_p^j s_p^l + \cdots\right]\,\,.
\label{eq:FT2}
\ea
The first term here is just the Fourier transform of the density field obtained by
performing a grid assignment with the nearest-grid-point (NGP) kernel. Similarly,
the coefficient of $-i k_j$ in the second term, $\sum_p e^{-i \vx_{g,p}\cdot\vk} s_p^j$,
is the Fourier transform of a Cartesian vector grid obtained by assigning the components
of the displacements within the cell weighted by the NGP kernel. The third term
correspondingly involves the Fourier transform of a Cartesian tensor grid.
\refeq{FT2} can be implemented efficiently numerically by performing the NGP
assignments in real space, and then summing the terms after transforming all grids
to Fourier space. In our implementation, we keep the zeroth- as well as linear-order
terms in $\vk\cdot\v{s}_p$, and refer to this as ``leading-order Fourier-Taylor'' assignment (while keeping only the zeroth order term in \refeq{FT2} corresponds to standard NGP assignment).

In order to test the grid assignment, we construct the matter density field from
the N-body particle output at $z=0$ and measure its power spectrum. We then take
the ratio of the power spectrum measured in a low-resolution grid ($256^3$) to a
reference result taken from a high-resolution ($1024^3$) CIC-assigned grid.
For a perfect sharp-$k$ filtered density field, the result should be unity for modes
below the cutoff. The result is shown in the right panel of \reffig{residual},
for both CIC and LO Fourier-Taylor kernels, where the Nyquist frequency of the low-resolution grid is $k_{\rm Ny}=0.4\iMpch$. Clearly, the Fourier-Taylor assignment is much closer
to the desired sharp-$k$ filter, with deviations within a few percent up to $0.8 k_{\rm Ny}$. This correspondingly reduces the grid resolution necessary to implement a desired cutoff $\L$, i.e. $k_{\rm Ny} \gtrsim 1.2\L$ is sufficient.

It is worth noting that, for the Fourier-Taylor assignment, the resulting density field is not guaranteed to satisfy $1+\d(\vx)>0$ everywhere. For the application in the EFT likelihood, this is not an issue, since there is no requirement on the positivity of $1+\d$. Similarly, the Fourier-Taylor assignment can be very useful whenever correlation functions of Fourier-space density fields are to be measured \cite{colombi/etal:2009}, and when the data are not regularly sampled \cite{Lavaux:2015tmi}.

\bibliographystyle{JHEP}
\bibliography{bibliography}

\end{document}